\documentclass[a4paper, 12pt]{article}
\usepackage{amsfonts,amsmath,amsthm,amssymb}
\usepackage{graphicx}
\usepackage{authblk}
\usepackage{hyperref}
\usepackage{microtype}
\usepackage{lipsum}
\usepackage[section]{placeins}
\usepackage{afterpage}
\usepackage{capt-of}
\usepackage{pifont}
\usepackage{bm}
\usepackage{pdfpages}
\usepackage{mdframed}
\usepackage{multirow}
\usepackage[ruled]{algorithm2e} 

\usepackage{algorithmic}
\usepackage{multicol}
\usepackage{multirow}
\usepackage{makecell}
\usepackage{rotating}
\usepackage{arydshln}
\usepackage[textsize=tiny]{todonotes}
\usepackage{diagbox} 
\usepackage{dsfont}
\usepackage{soul}
\usepackage{xspace}
\usepackage{bbding}

\DeclareMathOperator*{\argmax}{arg\,max}

\def\RR{{\mathbb R}}
\def\R{{\mathbb R}}

\def \p{{\mathrm p}}

\newcommand{\dikin}{\textsc{dikin\_walk}\xspace}
\newcommand{\vaidya}{\textsc{vaidya\_walk}\xspace}
\newcommand{\john}{\textsc{john\_walk}\xspace}

\usepackage{amsmath}
\usepackage[utf8]{inputenc}
\usepackage{soul}

\newcommand{\OO}{\mathcal{O}\xspace}
\newcommand{\sO}{\widetilde{\mathcal{O}}\xspace}

\newcommand{\volesti}{\href{https://github.com/GeomScale/volume_approximation}{\textcolor{blue}{\texttt{volesti}}}\xspace}
\newcommand{\eigen}{\href{http://eigen.tuxfamily.org}{\textcolor{blue}{\texttt{eigen}}}\xspace}
\newcommand{\boost}{\href{http://boost.com}{\textcolor{blue}{\texttt{boost}}}\xspace}

\usepackage{geometry}
\date{\quad}
\title{Randomized Control in Performance Analysis and Empirical Asset Pricing}
\author[1, 2]{Cyril Bachelard}
\author[3, 4]{Apostolos Chalkis}
\author[4, 5]{Vissarion Fisikopoulos}
\author[6, 4]{\\Elias Tsigaridas}

\affil[1]{\small{Faculty of Business and Economics (HEC)\\ Department of Operations\\ University of Lausanne, Switzerland}}
\affil[2]{QuantArea AG}
\affil[3]{Quantagonia}
\affil[4]{GeomScale org}
\affil[5]{National \& Kapodistrian University of Athens, Greece}
\affil[6]{Inria Paris and  IMJ-PRG\\ Sorbonne Universit\'e and Paris Universit\'e}

\begin{document}
\maketitle

\begin{abstract}

The present article explores the application of randomized control techniques in empirical asset pricing and performance evaluation. It introduces geometric random walks, a class of Markov chain Monte Carlo methods, to construct flexible control groups in the form of random portfolios adhering to investor constraints. The sampling-based methods enable an exploration of the relationship between academically studied factor premia and performance in a practical setting. In an empirical application, the study assesses the potential to capture premias associated with size, value, quality, and momentum within a strongly constrained setup, exemplified by the investor guidelines of the MSCI Diversified Multifactor index. Additionally, the article highlights issues with the more traditional use case of random portfolios for drawing inferences in performance evaluation, showcasing challenges related to the intricacies of high-dimensional geometry.

\end{abstract}

\newpage
\section{Introduction}
\label{sec:introduction}

This paper examines the use of randomized control techniques in empirical asset pricing and performance evaluation from various perspectives, including economic, mathematical, and computational considerations. The investigation addresses the role of controlled chance as a tool for performance evaluation and emphasizes the substantial influence of design on inference. A novel approach to both, the design and application of randomized controls is presented. This new methodology serves to evaluate the performance and risk drivers of systematic investment strategies, i.e., portfolios that exhibit certain quantifiable characteristics by construction. Examples of such inherent characteristics include exposures towards classical factors such as size\footnote{Stocks with lower market capitalization tend to outperform stocks with a higher market capitalization in the future.} \cite{bib:Banz1981}, value\footnote{Stocks that have a low price relative to their fundamental value, commonly tracked via accounting ratios like price to book or price to earnings outperform high-value stocks.} \cite{bib:RosenbergReidLanstein1985}, momentum\footnote{Stocks that have outperformed in the past tend to exhibit strong returns going forward.} \cite{bib:JegadeeshTitman1993} or quality\footnote{Stocks which have low debt, stable earnings, consistent asset growth, and strong corporate governance, commonly identified using metrics like return to equity, debt to equity, and earnings variability.} \cite{bib:AsnessFrazziniPedersen2019}. 

We introduce a way to construct random controls that account for investor constraints, including but not limited to upper bounds, short-selling restrictions, risk limits, and tolerances on factor exposures, which allows to analyze the relationship between portfolio characteristic and portfolio performance within stringent investment guidelines. The suggested approach roots in geometric random walks, a special class of (continuous) Markov Chain Monte Carlo (MCMC) methods. Further, a geometric perspective will help to categorize random controls into one of three groups based on the type of constraints that are to be mapped and on the notion of randomness. Given the portfolio context, a randomized control group is hereafter referred to as a random portfolio (RP). 

RPs are commonly utilized within the domain of performance evaluation, an area closely aligned with the intended application scope we are proposing. In the ensuing discourse, we shall integrate this domain. However, it will become apparent that employing RPs for purposes of performance comparisons is fraught with complexities. From our perspective, insufficient attention has been directed toward these challenges in the past. Conversely, RPs prove valuable for examining the relation between systematic portfolio features and performance within real-world scenarios. In particular, the suggested RP framework based on geometric random walk enables an examination of factor effects within a constrained setup and has the capacity to unveil non-linear and asymmetric relations, if such exist, not only between characteristics and returns but also between characteristics and risks. As such, RPs constitute a novel technique for exploring asset pricing puzzles that is more versatile than classical sorting-based approaches going back to (at least)~\cite{bib:FamaFrench1993}. Further, RPs can be designed to closely resemble a strategy of interest in all aspects, except for one particular characteristic in question, which enables an isolated analysis of the implication of that specific characteristic.

Formally, a RP is a real-valued random vector, denoted $\omega$, 
having a probability density function (pdf) $\pi$ with bounded support $\mathcal{K} \subset \mathbb{R}^n$. $\mathcal{K}$ is shaped by investor-imposed constraints and geometrically forms a convex body (which we will explore in detail later on). The components of $\omega$  
represent the proportions of potential monetary allocations to $n$ investable assets, i.e., the random portfolio weights. 

Our primary interest is on statistics derived from the random weights, i.e., the RP's risk, return and characteristic, where the latter typically comes in the form of a factor score. Performance statistics and characteristics are functions of the random weights and of the coefficients obtained from market observations.

Consider the following example. Let $\theta = (\mu, \sigma)$ be the vector of performance parameters, the mean $\mu$ and standard deviation $\sigma$ of a RP; both are (scalar) random variables. 
At any point in time $t$, we may estimate the parameters as $\mu_t = \langle\omega, r_t\rangle$, where $r_t$ is the vector of sample mean returns and $\sigma_t = \omega^{\top} \Sigma_t \omega$, where $\Sigma_t$ is the sample covariance matrix of securities' returns measured at $t$ over a specific period of stock market history. 

The distribution of $\theta_t$, estimated at a given point in time $t$, serves as a representation of the potential outcomes in terms of the specified performance measures, that could have been achieved over a preceding period, considering a pre-defined universe of investable assets and constraints imposed by investors. This set of counterfactual results and the corresponding probability measure provides a basis for two kinds of inferences. 

An option is to employ the performance distribution of the RP as a benchmark for comparing and statistically testing performance. It has been put forward as the null distribution of performance outcomes under the assumption of no skill, thereby framing performance evaluation as a hypothesis testing problem. This is the typical use case endorsed in the literature, see e.g., \cite{bib:Surz2006}, \cite{bib:Lisi2009}, \cite{bib:KimLee2016}. 
In the sequel, we demonstrate that this case subsumes the bootstrap approach of Efron~\cite{bib:Efron1979}, which appears frequently in the context of performance analysis, as a particular instance of a RP. The advantage of a general RP-based analysis over simple resampling procedures is that the former offers the possibility to incorporate investor constraints. This addresses limitations associated with classical benchmark or peer group comparisons which have faced criticism for their susceptibility to biases, such as the inclusion of investment styles that do not align with the strategy of interest or for constituting portfolios that fall outside the investor's specified constraints (see e.g., \cite{bib:Surz2006}).

The narrative supporting the RP-based performance evaluation approach primarily relies on the following two conjectures: (i) A RP forms an obvious choice of control for performance analysis as, by design, it incorporates no penchant to any investment strategy; any portfolio structure is just as likely to occur as any other, and (ii) For (i) to hold, random weights need to follow a uniform distribution over the domain of feasible allocations. Both claims are flawed; even though they are intuitively appealing and despite their many appearances in the literature (we refer the reader to the literature review in Section~\ref{sec:previous_work}). First, the idea that a RP is "analogous to an enumeration of all feasible portfolios" \cite{bib:KimLee2016} is misleading. Although the domain is defined such as to enclose all feasible portfolios, the probability measure over the domain strongly concentrates over a thin shell of typical allocations (the high-volume area of the domain) under a uniform distribution. This concentration phenomenon arises from the intricacies of high-dimensional geometry, which we will discuss in more details in the sequel. Second, the idea that a RP ``by construction incorporates no investment strategy, bias or skill" \cite{bib:DawsonYoung2003} is subject to interpretation and depends on whether one views the capitalization-weighted or the equal-weighted portfolio as the unbiased and unskilled reference (see the discussion in Section~\ref{sec:review_performance_analysis}).

These two drawbacks lead us to conclude that the use of RP-based statistical significance tests to evaluate the (over- or under-) performance of a particular investment strategy is not appropriate. As a purely descriptive tool, however, we consider RP-based analysis to be valuable, as exemplified in the subsequent Section.

The second use case, a novel concept in our understanding, pertains to exploit a RP for detecting and assessing asset pricing anomalies within real-world scenarios. Specifically, this entails exploring the interplay between factor exposure and the performance of portfolios that adhere to investor guidelines. At a time $t$, the factor exposure of the RP is given by $\gamma_{t,h} = \langle \omega, \beta_{t,h}\rangle$, where $\beta_{t,h}$ represents a cross-sectional vector of factor exposures of the firms in the investment universe. These asset-level exposures are (standardized) regression coefficients obtained by regressing the firm's return series on the series of a factor $f_h$ where the factor is a long-short portfolio formed by sorting stocks according to attribute $h$. 
Alternative to regression coefficients, company-specific characteristics could be used, e.g., fundamentals like a firm's market capitalization or it's balance sheet ratios. In accordance with the procedure of the index provider MSCI\footnote{\url{https://www.msci.com/}}, which serves as the basis for our practical application presented in the empirical part (Section \ref{sec:experiments}), we will sometimes employ the term (factor) \emph{score} to be generic. Ultimately, we are interested in estimating the conditional distribution of $\theta_t | \gamma_{t,h}$, which elucidates the interrelation between portfolio score and performance.

Factor scores can also form part of the constraints which define the RP, especially those not integral to the investigated investment style of interest. This grants the researcher complete control over the management and assessment of potentially confounding variables. In Section~\ref{sec:experiments}, we will employ this approach to examine the persistence of well-studied factor premia in a practical setting.

We would like to stress that our suggested use of RPs is not intended to establish causal relationships. Instead, it aims to identify statistical associations between variables and to evaluate the performance of different investment strategies under various conditions. We therefore pay attention not to confuse between causation and association. The latter is a statistical relationship between two variables, a conditional probability $\mathbb{P}(Y|X)$, while the former means that the exposure produces the effect. Our approach thus distinguishes from the typical application of randomized control in experimental designs like, for instance, pharmaceutical studies, which aim to isolate causal effects by splitting subjects into \emph{treatment} and \emph{control} groups.\footnote{As stressed in \cite{bib:LopezDePrado2022}, illuminating the causal nature of things requires more than just an experimental setup for testing but begins with a theory that outlines the causal mechanism(s) which are experimentally falsifiable.} 
In Section~\ref{sec:experiments}, we emphasize that our use of a randomized control is not to compare treated and non-treated groups, but rather to differentiate portfolios with and without a specific characteristic.

Also, it is not our intention to advocate RPs as a panacea to solve asset pricing puzzles. Rather, we view them as a valuable addition to the arsenal of statistical tools for performance related questions in financial economics. However, it is important to acknowledge the complexities involved in generating RPs that accurately reflect real-world constraints. In a first instance, RPs are theoretical constructs, and the key question is whether one can effectively determine the distribution of performance measures or portfolio characteristics for inferential purposes. In most cases of practical relevance, analytical solutions are not feasible, and we must resort to numerical techniques to obtain the desired results.

\subsection{Contributions}
\label{sec:contributions}


Our contributions within this article are fourfold. 
First, we clarify the inherent challenges associated with the conventional practice of utilizing RPs for performance assessment (Section \ref{sec:use_abuse_of_random_portfolios}). 
Second, we introduce an alternative use case focused on examining asset pricing anomalies within constrained scenarios. This use case is demonstrated in Sections~\ref{sec:use_abuse_of_random_portfolios} and \ref{sec:geometric_random_walks}. 
Third, we provide an extensive exposition of technical methodologies for generating RPs, covering both exact and approximate sampling-based solutions within a common geometric framework (Sections \ref{sec:random_portfolio_generation} and \ref{sec:experiments}). Specifically, we introduce geometric random walks, a class of continuous MCMC methods tailored for high-dimensional constrained scenarios from computational geometry, as an instrument to create a randomized control which is useful to evaluate asset pricing anomalies within constrained setups. We survey all existing random walks, examining the algorithms from both theoretical and practical perspectives, and include complexity results.
Lastly, we present an efficient open-source implementation in C++ with an interface in R (Sections \ref{sec:experiments} and \ref{sec:implementation}), facilitating performance analysis via RPs with ease. With this software, assuming the reader has access to the Wharton Research Data Services (WRDS), our empirical research is fully reproducible.

\subsection{Previous work}
\label{sec:previous_work}

Ever since the claim of the economist Burton Malkiel \cite{bib:Malkiel1973} that “a blindfolded monkey throwing darts at a newspaper’s financial pages could select a portfolio that would do just as well as one carefully selected by experts” the concept of a RP has been used to probe investment skill. Most prominently, the Wall Street Journal's Dartboard Contest, a monthly column published by the business newspaper between 1988 and 2002, put Malkiel's claim to the test by letting their staffers (acting as the allegoric monkeys) literally throw darts at a stock table, while investment experts picked their own stocks, always for a holding period of six months\footnote{Prior to January 1990, the holding period was for one month. The extension to six months was made to alleviate a possible bias from the price pressure resulting from the announcement effect.}. If nothing else, the game added another animal symbolism to the jargon at Wall Street, emblematic in the ongoing debate on active versus passive management and the underlying hypothesis on market efficiency. While the results of the game are not informative, the experimental design, or rather, its deficiencies contain meaningful learnings about the use of a randomized control for research in finance. Several academic studies have examined the game pointing out biases like expert’s tilt towards high risk stocks~\cite{bib:MetcalfMalkiel1994}, low dividend\footnote{Performances were computed on price series rather than on total return series, thus ignoring the effect of dividends and arguably incentivizing professionals to pick stocks with high growth opportunities.} yield stocks~\cite{bib:Liang1999} and high momentum stocks~\cite{bib:PettengillClark2001}.

From a procedural aspect, the dartboard game forms an educative example of what we call a naive RP (a formal definition follows in Section \ref{sec:naive_rps}) and sampling from it boils down to a bootstrap exercise in the spirit of \cite{bib:Efron1979}. Such bootstrap-like use of RPs is very common in the financial literature. One of the earliest reports we could find is \cite{bib:CohenPogue1967} who used a bootstrap-type of RP in the analysis of mutual fund performances. 


Dedicated articles to the topic of RP-based performance analysis are \cite{bib:Surz1994}, \cite{bib:Surz2006}, \cite{bib:DawsonYoung2003}, \cite{bib:Burns2007}, \cite{bib:Lisi2009}, \cite{bib:BillioEtAl2011}, \cite{bib:Stein2014}, \cite{bib:KimLee2016}  and \cite{bib:LeeEtAl2018}. All articles advocate the benefits of RP-based performance evaluations over traditional approaches and implicitely or explicitely promote the idea to view performance evaluation as a hypothesis test. In~\cite{bib:Surz2006} they argue that traditional performance evaluation methods used by the finance community, namely peer group and benchmark comparisons, suffer from inevitable biases and should be replaced by Monte Carlo approaches. In~\cite{bib:KimLee2016} they point towards inefficiencies of simulation-based methods and propose a closed-form expression for the probability distribution of the Sharpe ratio of a uniformly distributed RP. Another analytical procedure based on a geometric algorithm was suggested in \cite{bib:CalesChalkisEmirisFisikopoulos2018} and applied in \cite{bib:CalesChalkisEmiris2021} and
\cite{bib:ChalkisEtAl2021}, 
again imposing a uniform distribution for a otherwise unconstrained long-only RP. To our knowledge, there is currently no prior literature offering guidance on the creation of a RP with a well-defined distribution within a constrained domain.

Concerning geometric random walk methods, which we posit fill the identified gap, their origin traces to a substantial body of literature which has explored sampling methods for generating randomized approximations to the volumes of polytopes and other convex bodies \cite{DyerFrKa91, bib:LovaszSimonovits1990, bib:LovaszSimonovits1993,bib:CousinsVempala2016}. 
Geometric random walk methods possess a distinct lineage compared to the more conventional MCMC methods widely employed in finance, particularly within Bayesian frameworks for approximating posterior distributions. The uniqueness of geometric random walk methods lies in their specialization for approximating distributions characterized by a bounded support.

Geometric random walks have numerous application, including computational biology and medical statistics, where they play an important role in metabolic network studies \cite{bib:TianEtAl2009, bib:Nielsen2016, haraldsdottir2017chrr}. Additionally, geometric random walks prove useful in solving convex programs \cite{Bertsimas04, Kalai06} and mixed integer convex programs \cite{bib:HuangMehrotra2013}. Financial texts that use geometric random walk routines are sparse. In \cite{bib:Chiarawongse2012} they use a geometric random walk algorithm to optimize portfolios under qualitative input. To the best of our knowledge, the only text to use geometric random walk for performance analysis is \cite{bib:BachelardEtAl2023}.

The progress towards algorithms for volume computation, random sampling, and integration has developed deep connections between high-dimensional geometry and the efficiency of algorithms and shaped our understanding of convex geometry. Literature on those aspects include
\cite{bib:Vempala2005}, 
\cite{bib:CousinsVempala2016}, 
\cite{chen18},
\cite{Mangoubi19}. Further important work on geometric random walk sampling include 
\cite{smith84}, 
\cite{bib:LovaszSimonovits1990},
\cite{bib:LovaszSimonovits1993},
\cite{bib:KannanLovaszSimonovits1997} and 
\cite{bib:LovaszVempala2006}. 
References to the source literature of existing geometric random walk routines are given in Section \ref{sec:geometric_random_walks}.\\

The remainder of this article is structured as follows. Section \ref{sec:use_abuse_of_random_portfolios} contextualizes random portfolios within the realm of performance analysis and factor analysis. In Section \ref{sec:random_portfolio_generation}, we delve into the problem of RP generation from a geometric perspective. Following that, Section \ref{sec:geometric_random_walks} explores geometric random walk methods for constructing RPs tailored to address complex real-world scenarios. This Section provides an overview of possible algorithms, accompanied by descriptions of their properties and complexity, with a focus on their programmatic implementation. Subsequently, Section \ref{sec:experiments} presents empirical experiments using a geometric random walk-based RP to investigate the relationship between factor tilts and performance within the framework of the MSCI Diversified Multifactor index, which we consider representative of a typical setup. Finally, Section \ref{sec:conclusion} concludes.

\section{Use and abuse of Random Portfolios}
\label{sec:use_abuse_of_random_portfolios}

\subsection{A (brief) review of performance analysis}
\label{sec:review_performance_analysis}

To characterize RP’s in a performance evaluation and asset pricing context, we briefly review some classical performance analysis methods. 

Performance evaluation is inherently relative. 
The capitalization-weighted benchmark-relative perspective, arguably predominant in both industry and academia, has deep roots in economic theory, e.g., \cite{bib:Sharpe1964}, \cite{bib:Lintner1965}, \cite{bib:Mossin1966}, and forms the blueprint for classical performance analysis. In this context, the available tools try to identify sources of \emph{excess} returns and to attribute them to active bets undertaken by the portfolio manager. 

Holdings-based or transaction-based performance attribution tools in the line of \cite{bib:BrinsonFachler1985} (but applied to equity-only portfolios), building upon the work of \cite{bib:Dietz1966} and \cite{bib:BAI1968}, are widely employed in the industry. This is so because of their explanatory power to simultaneously outline the difference in the allocation structure between a portfolio and a benchmark by grouping stocks into easily interpretable categories like countries, sectors or currencies, and to quantify the individual contributions to overall performance coming from the allocation differences among and within the pre-defined categories. The \emph{determinants} of portfolio performance \cite{bib:BrinsonHoodBeebower1986} are however not to be understood in an etiological sense. Unless the categories, according to which the attribution is done, effectively correspond to deliberate bets undertaken by the portfolio manager which characterize the strategy, allocation and selection effects do not \emph{cause} over- or underperformance. They can be seen as residuals, meaning that the strategy may generate these effects while pursuing other objectives. They can only identify what we may refer to as the "causa proxima", the immediate cause, while the "causa remota", the remote and perhaps indirect cause, which is encoded in the strategy and ultimately leads to the performance delta, remains unexplored. 
For instance, the underperformance of a minimum variance strategy following the 2022 energy crisis might be attributed to the strategy's underweighting of the energy sector. However, it is not an inherent characteristic of minimum variance strategies to avoid investments in the energy sector. Rather, the strategy may not have selected energy-related firms due to the high volatility of stocks in that industry in response to the external shock. Thus, the cause of the performance difference is only superficially due to the sector performance but roots in the underlying variance minimizing mechanism that leads to the tendency of avoiding high-volatility assets regardless of their origin or sector.

Another school of performance evaluation, initiated by \cite{bib:FamaMacBeth1973}, subsumed under the term return- or factor-based models, uses regressions to break down observed portfolio returns into a part resulting from a manager’s ability to pick the best securities at a given level of risk and a part which is attributable to the dynamics of the overall market as well as to that of further risk factors which are recognized to explain security returns and are associated with a positive premium. This allows for an assessment as to whether the active performance is attributable to a particular investment style. 

While the Brinson-type of performance attribution is purely descriptive, the regression setup goes beyond simple performance measurement as it allows for statistical inference in terms of significance testing of the out- or underperformance (alpha) and the loadings on other return drivers (betas) and therefore provides a basis for normative conclusions (assuming that the usual assumptions for linear regression are met).  A skilled manager should be able to beat the market (factor) in a statistically significant manner after controlling for alternative betas. In this sense, skill is reserved to the active manager who is able to harvest a positive return premium not explained by known factors. Any benchmark replicating strategy, since it involves no active deviation from the capitalization-weighted allocation structure in a market, is therefore called passive and comes with performance expectations which should match the performance of a market index rather than trying to outperform it. 

This view, that the average investor possesses no skill, conflicts (at least semantically) with the concept of a naive RP as the null of no skill. This is because a naive RP averages on the equal-weighted portfolio which forms a natural alternative benchmark. From a portfolio selection perspective the equal-weighted portfolio is rightly called the naive benchmark since the strategy implies no views on market developments. It results as the optimal solution in a Markowitz framework when expectations on returns, variances and correlations are considered constant cross-sectionally. Yet, from an asset pricing perspective, equal weighting provides exposure to systematic investment styles, foremost to the low-size factor, and is thus not so naive after all. The lack of consideration of the different points of view has led to unnecessary misinterpretations in the past. Unnecessary because, as we will show, RPs can easily be constructed to reflect either reference points in expectation, the naive investor who bets on all companies in a universe equally, or the average investor who accounts for the the number of shares a company has issued\footnote{For the computation of total market value one may want to restrict calculations to shares that are free-floating, i.e., available to public trading, giving rise to a free-float adjusted market capitalization versus a non-adjusted capitalization considering all shares outstanding, i.e., also those that are held by company insiders and which are restricted to trade.} and the price they are traded for, i.e., firm capitalization (or actually any other reference point that could be relevant to an investor). In the forthcoming, we use the terms naive RP and basic RP to refer to either cases. A formal definition of the two concepts is given in Section \ref{sec:random_portfolio_generation} and a first  educative example follows shortly.


A third and more direct approach of performance analysis comes in the form of hypothesis tests for the difference in performance measures of two strategies; typically, the portfolio in question and a benchmark. For instance, Sharpe ratio tests building upon the work of \cite{bib:JobsonKorkie1981}, the correction of \cite{bib:Memmel2003} and the extension of \cite{bib:LedoitWolf2008} accounting for stylized facts of asset returns. However, at least since \cite{bib:KasakPohlmeier2019} it is known that these tests suffer from low power, that is, they are unlikely to identify superior performance in the data even if there is one.

A randomized procedure for performance analysis replaces the reference point with a reference set (equipped with the empirical probability measure). Using a RP is intuitively appealing as it is a model-free method. In principle, it eliminates the need for parameter estimation, thereby avoiding dependencies on extensive time series data and assumptions about the data generating process. Moreover, it offers flexibility in evaluating different performance measures while considering transaction costs and investment constraints. However, it relies on the assumption that the distribution of no-skill performance, crucial for hypothesis system evaluation, can be objectively approximated through sampling. We contend that such is not unequivocally feasible.

\subsection{RP-based performance evaluation - An example}
\label{sec:example_rp_based_performance_evaluation}

Let us illustrate the challenges of performance evaluation and the use as well as the potential for misuse of RP-based inference through a somewhat absurd, yet genuine, example. Imagine a portfolio manager who constructs a portfolio based on the number of times the letter 'Z' appears in the company names of constituents within the S\&P 500 index. Intuitively, this approach appears dubious and common sense would advise against entrusting this manager with our savings. However, when comparing a backtest of the manager's strategy over a 22-year period (from 2000-01-01 to 2022-12-31) to the capitalization-weighted parent index (i.e., the S\&P 500), the results show an annualized outperformance of $6.2$\% (based on daily geometric returns) and the Sharpe ratio test of Ledoit and Wolf (\cite{bib:LedoitWolf2008})\footnote{The test accounts for time series structures in the data by employing heteroscedasticity and autocorrelation consistent (HAC) estimates of standard error.} rejects the null hypothesis of equal Sharpe ratios at the 5\% tolerance level. What is going on? Is there anything special about the letter Z? Of course not. In fact, it happens to be the case that we could have chosen \emph{any} letter of the alphabet and the simulated out-of-sample\footnote{Our backtesting procedure ensures that at every point in time, only stocks that have been in the index at that point in time enter the portfolio selection (i.e., there is no look-ahead bias). The allocation is then held for three months, letting the weights float with total return (i.e., dividends are assumed to be reinvested) developments of the underlying stocks, until the portfolios are rebalanced.} backtest would have shown an outperformance.  We could even take the nonsense to the extreme and invert the strategies by investing in all assets except those with a particular letter in the company name. Again, all backtests outperform the benchmark. How should we make sense of this? 

Our absurd example is similar to the seemingly paradoxical results presented in \cite{bib:ArnottHsuKalesnikTindall2013} who find that the arguably nonsensical inverses of sensible investment strategies, i.e., strategies built from well-founded investment beliefs, which outperform the capitalization weighted benchmark, outperform even more. The cause is readily identified by the authors by a tilt towards the size and value factors meaning that both, sensible and senseless strategies outperform for the same (unintended) reasons. In particular, even randomly generated strategies, 
i.e., strategies generated from a Monte Carlo simulation (i.e., a RP approach), lead to outperformances for the same causes.
All these findings enforce the authors to conclude that “value and size arise naturally in non-price-weighted strategies and constitute the main source of their return advantage” and that, therefore, “a simple performance measure becomes an unreliable gauge of skill”.  

Also our inane letter strategies can be rationalized by their exposures to priced factors. For instance, when analyzed 
through the lens of the Fama-French-Carhart (FFC) 6-Factor model (\cite{bib:FamaFrench2015}, \cite{bib:Carhart1997}) one finds that the majority of factor loadings are statistically significant. Coefficients are predominantely positive for size and value as well as for the two quality-related factors --profitability and investment-- while all coefficients for momentum are negative. Market betas are, for the most part, below one and as low as $0.82$ for strategy Z (and $0.78$ when not controlling for other factors). However, alphas remain positive and significant for all strategies (at the 5\% level) except for strategy Q. This finding appears robust to the choice of factors, as evidenced by the qualitative consistency of alphas obtained from a regression on the $13$ factor themes suggested by Jensen, Kelly, and Pedersen (JKP) \cite{bib:JensenKellyPedersen2023}. Conversely, betas can differ substantially between the two models. Such discrepancies are to be expected, given the methodological disparities between the FFC and JKP models in constructing factors, utilizing underlying data, and selecting control factors.
\footnote{We found it interesting to study the (cross-sectional) correlation between FFC loadings and JKP loadings as they turn out to be surprisingly low. Over the analyzed period and for constituents of the S\&P 500, the average correlation among exposures under the two models are $0.72$ for SMB and size, $0.36$ for WML and momentum, merely $0.16$ for HML and value and are even negative $-0.08$ RMW and profitability and $-0.02$ for CMA and investment (the correlations are computed cross-sectionally and the average is taken over time).}
Nevertheless, the observed inferential disparity can be confusing, as in various instances, one model attributes outperformance to a positive exposure to a specific factor, while the other attributes the opposite, both with compelling statistical significance\footnote{Naturally, the regression setups should undergo thorough statistical analysis beyond a simplistic reliance on p-values, but unfortunately, such rigor is seldom practiced in reality.} (especially for profitability, investment, and momentum). To arrive at a conclusive understanding, it might be necessary to forsake the convenience of the regression approach and engage in a more intricate analysis of the fundamental data related to the stocks within the letter portfolios\footnote{For example, strategy J shows a clearly negative portfolio size score when calculated directly on the basis of company characteristics (we used the logarithm of market captitalization, standardized cross-sectionally to have mean zero), while the size exposure is positive under both FFC and JKP. For a long-only portfolio to have an exposure to a long-short factor does not necessarily mean that this translates to the portfolio having the corresponding characteristic (a portfolio with strong size exposure may nevertheless contain very large companies).}.

\begin{figure}[h!] 
    \centering
    \includegraphics[width=1\linewidth]{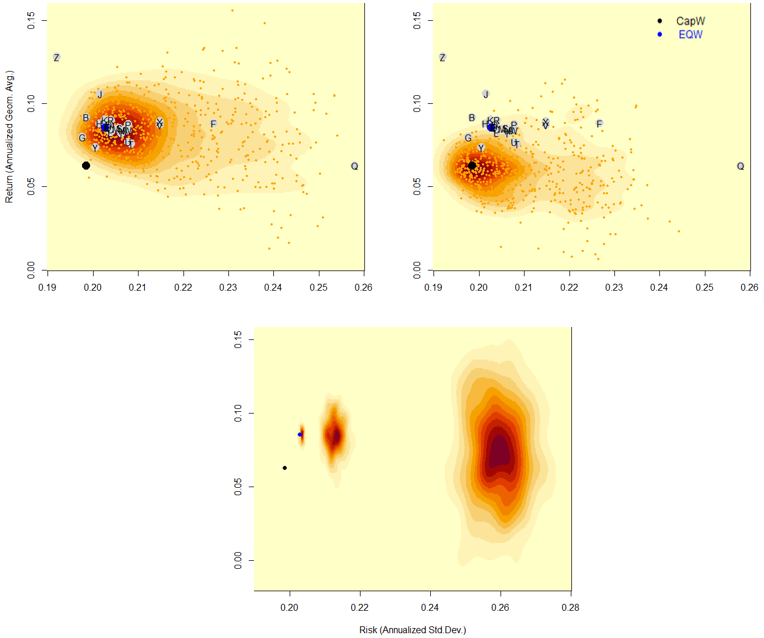}
    \caption{Risk-return distribution of different RPs (red to yellow level sets) together with the corresponding statistics for the 26 letter strategies (grey dots), realizations of the RPs (yellow small dots), the capitalization-weighted benchmark given by the S\&P 500 index (black dot) and the equal-weighted portfolio of the index constituents. In the top charts, RPs are constructed to center on the equal-weighted (left chart) and capitalization-weighted benchmark (right chart). The bottom chart displays the distributions of three RPs created through the dartboard game approach, selecting $4$, $30$, and all stocks from the investable universe from right to left.
    \label{fig:rp_risk_return_densities_alphabet_strategies}}
\end{figure}

Instead, let us explore whether an RP-based approach can provide clarity. The top left chart in Figure~\ref{fig:rp_risk_return_densities_alphabet_strategies} shows the bivariate risk-return distribution of a RP (red to yellow level sets) together with the corresponding statistics for the 26 letter strategies (grey dots), realizations of the RP (yellow small dots), the capitalization-weighted benchmark given by the S\&P 500 index (black dot) and the equal-weighted portfolio of the index constituents, rebalanced monthly (blue dot). 

\begin{figure}[h!] 
    \centering
    \includegraphics[width=1\linewidth]{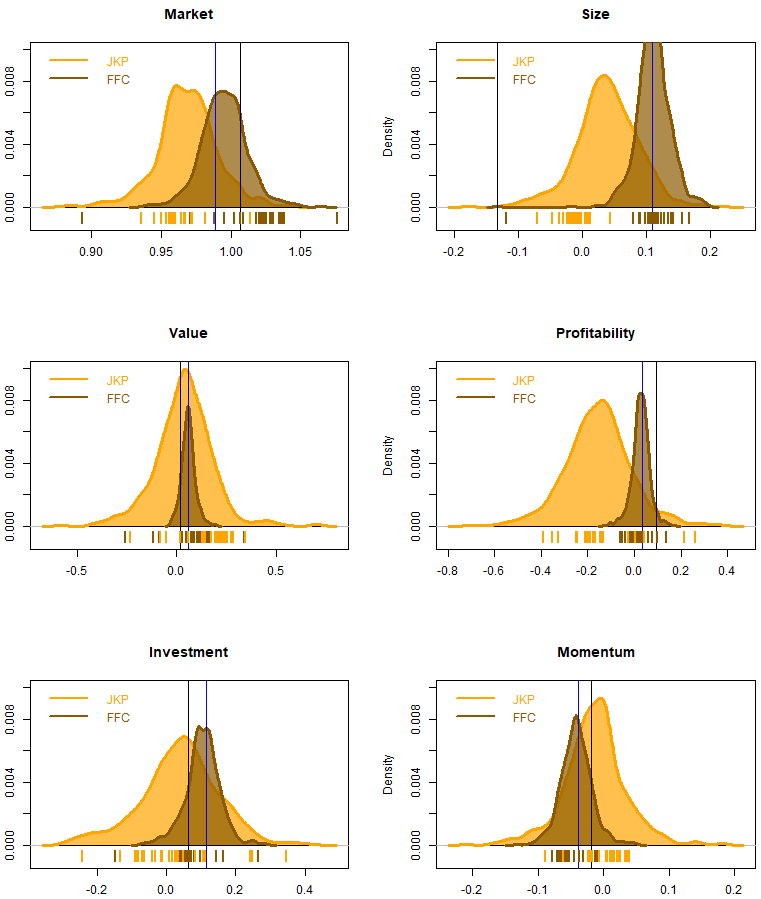}
    \caption{{\small Factor exposures
    \label{fig:factor_exposure_densities}}}
\end{figure}

Two immediate observations arise: Firstly, the performances of the letter strategies fall within the high-mass area of the RPs pdf, with Z and Q being outliers. Secondly, most of the mass of the RP return distribution is at levels above the return of the capitalization-weighted benchmark. The first observation suggests that any apparent performance or skill is consistent with chance and that the letter strategies could very well be instances of the RP. Analyzing the factor composition of the RP samples further reinforces this intuition. Figure~\ref{fig:factor_exposure_densities} shows the distributions of factor scores of RP samples under the FFC and the JKP models. The marks above the x-axis display the corresponding exposures of the letter strategies. While the divergence of the two models poses some concern, the consistency of the letter strategy scores with the RP distribution under either model individually paints a rather clear picture, which is that the outperformance of the letter strategies may be attributed to factor exposures that happen to be prevalent in the market. By contrast, the factor structure of the capitalization-weighted benchmark can be far-off the bulk of the RP's exposure distributions (this is evident for size where the benchmark must have a negative exposure by construction).

Unlike the letter-based performances, which mostly align with the outcomes of the RP, the capitalization-weighted benchmark exhibits a notable deviation. Here is where one has to be cautious not to jump to conclusions. If we were to statistically test the benchmark return based on the RP return distribution we would conclude that the average investor is rather unskilled since most investment decisions based on chance outperformed the market. This result would be in line with the findings of \cite{bib:ArnottHsuKalesnikTindall2013} as well as \cite{bib:ClareMotsonThomas2013} who locate the return of the capitalization-weighted benchmark far out in the left tail of a RP return distribution over a $49$ years (1964 - 2012) and $44$ years (1968 - 2011) period respectively. However, one should recall Sharpe's arithmetic of active management \cite{bib:Sharpe1991} that for every outperforming strategy there needs to be an underperforming one (relative to the capitalization-weighted benchmark). The fact that the RP outperforms the benchmark does not imply that any strategy, or even most strategies, outperform the benchmark. The outcome depends on the way the RP is constructed, circling back to the two previously mentioned issues in the RP narrative regarding the no-skill reference point and the misunderstanding that an RP covers the entirety of outcomes. 

Without delving too deeply into the subject of RP construction (a concise description follows in Sections \ref{sec:random_portfolio_generation} and \ref{sec:geometric_random_walks}), some qualitative aspects need addressing here. The RP depicted in the top left chart of Figure~\ref{fig:rp_risk_return_densities_alphabet_strategies} is not constructed as having a uniform distribution over the set of feasible assets. Such an approach would yield an inadequate representation of attainable performance, as it would consist solely of homogeneous and biased portfolios. Instead, it is crafted by convolving $26$ individual RPs, each centered on an equally-weighted portfolio while considering the number of assets in the specific letter portfolio at a given rebalancing date. 

As the number of assets considered decreases, the risk-return distribution of the RP widens, and the distribution's center shifts right, as depicted in the bottom chart of Figure~\ref{fig:rp_risk_return_densities_alphabet_strategies}. The chart displays three RP densities, constructed using a method commonly found in the literature. The procedure involves repeatedly sampling $m$ stocks uniformly from the set of investable stocks (here, with a set cardinality of 
$n = 500$) and forming equally-weighted portfolios. This process essentially constitutes an 
$m$-out-of-$n$ bootstrap. In the chart, we have used $m=\{500, 30, 4\}$ from left to right.

In the first case where all assets in the index are used, the density merely provides a measure of the uncertainty in the performance of the equal-weighted strategy\footnote{It can be understood as an estimate of the sampling distribution of the equal-weighted portfolio's performance statistics, or, if one would sample the asset space uniformly, as the posterior density of the risk-return parameters of the equal-weighted portfolio.}. This is because sampled portfolios are all very close to the equally weighed portfolio (in terms of the weights). In higher dimensions, sampling according to a uniform distribution does not mean that each possible portfolio will be sampled with equal probability. Quite the contrary. With overwhelming probability, samples will be drawn from the high-volume area of the sampling space, which forms a thin shell (the region between two concentric higher dimensional spheres of differing radii)\footnote{It can be shown that $\mathds{E}(||\omega||_2^2) = \frac{2n-1}{n^2}$ and that therefore, with $\bar{\omega}$ denoting the equal-weighted portfolio, $\mathds{E}(||\omega - \bar{\omega}||_2^2) = \frac{n^2-n}{n^3}$, which tends to zero with increasing $n$.}. Sparse or highly concentrated portfolios (such as the capitalization-weighted benchmark) are thus extremely unlikely to be sampled. 

In cases where $m < n$, as exemplified by $m=30$ and $m=4$, chosen to align with the selections made by \cite{bib:ArnottHsuKalesnikTindall2013} and in the dartboard game, respectively, the imposed sparsity leads to a broader dispersion of portfolio weights. This translates to a more extensive distribution of performance statistics. However, the resulting distributions remain relatively compact, covering only a small area of the space of feasible solutions, refecting the homogeneity of sampled portfolios. Therefore, such RPs do not form acceptable control groups. Utilizing these distributions for significance testing would likely result in overly optimistic rejections of the Null hypothesis.

Adapting the RP mechanism to center on the capitalization-weighted allocation leads to the risk-return distribution in the top right chart of Figure~\ref{fig:rp_risk_return_densities_alphabet_strategies}. This distribution is more consistent with Sharpe's arithmetic, having mode close to the performance of the market index. In contrast, the profile of the naive index significantly deviates from the center of mass. We could have further expanded or contracted the risk-return distribution by adjusting the variance in the weights, essentially reverse-engineering the random control to achieve a desired statistical test result. A dangerous game. We therefore conclude that a RP, as we have presented it so far, does not provide a statistically acceptable experimental design to probe skill (or lack thereof). Nevertheless, we would argue that, with some care in the construction of the RP, it can serve as a visual aid to characterize the dispersion of performance for non-elaborate strategies (such as letter-based investing). Hopefully, the geometric perspective which we advocate in the subsequent discussion offers some transparency in this context.

The utility of employing an RP lies in the ability to analyze the relationship between portfolio performance and portfolio characteristics. Let's consider the frequency of a letter in a company's name as a stock characteristic and examine whether there is a correlation between the returns of portfolios drawn from the RP and the characteristics of those portfolios\footnote{For each occurrence of a specific letter in the company name, a score of one is assigned (i.e., a company having letter 'A' appearing three times in it's name gets a score of three). The portfolio characteristics are computed as the weighted sum of stock-level characteristic-scores times the portfolio weights.}. If exposure to a particular letter, let's say 'Z', were a rewarded characteristic, this relationship should be reflected in the cross-section of RP samples as a correlation (or more broadly, a relation) between the loading on 'Z' and performance. This (cor-) relation should be present in both RPs used above, the one centered on the equal-weighted and the one centered on the capitalization-weighted benchmark. While there are simpler ways to dismiss letter-based weighting as a viable method for portfolio construction, our point with current example is that the RP approach lends itself to an analysis of the performance impact of any systematic portfolio formation approach. In Section \ref{sec:experiments}, we build on this approach to investigate recognized factors and their impact on performance. The analysis aims to detect genuine anomalies and reveal whether certain characteristics are still rewarded in the market when accounting for constraints that many investors need to adhere to.

\begin{figure}[h!] 
    \centering
    \includegraphics[width=0.8\linewidth]{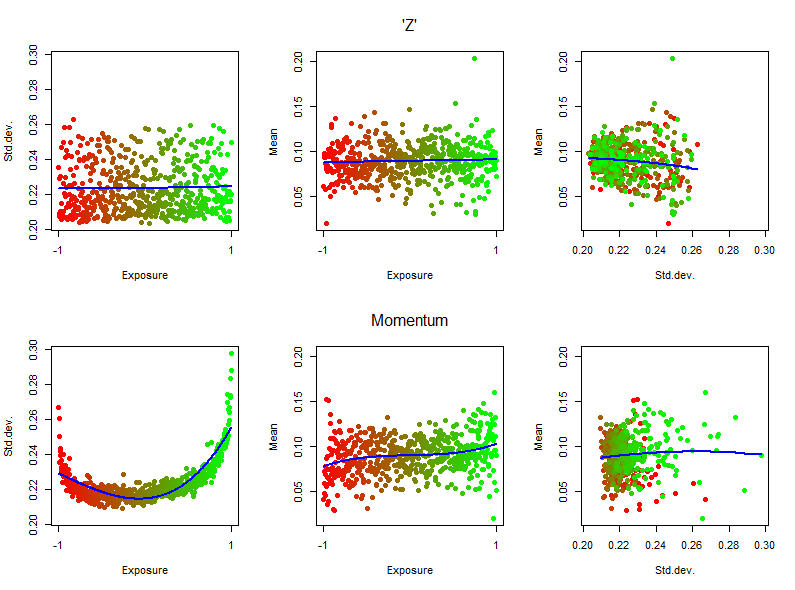}
    \caption{{\small The relation between portfolio performance and exposure towards towards the letter 'Z' in the company names of the portfolio constituents (first row) or towards the momentum factor (second row).
    \label{fig:Z_vs_Momentum}}}
\end{figure}

To finalize our example, Figure ~\ref{fig:Z_vs_Momentum} shows the scatter plot of performance versus characteristics of samples from the RP visualized in the left plot of ~\ref{fig:rp_risk_return_densities_alphabet_strategies}. Unsurprisingly, there is no observable correlation for characteristic 'Z', neither between return and characteristic nor between risk and characteristic. In contrast, the right panel of Figure \ref{fig:Z_vs_Momentum} illustrates the existence of a fairly strong relationship between momentum\footnote{we measure momentum by the cumulative return of the portfolio over the last 12 months, omitting the last month.} and performance at the portfolio level. The chart shows a non-linear (almost quadratic) pattern between risk and momentum. This implies that both, portfolios with the lowest and portfolios with the highest momentum scores exhibit high ex-post volatility, while volatility is low for intermediate-scoring portfolios. The correlation between portfolio's returns and momentum scores is positive (0.31) overall, with steeper segments at the low- and high-exposure endpoints, while the middle part appears to be relatively flat. The graph prompts us to contemplate whether the momentum anomaly would endure in a constrained setting where extreme exposures are no longer feasible due to constraints on asset weights and/or on additional factor exposures. This question is the focus of Section~\ref{sec:experiments}. Additionally, the Section will provide a detailed description of the backtesting methodology, which we have also applied here. For completeness, the two charts in the third row of Figure ~\ref{fig:Z_vs_Momentum} shows the risk-return patter for the RP samples. The color-coding from red (low exposure) to green (high exposure) helps to identify the unstructured and the partially structured nature of the Z-characteristic versus the momentum characteristic, respectively.\\

\section{Random Portfolio Generation}
\label{sec:random_portfolio_generation}

In this Section, we explore the design and construction of a RP. We propose a categorization of RPs into three groups of increasing complexity: naive, basic, and regularized. Among the regularized RPs, we distinguish between simply regularized (where sampling is easy) and generally regularized (where sampling is difficult).  The primary objective is to gain insights into the distribution of function values of a RP, such as its return. In certain cases, these distributions have an analytic expression. However, in general, analytic solutions do not exist and sampling-based approximations are necessary.

\subsection{Naive Random Portfolios}
\label{sec:naive_rps}

The arguably most intuitive case is to consider non-negative portfolio weights that sum up to one; these are the two defining properties of long-only portfolios. The geometric object which characterizes the space of long-only portfolios is the canonical simplex 
\begin{equation}
\label{eq:simplex_equation}
\mathcal{S} := \{ x \in \R^n \ |\ x_i \geq 0, \sum\nolimits_{i=1}^n x_i = 1, i=1,...,n \}.
\end{equation}

In practice, mutual funds frequently encounter legal constraints mandating full investment while prohibiting short selling. Consequently, the simplex constraint embodies a legal requirement to which they must adhere. Given its prevalence within the industry, our analysis will concentrate on this typical scenario, presuming the asset space to be defined by the standard simplex or a subset thereof, which arises when additional constraints are imposed. However, the geometric random walk routines which we propose in the sequel are by no means limited to the simplex case but readily extend to accommodate long-short applications. 

As for the definition of randomness, we begin with the (arguably) most intuitive case of a uniformly distributed RP, i.e., we focus on the uniform measure over the simplex. Hence, a naive RP may be given by $\omega \sim \mathrm{U}(\mathcal{S})$, which we parametrize it through a flat%
\footnote{Flat means that the elements of the parameter vector of the Dirichlet distribution, $\alpha$, are all equal to one.}
Dirichlet model; i.e., $\omega \sim \mathcal{D}(\alpha = \mathbf{1})$, where $\mathbf{1}$ is a vector of ones. The Dirichlet pdf is given by
\begin{equation}\label{eq:dirichlet_model}
    \mathcal{D}(\omega; \alpha_1,\dots,\alpha_n) \propto \prod_i \omega_i^{\alpha_i - 1},\  \text{where }\omega_i > 0,\ \sum_i \omega_i =1.
\end{equation}

This particular setup allows for an exact characterization of the distribution of linear statistics associated with a RP. For instance, the cumulative distribution function (cdf) of the return of a RP, denoted by $\mu$, has an expression as a ratio of volumes. I.e.,
\begin{equation}
\label{prop:exact_bb}
\mathds{P}(\mu \leq \gamma) = \frac{V(\mathcal{S} \cap \mathcal{H}(r_t, \gamma))}{V(\mathcal{S})}, 
\end{equation}
for some scalar $\gamma$, 
where $V$ denotes the volume operator and $\mathcal{H}(r_t, \gamma) = \{ h \in\mathbb{R}^n| r_t^{\top} h \leq \gamma \}$ is the half-space induced by a hyperplane with normal $r_t$ that is the empirical mean vector of the asset return distribution (see \cite{bib:Bachelard2024}).

The volume in the numerator of Eq.~\eqref{prop:exact_bb}
has a closed form expression, e.g., \cite{bib:Lasserre2015}. However, it has been observed that the evaluation of this expression is numerically unstable when the dimension is not small, $n \geq 20$, e.g., \cite{bib:CalesChalkisEmirisFisikopoulos2018, bib:CalesChalkisEmiris2021}.
To overcome this obstacle Chalkis et al \cite{bib:CalesChalkisEmirisFisikopoulos2018, bib:CalesChalkisEmiris2021}
suggest the use of an efficient and exact geometric algorithm,  
due to Varsi, see \cite{bib:Varsi1973} and \cite{bib:Ali1973},
for the evaluation of the volume(s). 
This is the preferable implementation strategy of naive RP applications. 
For the sake of convenience, we include the pseudocode for Varsi's algorithm in Appendix~\ref{sec:appendix_varsi}. 

Another form of a naive RP is given by the mechanism underlying the dartboard game. Dart throwing monkeys are easily digitalized by drawing counts $c = (c_1, ..., c_n)$ from a Multinomial distribution $\mathrm{Mult}(n, \p_1, ..., \p_n)$ with $n$ trials where, with uniform probabilities $\p_i = 1/n$, $i = 1,...,n$, which is exactly the setup underlying the classical bootstrap of \cite{bib:Efron1979}. Normalizing each count produces (RP) weights $\omega_i = c_i / n$. Unlike the Dirichlet model, the weight space is no longer given by the canonical simplex but consists of a discrete grid over the canonical simplex. As a result, the bootstrap distribution of any derived RP statistic is discrete. The grid node at the centroid of the simplex represents the equal-weighted portfolio and vertices represent single-asset portfolios. Grid nodes located on the boundary of the simplex form sparse portfolios, i.e., allocations where one or more position is exactly zero. 

The definition of randomness of the weights depends on the type of resampling plan one chooses in the bootstrap procedure. For instance, the specifications of the dartboard game, i.e., how many darts, $m$, are thrown at a list of how many company names, $n$, and whether a name can be hit multiple times or not (i.e., sampling with or without replacement), define the structure of the corresponding grid. In practical applications, resampling plans often involve drawing fewer than $n$ samples ($m < n$), corresponding to an $m$ out of $n$ bootstrap \cite{bib:BickelGotzeVanZwet1997}. In this case, the induced grid is biased toward the boundaries of the simplex, resulting in a wider dispersion of the bootstrap distribution of the statistic in question compared to the classical bootstrap RP version. 
The continuous analogue of this approach involves using a concentration parameter in the Dirichlet distribution such that $\sum_{i}^n \alpha_i < n$. By introducing a mapping $\lambda(n, m) = \frac{m-1}{n}$, the standard errors of the linear statistic of interest become equivalent under both $\omega \sim \frac{1}{n} \mathrm{Mult}(m, \p_1, ..., \p_n)$ and $\omega \sim \text{Dir}(\alpha \lambda(n, m))$ (see \cite{bib:Bachelard2024}). Notice that the flat Dirichlet model also describes a form of bootstrap version, namely the Bayesian bootstrap of \cite{bib:Rubin1981}.


To summarize, a \emph{naive} RP conforms to a classical or a Bayesian bootstrap scheme encompassing the $m$-out-of-$n$ bootstrap, where $m$ can be smaller, larger, or equal to $n$, and the case $\mathcal{D}(\alpha = \mathbf{1} \lambda)$, $\lambda > 0$ in the Bayesian paradigm\footnote{Notice that $\lambda > 1$ leads to a more concentrated distribution around the centroid of the simplex. On the other hand, $\lambda < 0$ pushes weights outwards towards the faces, edges and vertices of the simplex (that is compared to uniform density implied by the flat base case $\lambda = 1$).}. We call it naive because, in expectation, it recovers the naive benchmark, i.e., $\mathds{E}(\omega) = \bar{\omega}$, where $\bar{\omega}$ denotes the equally-weighted portfolio.



\subsection{Basic Random Portfolios}
\label{sec:basic_rps}

We call a RP basic if, like a naive RP, the geometric representation corresponds to a standard simplex. However, we also impose the condition that $\mathds{E}(\omega)$ is not equal to the centroid, i.e., one cannot recover the (naive) equally-weighted portfolio in expectation. This occurs whenever some form of asymmetry is introduced in the parametrization of the Dirichlet or the Multinomial model which breaks the symmetry in the distribution of the weights. By that we mean that paremeter elements cannot all be equal. For instance, setting $\alpha \propto \omega_{bm}$, where $\omega_{bm}$ is the capitilization-weighted allocation, generates a distribution of weights which has center of mass at $\omega_{bm}$. Although the full distribution of a linear RP statistic is no longer available in exact form under the non-flat parametrization, the central moments of the distribution still are (see \cite{bib:Bachelard2024}).


\subsection{Regularized Random Portfolios}
\label{sec:regularized_rps}

The simplex condition is typically not the only constraint that asset managers need to adhere to. When there are additional constraint,  it becomes quite challenging to obtain analytical results. These additional constraints often originate from regulatory requirements and are designed to limit the risk exposure to individual security issuers or groups of issuers\footnote{A common example is the UCITS 5/10/40 rule, which restricts single asset representation to no more than 10\% of the fund's assets and limits holdings exceeding 5\% to aggregate below 40\% of the fund's assets.}. As a result, upper bounds are imposed on the asset weights, either individually or collectively. 
Geometrically, these linear restrictions correspond to halfspaces that intersect the simplex, resulting in a polytope 
\begin{equation}
\label{eq:polytope_equation}
    \mathcal{P} := \{x\in\R^n\ |\ Ax\leq b\} \text{, for some }\ A\in\R^{m\times n} \text{ and } b\in\R^m .
\end{equation}

Furthermore, managers may impose additional constraints to prevent concentration, or to align with the benchmark allocations, e.g., by setting lower and upper bounds on country or sector exposures relative to the benchmark or in terms of variation of the return difference (tracking-error), to limit transaction costs, to control portfolio characteristics (e.g., factor exposure, sustainability criteria, risk metrics), or to satisfy other specific requirements.
From a sampling perspective, the mathematical characterization of constraints plays a crucial role, distinguishing between linear, quadratic, convex, non-linear, and other types. Certain risk measures like Value-at-Risk (VaR) are non-convex, meaning that, if a RP is subject to a maximum VaR constraint, it’s domain can no longer be represented by a common convex geometric body, which makes sampling extremely hard. Other measures of risk like variance or tracking error are quadratic and geometrically form ellipsoids 
\begin{equation}
\label{eq:ellipsoid_equation}
    \mathcal{E} := \{x\in\R^n\ |\ x^TEx \leq c \}, \text{ for some } E\in\R^{n \times n}, \text{ with } E\succeq 0, \text{ and } c\in\R_+ .
\end{equation}

Sampling from (the surface) of an ellipsoid is relatively straightforward. However, when the simplex condition must also be satisfied, sampling from the intersection of the simplex with an ellipsoidal surface becomes a non-trivial task (albeit possible, as demonstrated in \cite{bib:BachelardEtAl2023}).

Under general constraints, the cdf of a RP statistic is not tractable analytically and one has to turn to numerical methods (see Section \ref{sec:geometric_random_walks}). The following special case forms an exception.
Recently, \cite{bib:Bachelard2024} demonstrated that Varsi's algorithm can also be applied in the context of a shadow Dirichlet distribution (\cite{bib:FrigyikGuptaChen2010}), which is defined on a linearly constrained simplex. The shadow Dirichlet model allows for the consideration of linear regularizations of the weights, making it highly relevant in practical applications. If $\omega \sim \mathcal{D}(\alpha)$ and $M$ an $n \times n$ left-stochastic matrix (i.e., each column sums to one) of full rank, then $M \omega =: \tilde{\omega} \sim \mathcal{SD}(M, \alpha)$, where $\mathcal{SD}$ denotes the shadow Dirichlet model with pdf
\begin{align*}
    \mathcal{SD}(\tilde{\omega}, \alpha) &= \frac{1}{|\det(M)| \,\mathrm{B}(\alpha)} \prod_{i=1}^n (M^{-1}\tilde{\omega})_i^{\alpha_i-1}.
\end{align*}

The normalizing constant is the determinant of matrix $M$ times the normalizing constant of the standard Dirichlet distribution, i.e., the multinomial beta function $\mathrm{B}(\alpha) = \int_{\mathcal{S}_{n-1}} \prod_{i=1}^n \omega_i^{\alpha_i-1} d\omega = \frac{\prod_{i=1}^n \Gamma(\alpha_i)}{\Gamma(\alpha_0)}$, where $\Gamma(x) = \int_0^{\infty} t^{x-1} e^{-t} dt$ is the gamma function.

Therefore, in cases where linear constraints can be expressed using the mapping $M$, the distribution of linear functions of a random portfolio can be precisely obtained using Varsi's algorithm under the uniform measure. These cases are referred to as simply regularized. For all other cases where linear constraints cannot be directly accommodated, sampling methods need to be employed.

\begin{figure}[h] 
    \centering
    \includegraphics[width=0.8\linewidth]{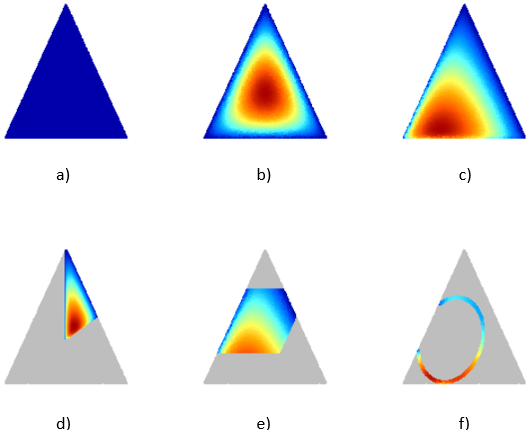}
    \caption{a) naive RP with a uniform density defined over the unit simplex, $\omega \sim \mathcal{D}(\mathbf{1})$.
    b) naive RP with a concentrated Dirichlet model, centered on the naive $1/n$ portfolio, $\omega \sim \mathcal{D}(\mathbf{1}\lambda)$, where $\lambda = 4$ (or $\lambda < 1$ for more mass towards the boundaries). 
    c) Basic RP with weights following a Dirichlet distribution, $\omega \sim \mathcal{D}(\alpha)$, where $\alpha = (0.5, 0.3, 0.2)$.
    d) Shadow of the Dirichlet distribution in c), denoted as $\omega \sim \mathcal{SD}(M, \alpha)$, where the weights are restricted by a monotonic matrix $M$ with a specific column structure. The constrained weights satisfy the ordering $\omega_1 > \omega_2 > ... > \omega_n$.
    e) Generally regularized RP defined by the intersection of the simplex with a polytope, resulting in a truncated distribution of the weights.
    f) Generally regularized RP defined by the intersection of the simplex with the boundary of an ellipsoid, also leading to a truncated distribution of the weights.}
    \label{fig:simplex_densities_2d}
\end{figure}

Figure \ref{fig:simplex_densities_2d} visualizes the different types of RPs for the case of $n=3$ assets. Subplot a) shows the uniform density of a naive RP defined over the unit simplex, i.e., $\omega \sim \mathcal{D}(\mathbf{1})$. Subplot b) also shows another naive RP since, like a), it also centers on the naive $1/n$ portfolio. However, the distribution of the weights is not uniform but follows a concentrated Dirichlet model with $\omega \sim \mathcal{D}(\mathbf{1}\lambda)$; $\lambda = 4$, giving the distribution more mass around the center. If we would have chosen the concentration parameter $\lambda < 1$, the density colors would be reverted (i.e., distributing more mass towards the boundaries of the simplex). Subplot c) shows a simple RP with $\omega \sim \mathcal{D}(\alpha)$; $\alpha = (0.5, 0.3, 0.2)$. d) visualizes the \emph{shadow} of the Dirichlet distribution in c), i.e., $\omega \sim \mathcal{SD}(M, \alpha)$, restricted by a mononotic $M$ which has $k-$th column $[0$ ... $0$ $1/(n - k + 1) . . . 1/(n - k + 1)]^{\top}$. The so constrained weights satisfy the ordering $\omega_1 > \omega_2 > ... > \omega_n$. Subplots e) and f) show instances of generally regularized RPs defined by the intersection of the simplex with a polytope in the former case, and by the intersection of the simplex with the boundary of an ellipsoid in the latter case. In both cases, the distribution of the weights is a truncated version of the basic RP in c).

\subsection{Sampling}
\label{sec:sampling}

The task of generating RP variates, i.e., the activity of obtaining realizations of a random composition having a specified distribution over a bounded domain, requires the implementation of a sampling engine. Ideally, the sampler is exact (results in random samples with exactly the desired distribution), efficient (in terms of storage space and execution\footnote{Execution time has two components: Set-up time and marginal execution time. Set-up time is the time required to do some initial computing depending on the particular problem and marginal execution time is the incremental time required to generate each observation.} time), robust (the algorithm is efficient for all parameter values), and not too complex (conceptually and also with respect to practical implementation). Whether these expectations can be met depends on the concrete problem specification, i.e., on the distributional assumption and the investment constraints which define the sampling space. 

To sample from a naive, a basic or a simply regularized RP one just has to sample a Dirichlet distribution with parameter vector $\alpha$. To do so, it is enough to sample the marginals from a Gamma distribution with shape parameter $a = a_i$ and fixed rate parameter $b = 1$ and then standardizing by the sum: $\omega_i = \frac{x_i}{\sum_{i=1}^n x_i}, \quad x_i \sim \mathrm{Gamma}( a = \alpha_i, b = 1 )$. Recall that under the uniform measure, the naive as well as the simply regularized RP allow for an exact solution of the distribution function for linear statistics via Varsi's algorithm. Therefore, sampling is actually not needed.

For a generally regularized RP, sampling is the only option. Under the uniform measure, the sampling problem forms an instance of the fundamental problem addressed in the seminal paper \cite{smith84}: Given a bounded $k$-dimensional body $\mathcal{K} \subset\mathbb{R}^n$, where $k \leq n$, find a way to efficiently sample pseudo-random points $(X_1, X_2, ..., X_d) \in \mathcal{K}$ such that $\mathds{P}(X \in A \subset \mathcal{K}) = V(A) / V(\mathcal{K})$ with $V$ denoting the $k$-dimensional content of $\mathcal{K}$. While \cite{smith84} considered the general case of sampling from a generic surface, here, we restrict analysis to sampling from a $\mathcal{K}$ which is either a polytope $\mathcal{P}$, the interior of an ellipsoid $\mathcal{E}$ or their intersection. 
Exact uniformity can only be obtained under specific circumstances which require either the applicability of transformation, composition or acceptance-rejection methods.

The \emph{transformation} technique maps uniformly distributed points from a hypercube $C$ with a smooth deterministic function $T$ onto $\mathcal{K}$ where $T(x)$ has to preserve uniformity. A necessary and sufficient condition for this is that the Jacobian of $T$ is constant over all $x \in C$~\cite{smith84}. In principle, this is a highly efficient method since there exist very efficient pseudo-random number generator to sample from $C$. The problem is just that $T$ is known only for a very limited class of regions $S$ like spheres or simplices, not so though for general polytopes. Theoretically, one could partition any bounded polytope into a finite union of simplices, and then apply the transformation from the hypercube to each simplex yielding the \emph{composite} technique. Although conceptually sound, the complexity of identifying the simplices is generally such that the approach is not tractable computationally.

The idea behind the \emph{acceptance-rejection} technique is to first find an enclosing set $K \supset \mathcal{K}$ for which efficient sampling algorithms exist, take samples from $K$ and either accept them if they lie within $\mathcal{K}$ or reject them, otherwise. Accepted points will be uniformly distributed in $\mathcal{K}$ because if a point is uniformly distributed within $K$, then it is conditionally uniformly distributed in $\mathcal{K}$ given it lies in $\mathcal{K}$. The problem here is that even when $K$ is chosen based on certain optimality conditions, like the smallest enclosing sphere, the number of trial points in $K$ needed to get a point in $\mathcal{K}$ grows, as Smith (\cite{smith84}) calls it 'explosively'\footnote{As an example, \cite{smith84} shows that when $\mathcal{K}$ is a hypercube and $K$ is a circumscribed sphere the expected number of points generated in $K$ needed to find one in $\mathcal{K}$ grows from 1.5 for $k = 2$ to $10^{30}$ for $k = 100$.}.

As the dimension grows, the (only viable) solution is to sample with geometric random walks. The next Section provides a survey of existing walks 
that we can use to sample from a RP and, ultimately, to address the finance problems of performance and factor analysis. 


\section{Geometric Random Walks}
\label{sec:geometric_random_walks}

Geometric random walk algorithms are a specific type of Markov chains that initiate from an interior point within a convex body $\mathcal{K}$. At each step, they transition to a neighboring point selected from a distribution dependent solely on the current position. The fundamental concept behind all geometric random walks is to generate a lengthy sequence of points, randomizing their order to render the sequence independent and identically distributed (i.i.d.) over $\mathcal{K}$. The complexity of the algorithms depends on its \emph{mixing time}, i.e., the number of steps required to bound the distance between the current and the stationary distribution, and on the complexity of the basic geometric operations performed at each step of the walk; the latter is termed \emph{per-step complexity}.

The problem of sampling from a bounded convex body $\mathcal{K}$ is closely related to the problem of approximating the volume of $\mathcal{K}$. The first celebrated result is given in~\cite{DyerFrKa91} where they sample approximately from the uniform distribution using a grid walk in $\mathcal{K}$. Since then a great effort has been devoted to geometric random walks; to mention a few seminal papers, in~\cite{smith84} they introduced the Hit-and-Run (HaR) algorithm, in~\cite{bib:KannanLovaszSimonovits1997} they crucially improve rounding and sampling results using Ball walk (BaW) and in~\cite{Lovasz06} they show fast mixing for HaR even when the random walk starts from a corner point in $\mathcal{K}$. Over the last 35 years various walks have been presented, each possessing distinct advantages and drawbacks. Some routines confine the sampling space to a polytope, while others are more generic, though often, there is a trade-off between flexibility and efficiency. Further, some walks are restricted to uniform sampling while others accommodate more general distributions, although they may need some pre-processing steps where uniform samples are required. The discriminating aspect among the various routines lies in their approach to take the next Markov step, i.e., in the choice of the direction, the step-length, the curvature of the trajectory and behavior when a boundary is hit. 

\begin{figure}[t!] 
    \centering
    \includegraphics[width=1\linewidth]{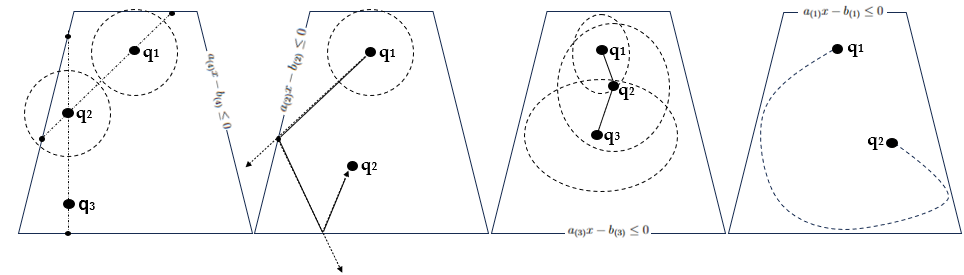}
    \caption{{\small Illustration of HaR, BiW, Dikin walk, HMC (from left to right)
    \label{fig:Illustration_GRWs}}}
\end{figure}

Figure ~\ref{fig:Illustration_GRWs} illustrates the mechanisms of four geometric random walks which are at the basis of various descendent methods. For HaR, the logic is as follows: Start with an arbitrary point $x_0$ inside the convex body. Then, generate a random direction vector $v$ with each component sampled independently from a standard normal distribution. Compute the intersections between the line defined by the point $x_0$ and the direction vector $v$ and the boundary of $\mathcal{K}$. Finally, choose a random point on the segment defined by the two computed boundary points and repeat the process. The choice of the random distribution on the segment needs to be chosen with respect to the target distribution (for instance, uniformly, if the target is to sample uniformly distributed points).
Billiard Walk (BiW)~\cite{bib:GryazinaPolyak2014} operates similarly to HaR, but it exclusively moves in the direction of $v$ and reflects the ray upon encountering the boundary whereby the exit angle matches the entry angle.  Both methods work on general $\mathcal{K}$, but BiW is limited to sampling from the uniform distribution. Dikin walk ~\cite{narayanan16} (and other ellipsoidal procedures like Vaidya and John walk) choose their steps by sampling uniformly from an ellipsoid centered at the current point whose shape and size are determined by the shape of $\mathcal{K}$ and the proximity of the current point to the boundary. Those algorithms are very efficient even in high-dimensional cases but only allow for uniform sampling from polytopes. Hamiltonian Monte Carlo (HMC)~\cite{Neal11} is a sophisticated technique that employs Hamiltonian dynamics to enhance sampling efficiency. It picks a random velocity according to a local distribution and then walks on a Hamiltonian trajectory, i.e., a trajectory which is given by the Hamiltonian dynamics, to obtain the next Markov point. So generated proposals exhibit less of a random-walk behavior, resulting in more effective and less correlated samples. Various versions of HMC exist, all allowing sampling from general log-concave distributions, though some versions (Riemannian HMC~\cite{kook22}) are limited to polytopes.

In the following we discuss important aspects that are common to all routines before moving to a detailed description of the individual procedures. We provide a survey of all existing routines. 
Readers seeking more in-depth understanding of the technicalities are encouraged to consult the source literature for a more comprehensive exploration.

\paragraph{Distribution support.}
\label{sec:distribution_support}
We assume that the support of the target distribution is either a full dimensional convex polytope given by a set of linear inequalities as Eq.~\eqref{eq:polytope_equation} or the intersection between a polytope and an ellipsoid (see Eq.~\eqref{eq:ellipsoid_equation}). In either case we may refer to the support as a convex body. 
In the case where the portfolio domain is given also by an additional set of equalities --which is always the case in our examples since we require portfolios to be fully invested (simplex condition)-- we transform it using an isometric transformation to obtain a full dimensional polytope and ellipsoid. In particular, let, for example, the set $\{x\in\R^n\ |\ Bx=0\}$ be defined by the corresponding additional equality constraints. Then, we compute an orthonormal basis of the null space of the matrix $B$ and we project the polytope and the ellipsoid onto the null space to obtain a full dimensional polytope and ellipsoid in the form of $\mathcal{P}$ and $\mathcal{E}$ respectively. Let us demonstrate an example using the Dirichlet model in Eq.~\ref{eq:dirichlet_model} truncated in the intersection between a convex polytope and an ellipsoid, i.e.,
\begin{equation}
\begin{split}
    \mathcal{D}(x; \alpha_1,\dots,\alpha_n) \propto \prod_i x_i^{\alpha_i - 1},\ & \text{where }x_i > 0,\ \sum_i x_i =1,\ Ax \leq b, x^TEx \leq c,\\ & A\in\R^{m \times n},\ b \in \R^m,\ E\in\R^{n \times n}\text{ pos.\ def., }c \in \R_+ .
\end{split}
\end{equation}
Notice that in this case $B = [1,\dots,1]^T$. 
Consider the transformation $\mathcal{T}(x) := N(x-x_0)$, where $N \in \R^{(n-1)\times n}$ is the matrix that generates the null space of $B$ and $x_0$ a feasible point in the support of $\mathcal{D}$. When we apply the transformation $\mathcal{T}$ on both $\mathcal{D}$ and its support we obtain the density, 
\begin{equation}
    [\mathcal{T} \circ \mathcal{D}] (y; \alpha_1,\dots,\alpha_n) \propto \prod_i (N^T_{(i, :)}y + x_{0i})^{\alpha_i - 1} ,
\end{equation}
since $x = N^Ty + x_0$. By replacing $x$ in the constraint equations one could also obtain the support of the transformed density $[\mathcal{T} \circ \mathcal{D}]$. Finally, we sample from the full dimensional body in the null space and we apply the inverse transformation to obtain the sampled portfolios in the initial space. 

In the sequel, we consider the case of sampling from a full dimensional convex set given by the intersection between a convex polytope and an ellipsoid,
\begin{equation}
\label{eq:body_K}
    \mathcal{K} = \mathcal{P} \cap \mathcal{E} .
\end{equation}

\paragraph{Computational oracles.}
Each geometric random walk uses certain sub-routines called \emph{oracles}. An oracle is an algorithm that answers a certain question that is needed for the implementation of a random walk. The oracle that every geometric random walk needs is the \emph{membership oracle} that answers if a given point belongs or not in $\mathcal{K}$. The simplest implementation of this oracle is to check the validity of the Eq.~(\ref{eq:polytope_equation}, \ref{eq:ellipsoid_equation}).

Another useful sub-routine is the \emph{boundary oracle} that computes the intersection between the boundary of $\mathcal{K}$, denoted $\partial \mathcal{K}$, and a ray that starts from a point in the interior of $\mathcal{K}$, let $\ell(t) := \{ x + t v\ |\ x\in\mathcal{K},\ v\in\R^n,\ t\in\R_+ \}$. To compute the intersection with $\partial \mathcal{P}$ we have to solve one linear equations per facet and keep the smallest positive root. We have,
\begin{equation}
\label{eq:boundary_oracle_polytope}
a_j^T(x + tv) = b_j \Rightarrow t = \frac{b_j - a_j^Tx}{a_j^Tv},\ j=[m],
\end{equation}
where $m$ is the number of facets and $a_j\in\R^n$ are the normal vectors defining the support hyperplane of each facet. The computations in Eq.~(\ref{eq:boundary_oracle_polytope}) costs $\OO(mn)$ operations\footnote{Here and in the remainder of the text we use Bachmann-Landau symbols to express computational complexity (in terms of oracle evaluations and arithmetic operations) as a function of the dimension of the geometric object. Additionally, the $\sO(\cdot)$ notation means that we are ignoring polylogarithmic factors.}. To compute the intersection with $\partial\mathcal{E}$ we have to solve the following equation and keep the smallest positive root,
\begin{equation}
    \label{eq:boundary_oracle_ellipsoid}
    (x + tv)^T E (x+tv) = c \Rightarrow \| v \|_2^2 t^2 + (2 x^TEv) t + (x^TEx - c) = 0 .
\end{equation}
That is a second order polynomial equation for which we have a closed form to compute its roots. The computations in Eq.~(\ref{eq:boundary_oracle_ellipsoid}) costs $\OO(n^2)$
operations.
Clearly, in the case of the body $\mathcal{K}$ we solve both Eq.~(\ref{eq:boundary_oracle_polytope}, \ref{eq:boundary_oracle_ellipsoid}) and keep the smallest positive root $t_+$. Then, $\ell(t_+) \in \partial\mathcal{K}$ is the intersection point. 

Some random walks, like e.g., BiW, use a sub-routine called \emph{reflection oracle}. This oracle computes the reflection of a ray $\ell(t)$ when the later hits the boundary of $\mathcal{K}$ so that the reflected ray continues in $\mathcal{K}$. Typically, this oracle is called after the boundary oracle which computes the boundary point $y$ of ray intersecting $\partial \mathcal{K}$. Let $s\in\R^n,\ \|s\|_2 = 1$ the normal vector defining the tangent hyperplane at $y$ and $v\in\R^n$ the direction vector of the ray $\ell(t)$ until it hits $\partial\mathcal{K}$. Then, the update rule to obtain the direction vector of the reflected ray is the following, 
\begin{equation}
    v \leftarrow v - 2\langle v,s\rangle s .
\end{equation}
Thus, the reflected direction can be computed after $\OO(n)$ operations, given the normal vector $s$.
When $\ell(t)$ hits a facet of $\mathcal{P}$ the vector $s$ is equal to the normal vector of that facet, i.e., equal to the normalized row of the matrix $A$ that corresponds to that facet. When $\ell(t)$ hits $\partial\mathcal{E}$ the vector $s$ is equal to $Ey/\| Ey \|_2$, where $y$ is the boundary point.

Last, certain random walks need to sample uniformly from special sets, namely from the boundary and/or the interior of the unit ball and the interior of a given ellipsoid. To sample from the boundary of the unit ball $\mathcal{B}_n$ we sample $n$ numbers $g_1,\dots ,g_n$ from the standard Gaussian distribution $\mathcal{N}(0,1)$ and then the vector $q=(g_1,\dots ,g_d)/\sqrt{\sum g_i^2}$ is uniformly distributed on the boundary of $\mathcal{B}_n$. Moreover, the point $u^{1/n}q$, where $u\in\R$ is uniformly distributed in $[0,1]$, is uniformly distributed in $\mathcal{B}_n$. Last, the point $Lx$, where $x$ is uniformly distributed in the boundary of $\mathcal{B}_n$ and $E=L^TL$, is uniformly distributed in the ellipsoid $\mathcal{E}$.

\paragraph{Computing an interior point.} To run a geometric random walk it is necessary to compute a point in the interior of the convex body we want to sample from. When the convex body is the convex polytope in Eq.~(\ref{eq:polytope_equation}) one could compute the largest ball inside $\mathcal{P}$, called \emph{Chebychev ball}. Then, by definition, the center of the ball lies inside $\mathcal{P}$. To compute the Chebychev ball we have to solve the following linear program,
\begin{equation}
\begin{split}
    & \max r \\
    & \text{subj.\ to: } a_i^T x + r\| a_i \|_2 \leq b_i , 
\end{split}
\end{equation}
where $a_i,\ i=[m]$ are the rows of the matrix $A$ in Eq.~(\ref{eq:polytope_equation}).
When the convex body is the intersection between a polytope and an ellipsoid, $\mathcal{K} = \mathcal{P} \cap \mathcal{E}$,
one could apply on $\mathcal{K}$ the transformation that maps the ellipsoid $\mathcal{E}$ to the unit ball $\mathcal{B}_n$ and consider the intersection of the transformed polytope with $\mathcal{B}_n$. Then, the largest ball inside the latter body can be computed by solving the following Second-Order Cone Program (SOCP),
\begin{equation}
\label{eq:socp}
\begin{split}
& \max\ r\\
& \text{subj.\ to: } a_i^Tx + r\|a_i\|_2 \leq b_i \\
& \hspace{1.65cm} \|x\|_2 \leq 1-r ,
\end{split}
\end{equation}
where the pairs $a_i,\ b_i$ define the facets of the transformed polytope. We can obtain a point in $\mathcal{K}$ by applying the inverse transformation on the computed center in Eq.~(\ref{eq:socp}).

\paragraph{Empirical convergence to the target distribution.}
In order to evaluate the quality of a sample as an accurate approximation of the target distribution, several convergence diagnostics \cite{Roy20} are available like potential scale reduction factor (PSRF)~\cite{Gelman92}, maximum mean discrepancy (MMD)~\cite{Gretton12} and the uniform tests~\cite{CousinsThesis17}. For a dependent sample, a powerful diagnostic is the effective sample size (ESS). It is the number of effectively independent draws from the target distribution that the Markov chain is equivalent to. In our empirical applications, we ensure that PSRF $< 1.1$ 
and ESS $> 0.95n$, where $n$ reflects the dimensionality of the problem (i.e., in general the number of assets minus the number of equality constraints).

\paragraph{Starting point.}

A crucial aspect for the efficiency of a random walk is its starting point, which needs to be a point in the interior of the convex body. When the starting point comes from a distribution close to the target distribution, then it is called a \emph{warm start}. The mixing time analysis in the literature usually requires a warm start to bound the rate of convergence. When the starting point is a fixed or a corner point it is called a \emph{cold start}. Regarding uniform sampling, in~\cite{LovVem} they give an algorithm that computes a warm start after $\sO(n^4)$ calls to the membership oracle of the input convex body.

In practical implementations the starting point usually is a "central" point as the Chebychev or the analytical center of the polytope~\cite{bib:CousinsVempala2016, Chalkis23, kook22}. A common practice is to allow for a user defined positive integer that corresponds to the number of Markov points to be ignored before the implementation starts to store the generated points. Several practical methods have been developed to compute a good starting point~\cite{Emiris14, ChalkisMetabolic21, ChePioCaz18, bib:CousinsVempala2016}.


\paragraph{Roundness of the distribution.}
Another key aspect affecting the performance of a random walk is the roundness of the target distribution, that is how close the covariance of the target distribution is to the identity matrix. Regarding uniform sampling, this is translated to body's roundness, measured by the ratio $R/r$~\cite{bib:KannanLovaszSimonovits1997,Lovasz06}. $R$ and $r$ are the radii of the largest and smallest ball centered at the origin that contains, and is contained, in $\mathcal{K}$, respectively; i.e., $rB_n \subseteq \mathcal{K} \subseteq RB_n$. 
Hence, before the actual sampling process is started, it is crucial to reduce $R/r$, i.e., to put $\mathcal{K}$ in a well-rounded position, where $R/r = \sO(\sqrt{n})$. This is particularly relevant for financial applications like the one we suggest in our empirical study in Section~\ref{sec:experiments} because lower and upper bounds on asset weights are typically tight, inducing a skinny polytope. 
 
A powerful approach to obtain well roundness is to put $\mathcal{K}$ in \emph{isotropic position}. In general, $\mathcal{K} \subset \RR^n$ is in isotropic position if the uniform distribution over $\mathcal{K}$ is in isotropic position, that is $\mathbb{E}_{X\sim \mathcal{K}}[X] = 0$ 
and $\mathbb{E}_{X\sim \mathcal{K}}[X^TX] = I_n$, where $I_n$ is the $n \times n$ identity matrix. Thus, to put $\mathcal{K}$ into isotropic position one has to generate a set of uniform points in its interior and apply to $\mathcal{K}$ the transformation that maps the point-set to isotropic position; then iterate this procedure until $\mathcal{K}$ is in $\OO(1)$-isotropic position \cite{ bib:CousinsVempala2016,Lovasz06}. In~\cite{Rudelson99} they prove that $\OO(\log(n))$ iterations and $\sO(n)$ uniformly distributed points per iteration suffice to achieve isotropic position. There are several algorithms based on this routine~\cite{bib:KannanLovaszSimonovits1997, LovVem, Jia21}. In~\cite{Jia21} they build upon~\cite{Chen21} to provide the best algorithm so far that puts a convex body in isotropic position after $\sO(n^3)$ membership oracle calls. 
The practical method in~\cite{ChalkisMetabolic21} brings a convex body in near-isotropic position by using BiW with multiple starting points for uniform sampling in each phase. It successfully rounds convex polytopes in a few thousand dimensions. 

An alternative notion of well roundness is the John position. A convex body in John position has a sandwiching ratio of $R/r = \sO(n)$ which is worse than that of isotropic position. To put the body in John position one needs to compute the maximum volume ellipsoid (MVE) in it and apply to the body the transformation that maps the ellipsoid to the unit ball. To our knowledge there are specialized results only for the case of a convex polytope $\mathcal{P}$. In~\cite{Nemirovski99, Anstreicher02} they independently give an algorithm that computes the John ellipsoid of $\mathcal{P}$ in $\sO(m^{3.5})$ operations. Interestingly, in~\cite{Khachiyan93} 
they provide a linear time transformation of the MVE problem by computing a minimum volume enclosing ellipsoid (MVEE) of a set of points. Thus, the algorithms in~\cite{Kumar05, Todd07} that solve the MVEE problem can be used to compute the John ellipsoid after $\OO(mn^3/\epsilon)$ operations. The practical method in~\cite{Zhang03} to compute the MVE has been used in~\cite{haraldsdottir2017chrr} to bring convex polytopes of thousands of dimensions in John position. However, the practical method in~\cite{ChalkisMetabolic21} obtains, in almost the same runtime, both a better sandwiching ratio than~\cite{Zhang03} --as it brings the polytope to a near isotropic position-- and a uniformly distributed sample in $\mathcal{P}$. Last, in~\cite{Cohen19, Song2022} they provide algorithms to compute the John ellipsoid in the special case of a centrally symmetric convex polytope achieving near optimal performance. However, those algorithms can not be used for the purpose of rounding any convex polytope. For an overview in rounding a convex body we refer to Table~\ref{tab:rounding_body}.

\begin{table}[t!]
\centering
\begin{tabular}{|ccccc|}\hline
\multicolumn{1}{|c}{Year \& Authors} & \multicolumn{1}{c}{Type of rounding}  & \multicolumn{1}{c}{Total cost} & Polytope $\mathcal{P}$ & Convex body $\mathcal{K}$ \\ \hline\hline
1997~\cite{bib:KannanLovaszSimonovits1997} & Isotropic position  & \makecell{$\sO(n^5)$\\ memb.\ calls}  & \ding{51} & \ding{51} \\\hline
1999~\cite{Nemirovski99, Anstreicher02} & John position & \makecell{$\sO(m^{3.5})$ \\ operations} & \ding{51} & \ding{55} \\\hline
2003~\cite{Zhang03, haraldsdottir2017chrr} & $^*$John position & ?? &  \ding{51} & \ding{55}\\\hline
2005~\cite{Kumar05, Todd07} & John position & \makecell{$\OO(mn^3/\epsilon)$ \\ operations} & \ding{51} & \ding{55} \\\hline
2006~\cite{LovVem} & Isotropic position &  \makecell{$\sO(n^4)$ \\ memb.\ calls} & \ding{51} & \ding{51} \\\hline
2016~\cite{bib:CousinsVempala2016} & $^*$Isotropic position & ?? & \ding{51} & \ding{51} \\\hline
2021~\cite{Jia21} & Isotropic position &  \makecell{$\sO(n^3)$ \\ memb.\ calls} & \ding{51} & \ding{51} \\\hline
2021~\cite{ChalkisMetabolic21} & $^*$Isotropic position & ?? & \ding{51} & \ding{51} \\\hline
\end{tabular}
\caption{Overview of the algorithms to round a convex body.\ \\ $^*$Practical methods.
\label{tab:rounding_body}}
\end{table}

To round a log-concave distribution with density function $\pi$, $R$ is chosen to bound the expected squared distance of the random variable from the centroid of $\pi$, i.e., $R^2 \geq \mathbb{E}_{\pi}(|x-z_{\pi}|^2)$, where $z_{\pi}$ is the centroid. We say that a log-concave density function is well-rounded if $R/r = \OO(\sqrt{n})$, where $r$ is the radius of the ball contained in a level set of $\pi$ of constant probability.

In~\cite{LovaszVempala03} they provided the first algorithm to round a log-concave distribution $\pi \propto e^{-f(x)}$ after $\sO(d^5)$ membership oracle calls, where $f:\RR^n\rightarrow\RR$ is a convex function. They introduce an iterative algorithm that brings to isotropic position a certain level set of the density function, showing that it suffices to bring it to a near isotropic position. The algorithm can use either HaR or BaW.
In~\cite{bib:LovaszVempala2006} they give an algorithm that uses HaR and rounds a log-concave distribution after $\sO(n^{4.5})$ membership oracle calls; they generalize the algorithm in~\cite{Kalai06} that is specialized for the exponential distribution. Moreover, if a near-optimal point of $\pi$ is given they provide a multi-phase algorithm that rounds $\pi$ after $\sO(n^4)$ membership oracle calls. The main idea in~\cite{Kalai06, bib:LovaszVempala2006} is to consider the density function,
\begin{equation}
    \pi_{\beta} \propto e^{-\beta f(x)},\ x\in\mathcal{K},\ \beta\in\R_+
\end{equation}
where the parameter $\beta > 0$ controls the variance of $\pi_{\beta}$. They introduce a new multi-phase algorithm where in each phase they set a different value of $\beta$; the last phase sets $\beta = 1$ and samples from $\pi$. 
Initially, the algorithm invokes the rounding algorithm outlined in~\cite{LovVem} to obtain a warm start for the uniform distribution over $\mathcal{K}$. In the initial phase, it samples for $\beta = 0$, representing the uniform distribution. As the algorithm progresses to the $j$-th phase, the parameter $\beta_j$ incrementally increases to ensure that a sampled point from $\pi_{\beta{j-1}}$ serves as a warm start for $\pi_{\beta_j}$ in accordance with the $L_2$ norm on density functions. Furthermore, the sample generated in the $j$-th phase is utilized to estimate the covariance matrix for the subsequent phase. Upon reaching $\beta = 1$, the algorithm concludes by applying a linear transformation to both $\pi$ and $\mathcal{K}$, converting the ellipsoid defined by the approximated covariance into the unit ball. Consequently, the resulting density function is well-rounded. Lastly, it is important to state that these rounding results on constrained, log-concave density functions, highlight the importance of HaR since it can mix starting from $L_2$ warm starts or even from cold start, which can be exploited to reduce the number of phases we need to round $\pi$.

\begin{table}[t!]
\footnotesize
\centering
\begin{tabular}{|ccccc|}\hline
\multicolumn{1}{|c}{Year \& Authors} & \multicolumn{1}{c}{Random walk}  & \multicolumn{1}{c}{Mixing time} & \makecell{Cost per \\ step in $\mathcal{P}$} & \makecell{Cost per \\ step in $\mathcal{K}$} \\ \hline\hline
2006~\cite{Lovasz06} & Hit-and-Run &  $\sO(n^2(R/r)^2)$ & $\OO(mn)$ & $\OO(mn + n^2)$ \\\hline
2014~\cite{bib:GryazinaPolyak2014} & Billiard walk & ?? & $\OO(\rho mn)$ & $\OO(\rho (mn +n^2))$ \\\hline
2016~\cite{LeeVemGeodesic} & Geodesic walk & $\sO(mn^{3/4})$ & $\OO(mn^{\omega-1})$ & -- \\\hline
2016~\cite{narayanan16} & Generic Dikin Walk &  $\sO(nm+n^3)$ & $\OO(mn^{\omega-1})$  & $\OO(mn^{\omega-1} + n^{\omega+1})$ \\\hline
2017~\cite{chen_vaidya_17} & Vaidya walk & $\sO(m^{1/2}n^{3/2})$ & $\OO(mn^{\omega-1})$ & -- \\\hline
2018~\cite{chen18} & Approximate John walk & $\sO(n^{2.5})$ & $\OO(mn^{\omega-1})$ & -- \\\hline
2018~\cite{bib:GustafsonNarayanan2018} & John's Walk & $\sO(n^7)$ & $\OO(mn^4 + n^8)$ & -- \\\hline
2018~\cite{Lee18c} & Uniform Riemannian HMC & $\sO(mn^{\frac{2}{3}})$ & $\OO(mn^{\omega-1})$ & -- \\\hline
2019~\cite{bib:KannanLovaszSimonovits1997} & Ball walk &  $\sO(n^2R^2/r^2)$ & $\OO(mn)$ & $\OO(mn + n^2)$ \\\hline
2020~\cite{Laddha20b} & Dikin walk & $\sO(mn)$ & $\OO(nnz(A) + n^2)$ & -- \\\hline
2020~\cite{Laddha20b} & Weighted Dikin Walk & $\sO(n^2)$  &  $\OO(mn^{\omega-1})$  & -- \\\hline
2021~\cite{ChalkisMetabolic21} & Multiphase Billiard walk & ?? & $\OO((\rho + n)m)$ & $\OO(\rho (mn + n^2))$ \\\hline
2021~\cite{bib:LeeVempala2018, Mangoubi19, Chen21} & $^\dagger$Ball walk & $\sO(n^{2})$ & $\sO(m)$ & $\OO(mn + n^2)$ \\\hline 
2022~\cite{Chen2022} & $^\dagger$Hit-and-Run &  $\sO(n^2)$ & $\OO(mn)$ & $\OO(mn + n^2)$ \\\hline
2022~\cite{kook22, kook23} & Riemannian HMC & $\sO(mn^3)$ & $\OO(n^{3/2})$ & -- \\\hline
2023~\cite{Laddha23} & Coordinate Hit-and-Run &  $\sO(n^9(R/r)^2)$ & $\OO(m)$ & $\OO(m + n)$\\\hline
\end{tabular}
\caption{Overview of all known geometric random walks for uniform sampling. $n$ is the dimension of the convex body; $m$ is the number of facets of the  polytope; $D$ is the diameter of the convex body; $\omega$ is the matrix multiplication exponent, i.e., $\omega \approx 2.37$; $R$ is the radius of the smallest enclosing ball and $r$ the radius of the largest ball enclosed in the convex body; $\rho$ is the number of boundary reflections; $nnz(\cdot)$ gives the number of non-zero elements of a matrix. \\
$^\dagger$ For a convex body in isotropic position;
for Ball walk in the case of an isotropic convex polytope the (amortized) cost per step is $\sO(m)$ when an $\OO(1)$-warm start in the 
interior of the polytope is given. \label{tab:random_walks_uniform}
}
\end{table}

\paragraph{Best mixing times.}
Regarding uniform sampling from a general convex body, best mixing time is achieved by HaR and Ball walk (BaW, see Section \ref{subsec:ball_walk}, that is $\sO(n^2R^2/r^2)$ steps for a body in general position with sandwiching ratio equal to $R/r$. However, one could bring the body to a near isotropic position, as a preprocessing step, after $\sO(n^3)$ membership oracle calls. Then, both HaR and BaW can generate an almost uniformly distributed point after $\sO(n^2)$ steps. Then, the efficiency per almost unifromly distributed point depends on which oracle is cheaper (i.e., the membership or the boundary oracle). In the case of sampling uniformly from a bounded convex polytope $\mathcal{P}$, we have that the number facets $m \geq n + 1$ to guarantee boundness. Therefore, for a convex polytope in a general position, the best mixing time is given between Weighted Dikin walk~\cite{Laddha20b} and Riemannian HMC~\cite{Lee18c} while their mixing times do not depend on $R/r$. Since they have the same cost per step, which is the most efficient random walk depends on the number of facets $m$. Let $m=\OO(n^k)$ where $k\geq 1$. For $k\leq \frac{4}{3}$ the most efficient random walk is Riemannian HMC and for $k>\frac{4}{3}$ Weighted Dikin walk achieves a smaller mixing time. However, if one brings $\mathcal{P}$ to a near isotropic position, the most efficient option is BaW since its mixing time is $\sO(n^2)$ (same with HaR and Weighted Dikin walk) and its cost per step is the smallest among all uniform samplers, i.e., $\sO(m)$ steps.

Considering non-uniform sampling, there are known bounds only for the case of general log-concave distributions. The geometric random walk that achieves the best mixing time is BaW. In particular, for a log-concave in a general position BaW mixes after $\sO(n^2D)$ steps where $D$ is the diameter of the support of the target distribution~\cite{Lee17}. If the target distribution is in isotropic position then BaW mixes after $\sO(n^2)$ steps. However, when the starting point is a corner point BaW is known to mix slowly, while HaR is known to mix rapidly from any given interior point, even a corner point~\cite{bib:LovaszVempala2006}.
However, both Reflective and Riemannian HMC have shown superior performance in practice~\cite{Chalkis23b, kook22}. For a quick overview on the random walks that allow for non-uniform sampling we refer to the Table~\ref{tab:random_walks_logconcave}.

\begin{table}[t!]
\small
\centering
\begin{tabular}{|ccccc|}\hline
\multicolumn{1}{|c}{Year \& Authors} & \multicolumn{1}{c}{Random walk}  & \multicolumn{1}{c}{Mixing time} & \makecell{Cost per \\ step in $\mathcal{P}$} & \makecell{Cost per \\ step in $\mathcal{K}$} \\ \hline\hline
1987~\cite{smith84} & Coordinate Hit-and-Run &  ?? & $\OO(m)$ & $\OO(m + n)$\\\hline
2006~\cite{bib:LovaszVempala2006} & Hit-and-Run &  $\sO(n^2R^2/r^2)$ & $\OO(mn)$ & $\OO(mn + n^2)$ \\\hline
2018~\cite{ChePioCaz18} & Exact HMC with reflections & ?? & $\OO(\rho \phi)$ & $\OO(\rho\phi)$ \\\hline
2019~\cite{Lee17} & Ball walk & $\sO(n^2D)$  & $\OO(mn)$ & $\OO(mn + n^2)$ \\\hline
2021~\cite{bib:LeeVempala2018, Chen21} & $^\dagger$Ball walk & $\sO(n^2)$  & $\OO(mn)$ & $\OO(mn + n^2)$ \\\hline

2022~\cite{kook22, kook23} & Riemannian HMC & $\sO(mn^3)$ & $\OO(n^{3/2})$ & -- \\\hline
2023~\cite{Chalkis23b} & Reflective HMC & ?? & $\OO(Lmn)$ & $\OO(L(mn+n^2))$ \\\hline
\end{tabular}
\caption{Overview of all known geometric random walks for general log-concave sampling. $\phi$ is the cost to compute the intersection between the Hamiltonian trajectory and the boundary of the convex body. $D$ is the diameter of the support of the distribution. $L$ is the number of leapfrog steps in a HMC step. For the rest parameters we refer to the Table~\ref{tab:random_walks_uniform}.\\
$^\dagger$ For a log-concave distribution in isotropic position.
\label{tab:random_walks_logconcave}}
\end{table}

\paragraph{Geometric random walks in practice.}
Considering uniform sampling in practice, there are several areas of application with the most effort being devoted to sampling steady states of metabolic networks in biology~\cite{haraldsdottir2017chrr, ChalkisMetabolic21, HOPS20} and practical volume calculation of convex polytopes~\cite{Emiris14, bib:CousinsVempala2016, Chalkis23}. Until recently the dominant paradigm for random walks was Coordinate Direction Hit-and-Run (CDHR)~\cite{kaufman98}, which is a version of HaR that uses directions parallel to the axes in each step instead of random ones. The mixing time of CDHR for uniform sampling has been bounded by $\sO(n^9(R/r)^2)$~\cite{Laddha23, Narayanan20} which is the worst bound among uniform geometric random walk samplers. Interestingly, experiments~\cite{Emiris14, haraldsdottir2017chrr} indicate that CDHR achieves a similar mixing rate with both HaR and BaW. Thus, the faster step of CDHR compared to HaR's and BaW's---$\OO(m)$ vs. $\OO(mn)$---is the main reason why CDHR overshadowed, until recently, all other random walks in practical computations on convex polytopes. However, in~\cite{ChalkisMetabolic21} a new multi-phase sampler based on BiW 
proved to be much faster in practice than CDHR for uniform sampling over convex polytopes. 
Moreover, in~\cite{Chalkis23} they again show the superior performance of BiW against HaR and CDHR by developing a new practical algorithm to approximate volume of polytopes. 
The Riemannian HMC in~\cite{kook22} is specialized for convex polytopes and its mixing rate does not depend on the roundness of the polytope. In their experiments on metabolic networks they show that it is faster than CDHR for convex polytopes in a general position. However, there are not any experimental results that compare this Riemannian HMC sampler with the multi-phase sampler in~\cite{ChalkisMetabolic21} for rounded and non-rounded polytopes. 

Regarding non-linear convex bodies, in~\cite{Chalkis22} they provide computational oracles to implement several geometric random walks for the case of spectrahedra (the feasible region of Semidefinite Programs). Their experiments show that BiW outperforms HaR, CDHR and BaW despite the fact that its cost per step is bigger comparing to all of the rest cost per steps. Regarding non-uniform sampling from non-linear convex bodies, in~\cite{Chalkis23b} they show that the Reflective HMC is the best performer for isotropic log-concave distributions. Last, for the case of log-concave sampling over a convex polytope again in~\cite{Chalkis23b} they show that the Reflective HMC outperforms both HaR and CDHR. The Riemannian HMC in~\cite{kook22} appears to have a relatively small cost per step and exhibits mixing times seemingly unaffected by the roundness of the distribution. These characteristics suggest that it could serve as a competitive random walk for this particular scenario. Nevertheless, there is currently a lack of empirical experiments evaluating its performance specifically on log-concave distributions.

\subsection{Ball walk (BaW)}
\label{subsec:ball_walk}

BaW can be seen as a special case of Metropolis Hastings~\cite{Hastings70} for constrained sampling in convex domains. However, BaW was first introduced for uniform sampling over convex bodies in~\cite{bib:KannanLovaszSimonovits1997}. They used BaW for volume approximation and rounding convex bodies. They prove a bound of $\sO(n^2R^2/r^2)$ on mixing time, that is $\sO(n^3)$ for an isotropic convex body. For a general isotropic log-concave distribution $\pi$ constrained in a convex body the mixing time of BaW is $\sO(n^2/\psi(\pi)^2)$, where $\psi(\pi)$ is the \emph{isoperimetric coefficient} of $\pi$~\cite{bib:KannanLovaszSimonovits1997}. 
In~\cite{bib:KannanLovaszSimonovits1997} they bound $\psi(\pi)$ from below with $n^{1/2}$, that results to a mixing time of $\sO(n^3)$ for isotropic log-concave distributions. Improving the bound on $\psi(\pi)$ it was a very important open problem in convex geometry for years, while bounding $\psi(\pi)$ by a constant was a famous problem known as \emph{Kannan-Lovasz-Simonovits (KLS) conjecture}~\cite{bib:LeeVempala2018}. Moreover, it is known~\cite{Lee17} that, if KLS conjecture is true, the mixing time of ball walk is $\sO(n^2)$ when $\delta = \OO(1/\sqrt{n})$, while if $\delta$ is larger then the rejection probability in each step becomes too high and the mixing time increases. Thus, a lower bound of $\Omega(n^2)$ is valid for the mixing time of BaW.

In~\cite{Lee17} they improved the bound on $\psi(\pi)$ to $n^{1/4}$ which leads to a mixing time of $\sO(n^{2.5})$ for isotropic log-concave distributions. They also extend this result and prove that the mixing time of BaW is $\sO(n^2D)$, where $D$ is the diameter of the convex body. Finally, in~\cite{Chen21} they give an almost constant bound on $\psi(\pi)$, meaning that for increasing dimension $n$ the mixing time of BaW is $\sO(n^2)$ for isotropic log-concave distributions -- which is also a tight bound on its mixing time. Interestingly in~\cite{Mangoubi19}, for the case of uniform distribution and an isotropic convex polytope they provide an implementation of BaW's step that costs $\sO(m)$ operations after the very first step that costs $\OO(mn)$ operations.
Thus, the total cost to generate an almost uniformly distributed point drops by a factor of $n$ in the case of an isotropic convex polytope.

\paragraph{Step overview.}
In each step, BaW picks uniformly a point $y$ from a ball of radius $\delta$ and centered on the current Markov point $p$. Then, if the point lies outside of $\mathcal{K}$ the random walk stays on $p$; otherwise, it applies a Metropolis filter to decide moving to $y$ or not. In theory the radius $\delta$ should be around $\sO(1/\sqrt{n})$~\cite{bib:KannanLovaszSimonovits1997, CousinsV14} to achieve the optimal mixing time for log-concave sampling. Since the implementation of BaW requires only membership oracle calls the cost per step is $\OO(mn)$ in the general case. In the case of uniform sampling from an isotropic polytope $\mathcal{P}$ in~\cite{Mangoubi19} they reduce the cost per step to $\sO(m)$ operations by conducting the following analysis: First, for all the sampled points the starting point is always the $\OO(1)$-warm start given as input. Then, they employ the property that any Markov chain whose stationary distribution is close to the uniform distribution on the polytope will spend on average at least $\frac{1}{n}$ of its time a distance $\frac{1}{n}$ from the boundary of the polytope. They compute a lower bound on the distance of the current Markov point to each one of the facets. Thus, they check a certain linear constraint $a_i^Tp \leq b_i$ only if the lower bound on the distance of the point $p$ to the $i$-th facet is $\OO(\frac{1}{n})$. Finally, they prove that with at least probability $1-\epsilon$ the BaW will not violate any linear constraint, where the small $\epsilon > 0$ is a parameter used to bound the size of BaW's step towards any facet. 

\begin{algorithm}[t!]
  \caption{Ball\_walk$(\mathcal{K}, p, \pi, \delta)$}
	\label{alg:ball_walk}
	\SetKwInOut{Input}{Input}
	\SetKwInOut{Output}{Output}
    \SetKwRepeat{Do}{do}{while}
	
    \Input{Convex body $\mathcal{K}$; point $p\in\mathcal{K}$, $\pi$ probability distribution function over $\mathcal{K}$.}

    
	\Output{A point $q$ in $\mathcal{K}$}
    
    \BlankLine
    	Pick a uniform random point $y$ from the ball of radius $\delta$ centered at $p$\;
		\textbf{if }$y\notin \mathcal{K}$ \textbf{then }$q\leftarrow p$ \textbf{end}\\
        \ElseIf {$y\in \mathcal{K}$} {
            $\alpha = \min \bigg\{ 1, \frac{\pi(y)}{\pi(p)} \bigg\}$\;
            $u\sim \mathcal{U}[0,1]$\;
            \textbf{if }$u \leq \alpha$ \textbf{then }$q\leftarrow y$ \textbf{else }$q\leftarrow p$\;
    	}
    \textbf{return} $q$\;
\end{algorithm}

\paragraph{Practical performance.}
In~\cite{bib:CousinsVempala2016} they provide a BaW implementation for spherical Gaussian sampling to approximate the volume of a polytope. The mixing time in practice is --as expected-- $\sO(n^2)$. After an experimental evaluation, they set the radius in each step to $\delta = 4r/\max\{1,\alpha\}n$, where $\alpha = 1/2\sigma$ and $r$ is the radius of the largest inscribed ball in the convex body. However, due to constants the number of steps BaW needs to mix in practice is higher that the number of steps that both HaR and CDHR need. Thus, computations carried out with HaR are faster by a constant than the same computations carried out by BaW. CDHR is faster as the dimension increases than BaW since it has a smaller cost per step. Considering uniform sampling, BaW's mixing time in practice is also $\sO(n^2)$~\cite{Chalkis21} while the radius of each step is $\delta = 4r/\sqrt{n}$.

\paragraph{Software.}
BaW for Gaussian sampling is implemented in the MATLAB library \texttt{cobra}, based on the results in~\cite{bib:CousinsVempala2016}. The implementation is replicated in C++ library \texttt{volesti}. There exist two C++ implementation for uniform sampling; that one in \texttt{volesti} and a second based implemented in the C++ package \texttt{PolytopeWalk}~\cite{PolytopeWalk21}, which is based on the algoroithms given in~\cite{chen18}. Regarding the smaller cost per step in the uniform case, to our knowledge, there is not any practical implementation that takes advantage of this result to obtain a faster software based on BaW uniform polytope sampler in~\cite{Mangoubi17}.

\subsection{Hit-and-Run (HaR)}

HaR was first defined by Smith in~\cite{smith84} for uniform sampling from bounded convex bodies, and extended in~\cite{Smith93} for any density function over a (non-)convex body. HaR's mixing time has been studied extensively over the past decades. In~\cite{bib:LovaszVempala2003, Lovasz06} they show that HaR mixes after $\sO(n^2(R/r)^2))$ steps for log-concave distributions (including the uniform distribution) and for both a warm and a cold start. That implies a mixing time of $\sO(n^3)$ for rounded target distributions. In the case of sampling uniformly an isotropic convex body, in~\cite{Chen2022} they prove that its mixing time is $\sO(n^2)$ building upon the near optimal bound in KLS conjecture. There have been developed several variants of HaR, i.e., in~\cite{kaufman98} they study optimal direction choices instead of picking a uniformly random direction in each step. It turns out that the optimal direction choice is to pick a direction from the Gaussian distribution $\mathcal{N}(0,\Sigma)$ where $\Sigma$ is the covariance of the target log-concave distribution; that is equivalent of sampling from an isotropic distribution~\cite{Kalai06}. 

An important property of HaR it is that it mixes from a warm start that comes from a distribution with a bounded $L_2$ norm from the target distribution. This allows to develop fast multi-phase polynomial randomized algorithms for several important problems in computational statistics and optimization as one would need a smaller number of phases comparing to random walks that mix from an $\OO(1)$ warm start, i.e., BaW. Thus, HaR is the random walk of choice to round a log-concave distribution, to approximate an integral over a convex set or solving a  convex problem~\cite{bib:LovaszVempala2006}. 
Regarding additional applications, a specific version, i.e.\ \emph{artificial centering hit and run}, have been widely used to study constrained based models in biology \cite{Megchelenbrink14, Saa16}. HaR has been used for volume computation~\cite{LovVem}, sampling steady states of metabolic networks~\cite{Megchelenbrink14, Saa16}, convex optimization~\cite{Kalai06} and other applications~\cite{Berbee1987, haraldsdottir2017chrr}. Its practical performance established this random walk as one of the main paradigms for constrained log-concave sampling.

\paragraph{Step overview.}
At each step the algorithm picks a uniformly distributed unit vector that defines a line through the current Markov point and computes the intersection of the line with $\partial\mathcal{K}$ by calling the boundary oracle of $\mathcal{K}$ obtaining a segment in the interior of $\mathcal{K}$. Then, the next Markov point is chosen according to $\pi_{\ell}$, i.e., the constraint of $\pi$ over that segment. To sample from $\pi_{\ell}$ one could use a univariate Metropolis Hastings algorithm for $\OO(1)$ iterations, or perform exact sampling if there is a closed form of $\pi_{\ell}$ at hand.


\paragraph{Practical performance.}
Extended experiments~\cite{bib:CousinsVempala2016, haraldsdottir2017chrr} have shown that HaR mixes after $\sO(n^2)$ steps for well-rounded convex polytopes; this result agrees with the latest theoretical bounds on the mixing time~\cite{Chen2022}. Moreover, the experimental results in~\cite{bib:CousinsVempala2016} show a convergence after $\sO(n^2)$ steps for the Gaussian distribution. Interestingly, the constant in the $\sO(\cdot)$ notation is many orders of magnitude smaller than in theory. For example, in~\cite{haraldsdottir2017chrr} they propose a step of $8n^2$.

\paragraph{Software.}
The current (competitive) implementations of HaR include, the C++ library \texttt{volesti}~\cite{Chalkis21,volesti24}, the MATLAB library \texttt{cobra}~\cite{cobraToolbox}, the C++ library \texttt{HOPS}~\cite{HOPS20}, as well as the R packages \texttt{HitAndRun}~\cite{HitAndRunCranPkg} and \texttt{walkr}~\cite{walkr20}. The most efficient ones are those in \texttt{volesti} and \texttt{HOPS} as they are both optimized C++ implementations.

\begin{algorithm}[t!]
  \caption{Hit\_and\_Run$(\mathcal{K},p,\pi)$}
	\label{alg:hnr}
	\SetKwInOut{Input}{Input}
	\SetKwInOut{Output}{Output}
\Input{Convex body $\mathcal{K}\subset\RR^n$, point $p\in \mathcal{K}$, $\pi$ probability distribution over $\mathcal{K}$}
\Output{A point $q\in \mathcal{K}$}
\BlankLine
Pick a line $\ell$ through $p$\;
\textbf{return} a random point $q$ on the chord $\ell \cap \mathcal{K}$ chosen from the distribution $\pi_{\ell}$\;
\end{algorithm}

\subsection{Coordinate Directions Hit-and-Run (CDHR)}

CDHR can be used to sample from any distribution $\pi$ over a convex body $\mathcal{K}$. It was first introduced for uniform sampling in~\cite{smith84}. Regarding applications, CDHR remained in the dark for years, with a few exceptions~\cite{Berbee1987} until it used in practical volume approximation algorithms~\cite{Emiris14, bib:CousinsVempala2016} for either uniform or spherical Gaussian sampling. Interestingly, in both cases CDHR outperforms HaR. This success story was followed by~\cite{haraldsdottir2017chrr}, where CDHR was used to uniformly sample steady states of metabolic networks, where the sampled region is a convex polytope $\mathcal{P}$. In~\cite{haraldsdottir2017chrr} they combine CDHR with a preprocessing step that brings the polytope into John position. They call this algorithmic pipeline as Coordinate Hit-and-Run with Rounding (CHRR). Since then, CDHR was considered as the main paradigm for sampling from polytopes until the development of a new implementation of BiW as we present in Section~\ref{subsec:billiard_walk}. Considering sampling from non-linear constrained regions, in~\cite{Chalkis22} they implement CDHR for log-concave sampling from the feasible region of Semidefinite programs. In their experimental analysis they show that CDHR outperforms HaR but its performance is worse than BiW in the case of uniform sampling. The success of CDHR in practice increased the amount of effort spent from researchers to study its mixing time with~\cite{Laddha23, Narayanan20} giving the first upper bound guarantees for uniform sampling. 

\paragraph{Step overview.}
In each step we uniformly pick an orthonormal vector $e_k\in\RR^n$ that defines the line $\ell$ passing thought the current Markov point $p$ (see Alg.~\ref{alg:cdhr}). When the sampled body is a convex polytope, the boundary oracle in the very first step costs $O(md)$ as showed above. For the step $j$-th step with $j>1$, to compute the intersection between $\ell$ and $\partial \mathcal{P}$ we compute th following roots,
\begin{equation}
    a_i^T(p + te_k) = b_i \Rightarrow t = \frac{b_i - a_i^Tp}{a_i^Te_k} ,\ i \in[m],
\end{equation}
and we keep the smallest positive and the largest negative. Notice that $a_i^Tp$ is already computed in the $(j-1)$-th step and $a_i^Te_k$ takes $1$ operation. Sampling from the constrained $\pi_{\ell}$ takes $\sO(1)$ operations and, moreover, we have to update just one coordinate per step. Thus, the amortized cost per step of CDHR for polytopes is $O(m)$. It has the smallest per-step cost among all the known geometric random walks together with Ball walk in the case of convex polytopes. 

When the sampled body is an intersection between a polytope and an ellipsoid $\mathcal{K} = \mathcal{P} \cap \mathcal{E}$ the cost of computing the intersection of the line $\ell$ with the boundary of the ellipsoid $\mathcal{E}$ is added to the cost per step of CDHR. For this we have,
\begin{equation}
    \label{eq:ellipsod_cdhr_bnd_oracle}
    (p + te_k)^TE(p + te_k) = c \Rightarrow E_{kk}t^2 + (2p^TEe_k)t - c = 0 ,
\end{equation}
where the dominant computation is the dot product between $p$ and the $k$-th column of the matrix $E$ which takes $\OO(n)$ operations. Then, the real roots in the Eq.(\ref{eq:ellipsod_cdhr_bnd_oracle}) correspond to the intersection $\ell \cap \partial\mathcal{K}$. Thus, the amortized cost per step of CDHR for sampling $\mathcal{K}$ is $\OO(m+n)$.

\begin{algorithm}[t!]
  \caption{Coordinate\_Directions\_Hit\_and\_Run$(\mathcal{K},p,\pi)$}
	\label{alg:cdhr}
	\SetKwInOut{Input}{Input}
	\SetKwInOut{Output}{Output}
\Input{Convex body $\mathcal{K}\subset\RR^n$, point $p\in \mathcal{K}$, $\pi$ probability distribution over $\mathcal{K}$}
\Output{A point $q\in \mathcal{K}$}
\BlankLine
Pick a standard orthonormal vector $e_i,\ i\in[n]$\;
Let the line $\ell = \{x\in\RR^n\ |\ x = p + te_i,\ t\in\RR\}$ through $p$\;
\textbf{return} a random point $q$ on the chord $\ell \cap \mathcal{K}$ chosen from the distribution $\pi_{\ell}$\;
\end{algorithm}

\paragraph{Practical performance.}
Extended experiments~\cite{bib:CousinsVempala2016,haraldsdottir2017chrr} have shown that CDHR mixes after $\sO(n^2)$ steps achieving the same mixing rate with HaR and BaW. In~\cite{haraldsdottir2017chrr} they propose a step of $8n^2$ to sample from polytopes in John position. Since CDHR achieves the same mixing rate with HaR in practice, it outperforms HaR because of its cheaper cost per step.

\paragraph{Software.}
CDHR is implemented in the C++ library \texttt{volesti}, the MATLAB library \texttt{cobra} and the C++ library \texttt{HOPS}. The most efficient ones are those in \texttt{volesti} and \texttt{HOPS} as they are both optimized C++ implementations.

\subsection{Billiard walk (BiW)}
\label{subsec:billiard_walk}

BiW method is designed for sampling exclusively from the uniform distribution over a convex (or non-convex) body. It was first introduced in~\cite{MeerscheXsample} and later independently by Polyak and Gryazina in~\cite{bib:GryazinaPolyak2014} 
where they prove asymptotic convergence for convex and non-convex bodies. Nevertheless, the mixing time is remains an open problem. In both~\cite{MeerscheXsample, bib:GryazinaPolyak2014}, the authors experimentally showcase that the mixing rate of BiW surpasses that of HaR for a limited class of bodies.

To our knowledge, considering applications that require uniform sampling, BiW has been used 
to address important applications in~\cite{ChalkisMetabolic21, Chalkis23}. 
In~\cite{ChalkisMetabolic21} they introduced an enhancement in the cost per step of BiW for convex polytopes, demonstrating superior performance over both HaR and CDHR when sampling the largest human metabolic network known at that time. The authors also present a multi-phase sampling pipeline that enhances the mixing rate of BiW from one phase to the next, ultimately providing both a rounded body and a uniformly distributed sample in the initial body upon termination. 
In~\cite{Chalkis23}, the improved version of BiW is utilized to estimate polytope volumes in thousands of dimensions for the first time.

\begin{algorithm}[t!]
  \caption{Billiard\_Walk$(\mathcal{K}, p, \rho, \tau)$}
	\label{alg:reflectionop}
	\SetKwInOut{Input}{Input}
	\SetKwInOut{Output}{Output}
    \SetKwRepeat{Do}{do}{while}
	
    \Input{Convex body $\mathcal{K}$; point $p \in \mathcal{K}$; upper bound on the number of reflections $\rho$; length of trajectory parameter $\tau$.}

    
	\Output{A point $q$ in $\mathcal{K}$}
    
    \BlankLine
    	Length of the trajectory $L \leftarrow -\tau\ln\eta$, $\eta\sim \mathcal{U}(0,1)$\;
    	Current number of reflections $i\leftarrow 0$\;
    	Initial point of the step $p_0\leftarrow p$\;
    	Uniformly distributed direction $v\in\mathcal{S}_{n-1}$\;
    	\Do{$i\leq \rho$} {
    		Segment $\ell \leftarrow \{ x_i + t v_i, 0 < t \leq L \}$ \;
    		\If{$\partial P\cap\ell=\emptyset$} {
    			$p_{i+1} \leftarrow p_i + L v_i$\;
                $i \leftarrow i+1$\;
    			\textbf{break}\;
    		}%
    		$p_{i+1} \leftarrow \partial \mathcal{K}\cap\ell$\;
    		The inner vector, $s$, of the tangent plane on $p$, s.t.\ $||s|| = 1$\;
    		$L \leftarrow L - |P\cap\ell|$\;
    		$v_{i+1} \leftarrow v_i - 2\langle v_i,s \rangle s$\;
    		$i \leftarrow i+1$\;
    	}
        $q \leftarrow p_i$\;
    	\If{$i=\rho+1$} {
    		$q \leftarrow p_0$\;
    	}
    	
    \textbf{return} $q$\;
\end{algorithm}

\paragraph{Step overview.}
BiW exploits linear trajectories and boundary reflections. In each step it uniformly picks a unit direction as HaR. The random walk being at $p$ moves only forward. When the ray hits the boundary it is reflected, while it might be further reflected $\rho$ times in total. The step is completed when traveling for a length $L$ picked from a fixed distribution. If the trajectory is reflected more than $\rho$ times before travelling length $L$, the random walk stays on $p$.

In the case of a convex polytope $\mathcal{P}$ in~\cite{ChalkisMetabolic21} they reduce the cost per step by cinducting the following analysis: To compute the intersection of the ray $\ell:=\{p+tv,\ t>0 \}$ with $\partial \mathcal{P}$ in the very first step, BiW uses the boundary oracle of $\mathcal{P}$ which costs $\OO(mn)$ operations. Let $p_i$ and $v_i$ the point and the direction vector of each reflection in a BiW step, while $i>1$. To compute the next intersection we have to solve, 
\begin{equation}\label{eq:billiard_implementation1}
a_j^T(x_i + tv_i) = b_j \Rightarrow a_j^T(x_{i-1} +  t_{i-1}v_{i-1}) + ta_j^T(v_{i-1}-2\langle v_{i-1},a_k\rangle a_k) = b_j,\ j=[m]
\end{equation}
and
\begin{equation}\label{eq:billiard_implementation2}
v_{i+1} = v_i -2\langle v_i,a_l\rangle a_l,
\end{equation}
where $a_k,\ a_l$ are the normal vectors of the facets that $\ell$ hits at iteration $i-1$ and $i$ respectively. Index $l$ is computed by solving all the $m$ equations in~(\ref{eq:billiard_implementation1}) while each one of them is solved after $O(1)$ operations as we use the stored values from the previous iteration. Also the inner product $\langle v_i,a_l\rangle$ in Eq.~(\ref{eq:billiard_implementation2}) is stored in the previous iteration. After computing all $\langle a_i,a_j\rangle$ as a preprocessing step, which takes $m^2d$ operations, the per-step cost of BiW is $\OO((d+\rho)m)$.

In the case of the convex body $\mathcal{K}$ (a polytope intersected with an ellipsoid) the computations in Eq.~(\ref{eq:billiard_implementation1}) can not be performed after $\OO(m)$ operations as in the case of polytope. In particular, the direction $v_{i-1}$ might be reflected from the ellipsoid's boundary and thus, a constant number of dot products must be computed in the $j$-th equation resulting to a total of $\OO(mn)$ operation in Eq.~(\ref{eq:billiard_implementation1}). Since the intersection between a ray and the boundary of ellipsoid is computed after $\OO(n^2)$ operations, the total cost per step in this case is $\OO(\rho (mn + n^2))$.

\paragraph{Practical performance.}
In~\cite{Chalkis23} they experimentally show that BiW mixes after $\sO(1)$ steps for well-rounded convex bodies. Moreover, in both~\cite{ChalkisMetabolic21, Chalkis23} they show that setting $\rho = \OO(n)$ suffices to both achieve this mixing rate and to avoid pathological cases where the trajectory stacks in corners. The latter implies that in practice BiW achieves a $\sO(1)$ mixing time while the cost per step is $\OO(mn)$ for well-rounded polytopes. In the case of the intersection between a polytope and an ellipsoid the cost per step becomes $\OO(n^2)$.

\paragraph{Software.}
BiW is implemented in the C++ library \texttt{volesti} that provides both the improved version of BiW and the multi-phase sampling method given in~\cite{ChalkisMetabolic21}. 
BiW is also implemented in the R package \texttt{Xsample}~\cite{MeerscheXsample} where it was first introduced as a random walk.


\subsection{Dikin walk}

Dikin walk can be used only for uniform sampling from a convex body; here we consider only the case of a convex polytope $\mathcal{P}$. It was first defined in~\cite{bib:KannanNarayanan2012} where they employ interior point methods based on Dikin ellipsoids, which by definition are enclosed in $\mathcal{P}$. Dikin walk is invariant under affine transformations of the polytope, meaning the mixing rate does not depend on the roundness of the polytope, i.e., the sandwiching ratio $R/r$, as the previously presented random walks. In~\cite{bib:KannanNarayanan2012} they bound the mixing time by $\sO(mn)$ steps starting from a warm start (in~\cite{Sachdeva2016} they provide a shorter proof of a Gaussian variant). When it starts from the Chebychev center they prove that $\sO(mn^2)$ steps require to converge to the target distribution. Moreover, they use Dikin walk to solve linear programs after $\sO(mn^2)$ steps. To define the Dikin ellipsoids they use the Hessian of the simple log-barrier of the polytope,
\begin{equation}\label{eq:hessian_barrier}
H(x) = \sum_{1\leq i \leq m}\frac{a_ia_i^T}{(1-a_i^Tx)^2},\ x \in \mathcal{P} .
\end{equation}

In~\cite{Laddha20b} they introduce the notion of strong self-concordance of a barrier function. They prove that Dikin walk mixes in $\sO(n\hat\nu)$ steps from a warm start in a convex body using
a strongly self-concordant barrier with symmetric self-concordance parameter $\hat\nu$. They show that the  strong
self-concordance of Lee-Sidford barrier is bounded from the the standard self-concordance parameter $\nu = \OO(n)$. Thus, they prove that Dikin walk, in a polytope $\mathcal{P}$, with Lee-Sidford barrier mixes after $\sO(n^2)$ steps.

Finally, in~\cite{narayanan16} they extend the results in~\cite{bib:KannanNarayanan2012} for an intersection between a polytope and a spectrahedron, that is the feasible region of a Semidefinite Program, i.e., the convex set given by a Least Matrix Inequality (LMI). They prove that Dikin walk mixes after $\sO(mn + n\nu_h)$, where $\nu_h$ is the size of the matrices in LMI.

\paragraph{Step overview.}
Dikin walk uses the Hessian of a convex barrier function to define an ellipsoid centered at the current Markov point, which is contained in the body. In each step picks a uniformly distributed point in the ellipsoid. The new point is accepted with a probability that depends on the ratio of the volumes of the ellipsoids centered at the two points. The latter filter is applied to guarantee the convergence to the uniform distribution. 

Since the implementation of each step requires to compute the volumes of two ellipsoids, which involves the computation of the determinant of a positive matrix, the cost per step of Dikin walk is $\OO(mn^{\omega - 1})$. In~\cite{Laddha20b} they show that the step of Dikin walk with the simple log-barrier can be implemented after $\OO(\text{nnz}(A) + n^2)$, where $\text{nnz}(\cdot)$ denotes the number of non-zero elements in matrix $A$.

\begin{algorithm}[t!]
  \caption{\dikin$(\mathcal{P}, p, H(\cdot), r)$}
	\label{alg:reflectionop}
	\SetKwInOut{Input}{Input}
	\SetKwInOut{Output}{Output}
    \SetKwRepeat{Do}{do}{while}
	
    \Input{polytope $\mathcal{P}$; positive definite matrix $H(x)$ for each point point $x\in\mathcal{P}$; $p\in\mathcal{P}$; radius $r$.}

    
	\Output{A point $q$ in $\mathcal{P}$}
    
    \BlankLine
    Let the ellipsoid $E_p(r) = \{ x \in\RR^n\ |\ (x-p)^TH(p)(x-p)\}$\;
    Pick $y$ uniformly from $E_p(r)$\;
    $\alpha = \min\bigg\{ 1, \frac{vol(E_p(r)}{vol(E_y(r))} \bigg\}$\;
    $u\sim \mathcal{U}[0,1]$\;
    \textbf{if }$u \leq \alpha$ \textbf{then }$q\leftarrow y$ \textbf{else }$q\leftarrow p$\;
    \textbf{return} $q$\;
\end{algorithm}

\paragraph{Practical performance.}
In~\cite{PolytopeWalk21} they implement Dikin, Vaidya and John walk as given in~\cite{chen18}. They compare them by performing experiments on well-rounded polytopes. In most of the cases Dikin walk beats both Vaidya and John walk. However, no extensive experimental evaluation of Dikin walk took place until now. Moreover, to our knowledge, there is not any important application that was addressed by Dikin walk.

\paragraph{Software.}
Dikin walk is implemented in the C++ package \texttt{PolytopeWalk}. The same implementation is integrated in \texttt{volesti}. Last, there is an R implementation provided by the R package \texttt{walkr}~\cite{walkr20}.

\subsection{Vaidya walk}

Vaidya walk is closely related to Dikin walk. It can be used to sample from the uniform distribution over a convex polytope $\mathcal{P}$. The point of difference is the ellipsoid it uses in each step. Instead of a Dikin it uses a Vaidya ellipsoid centered on the current Markov Point which is is also included in the polytope $\mathcal{P}$~\cite{vaidya96}. 
In \cite{chen18} they prove an upper bound $\sO(m^{1/2}n^{3/2})$ for the mixing time of Vaidya walk starting from a warm start. Notice that comparing Vaidya with Dikin walk appears a trade-off between the number of facets $m$ and the dimension $n$. Vaidya walk is more efficient than Dikin walk when $m \gg n$, but when $m$ is small the mixing rate of Dikin walk outperforms that of Vaidya walk. 

Vaidya walk uses a weighted version of the simple log-barrier function. When using the simple log-barrier, Dikin walk's efficiency relies on the representation of $\mathcal{P}$ and in particular in the number of redundant inequalities. A polytope $P$ remains unchanged if we add any number of redundant inequalities, but on the other side, they have heavy effect on the Hessian of $f$ and thus, reducing significant the size of Dikin ellipsoid and affecting crucially the mixing time of Dikin walk. The main idea behind Vaidya walk is to deal with this problem and try to improve the dependence of Dikin walk on the number of facets $m$. An intuitive thought that Vaidya walk follows is to use a barrier with unequal weights for the logarithmic terms in the simple log-barrier function or, equivalently, a matrix with unequal weights in Eq.~(\ref{eq:hessian_barrier}) to define the ellipsoid in each step of the random walk. Such modifications has been used in optimization \cite{vaidya96, lee14} providing significant improvements.

\begin{algorithm}[t!]
  \caption{\vaidya$(P, p, r)$}
	\label{alg:vaidya}
	\SetKwInOut{Input}{Input}
	\SetKwInOut{Output}{Output}
    \SetKwRepeat{Do}{do}{while}
	
    \Input{polytope $\mathcal{P}$; point $p\in\mathcal{P}$; radius $r$.}

    
	\Output{A point $q$ in $\mathcal{P}$}
    
    \BlankLine
    	$C\sim$ fair coin\;
        
    	\textbf{if }$C$ head \textbf{then } do nothing, go to the next step\;
    	Pick $\xi\sim \mathcal{N}(0,I_n)$\;
    	Propose $y = p + \frac{r}{(mn)^{1/4}}V_p^{-1/2}\xi$\;
        \textbf{if }$y\notin \mathcal{P}$ \textbf{then }$q\leftarrow p$ \textbf{end}\\
        \ElseIf {$y\in \mathcal{P}$} {
            $\alpha = \bigg\{ 1, \frac{g_y(p)}{g_p(y)} \bigg\}$\;
            $u\sim \mathcal{U}[0,1]$\;
            \textbf{if }$u \leq \alpha$ \textbf{then }$q\leftarrow y$ \textbf{else }$q\leftarrow p$\;
    	}
    \KwRet $q$\;
\end{algorithm}

\paragraph{Step overview.}

Vaidya walk instead of uniformly distributed proposals from state-dependent ellipsoids it uses a Gaussian proposal at the current Markov point $p$ of the random walk and the point-depended covariance being,

\begin{equation}
V_p = \sum_{i=1}^m(\sigma_{p,i} + \beta_{\nu})\frac{a_i^Ta_i}{(1-a_i^Tp)^2}
\end{equation}
where $\beta_{\nu} = n/m$ and the scores,
\begin{equation}
\sigma_p := \bigg( \frac{a_1^T H(p)a_1}{(1-a_1^Tp)^2},\dots ,\frac{a_m^T H(p)a_m}{(1-a_m^Tp)^2} \bigg)
\end{equation}

A single step of Vaidya walk with radius parameter $r>0$ is given by the pseudocode of Alg.~\ref{alg:vaidya}. With $g_p(z)$ we write the probability density function of the Gaussian with covariance $V_p$ and centered at $p\in \mathcal{P}$,
\begin{equation}
g_p(z) = \sqrt{\det V_p}\bigg( \frac{mn}{2\pi r^2} \bigg)^{n/2} \exp \bigg( -\frac{mn}{2r^2}(z-p)^TV_x(z-p)\bigg)
\end{equation}
%



\paragraph{Practical performance.}
In~\cite{chen18} they compare Vaidya walk against Dikin and John walk by performing experiments on well-rounded polytopes. In most of the cases Vaidya walk beats John walk due to a cheaper cost per step. However, no extensive experimental evaluation of Vaidya walk took place until now. 

\paragraph{Software.}
Vaidya walk is implemented in the C++ package \texttt{PolytopeWalk} which is based on the algorithms given in~\cite{chen18}. The same implementation is integrated in \texttt{volesti}.

\subsection{John walk}
John walk was first introduced for uniform sampling from a convex polytope in~\cite{bib:GustafsonNarayanan2018} and latter an improved version was given in~\cite{chen18} introducing Approximate John walk. It is also relevant to Dikin walk and it goes one step further than Vaidya walk to establish a sublinear dependence on the number of facets $m$ in the mixing time. In~\cite{bib:GustafsonNarayanan2018} they compute the exact largest ellipsoid centered on the current Markov point $p$, namely the exact John ellipsoid. Their analysis gives a mixing time of $\sO(n^7)$ steps. Interestingly, in~\cite{chen18} they employ the analysis in~\cite{lee14} that use John's weights to improve the performance of interior point methods to solve linear programs by employing approximate John ellipsoids. In~\cite{chen18} they integrate that analysis to improve the mixing time of John walk and prove a bound of $\sO(n^{2.5})$ steps which depends logarithmic on the number of facets $m$.

\paragraph{Step overview.}
In each step being at a point $p\in \mathcal{P}$ John walk approximates the largest inscribed ellipsoid in $\mathcal{P}$ centered at $p$ and employs the covariance matrix that defines the ellipsoid to compute a Gaussian proposal. The inverse covariance matrix underlying the John walk is given by,

\begin{equation}
J_p = \sum_{i=1}^m \zeta_{p,i}\frac{a_ia_i^T}{(1-a_i^Tp)^2} .
\end{equation}
the weight vector $\zeta_p\in\RR^m$ is obtained by solving the convex program,

\begin{equation}\label{eq:john_weights}
\zeta_p = \argmax\limits_{w\in\RR^m}\bigg\{ \sum_{i=1}^m w_i - \frac{1}{a_J}\log\det (A^TS_p^{-1}W^{a_J}S_p^{-1}A) - \beta_J \sum_{i=1}^m\log w_i \bigg\},
\end{equation}
where $\beta_J := d/2m$, $a_J := 1 - 1/ log_2 (1/\beta_J)$, matrix $W=\text{diag}(w)$ and $S_p$ the slackness matrix at $p$. The convex program of Eq.~(\ref{eq:john_weights}) was introduced in \cite{lee14} and $\zeta_p$ are called approximate John weights. It is also closely related with the problem of computing the largest ellipsoid in $P$ centered at $p$, which was first studied by John \cite{john48}. However, John walk step is costlier than that of Dikin and Vaidya walk by a constant factor~\cite{chen18}. 


\begin{algorithm}[t!]
  \caption{\john$(\mathcal{P}, p, r)$}
	\label{alg:john_walk}
	\SetKwInOut{Input}{Input}
	\SetKwInOut{Output}{Output}
    \SetKwRepeat{Do}{do}{while}
	
    \Input{polytope $\mathcal{P}$; point $p$; radius r of the John ellipsoid.}

    
	\Output{A point $q$ in $\mathcal{P}$}
    
    \BlankLine
    	$C\sim$ fair coin\;
    	\textbf{if }$C$ head \textbf{then } do nothing, go to the next step\;
    	Propose $y \sim \mathcal{N}\bigg( x,\frac{r^2}{(d)^{3/2}}J_x^{-1}\bigg)$\;
    	\textbf{if }$y\notin \mathcal{P}$ \textbf{then }$q\leftarrow p$ \textbf{end}\\
        \ElseIf {$y\in \mathcal{P}$} {
            $\alpha = \bigg\{ 1, \frac{g_y(x)}{g_x(y)} \bigg\}$\;
            $u\sim \mathcal{U}[0,1]$\;
            \textbf{if }$u \leq \alpha$ \textbf{then }$q\leftarrow y$ \textbf{else }$q\leftarrow p$\;
    	}
    \KwRet $q$\;
\end{algorithm}


\paragraph{Practical performance.}
In~\cite{chen18} they compare Vaidya walk against Dikin and John walk by performing experiments on well-rounded polytopes. Both Dikin and Vaidya presented to achieve a better performance than John walk. However, no extensive experimental evaluation of John walk took place until now. 

\paragraph{Software.}
John walk is implemented in the C++ package \texttt{PolytopeWalk}. The same implementation is integrated in \texttt{volesti}.


\subsection{Hamiltonian Monte Carlo (HMC)}

HMC can be used to sample from any probability density function $\pi(\cdot)$ in $\RR^n$. It simulates a particle moving in a conservative field determined by $-\log\pi(p)$ and $-\nabla\log\pi(p)$. Being at $p\in \mathcal{K}$, HMC introduces an auxiliary random variable $v\in\RR^d$, called the \emph{momenta}, and generates samples from the joint density function
\begin{equation}
\pi(p,v) = \pi(v|p)\pi(p) .
\end{equation} 
Then, marginalizing out $v$ recovers the target density $\pi(p)$. Usually we consider the momenta $v$ to follow the standard Gaussian distribution $\mathcal{N}(0,I_n)$. Then, the probability density function,
\begin{equation}
    \pi(p,v) \propto e^{-H(p,v)},
\end{equation}
where $H(p,v) =  -\log\pi(p,v) =  -\log\pi(p) + \frac{1}{2}\vert v\vert^2$ is the corresponding \emph{Hamiltonian function}. In each step being at $p$, HMC draws a value for the momentum, $v\sim\mathcal{N}(0,I_n)$.  Then, $(p,v)$ is given by the Hamilton's system of ordinary differential equations (ODE),
\begin{equation}
    \label{eq:ode_hmc}
	\begin{split}
	\frac{dp}{dt} = \frac{\partial H(p,v)}{\partial v}\\
	\frac{dv}{dt} = -\frac{\partial H(p,v)}{\partial p} 
	\end{split}
	\quad\Rightarrow\quad
	\left\{
	\begin{array}{lll}	
	\frac{dp(t)}{dt} = v(t)\\
	\quad\\
	\frac{dv(t)}{dt} = -\nabla\log\pi(p)\\
	\end{array} 
	\right. 
\end{equation}
The solution of the ODE in Eq.~(\ref{eq:ode_hmc}) gives a trajectory $p(t) \in \R^n$, where, ideally, the next Markov point is randomly picked from it. When a closed form solution is not known, the ODE is solved by leapfrog method~\cite{} so that the convergence is guarantee. Thus, to discretize the Hamiltonian Dynamics we use the leapfrog integrator,
\begin{equation*}
    \hat v_{i + 1} = v_i + \frac \eta 2 \nabla \log(\pi(p_i)), \quad p_{i + 1} = p_i + \eta \hat v_{i + 1}, \quad v_{i + 1} = \hat v_{i + 1} + \frac \eta 2 \nabla\log(\pi(p_{i+1})) ,
\end{equation*}
where $\eta > 0$ is the leapfrog step. The new position is accepted with a probability obtained by a regular Metropolis filter. In general, the step length and the number of steps until we obtain the next Markov point are inputs. However, there are several heuristics to determine those two parameters~\cite{Gelman14, Chalkis23b} to improve the mixing rate of the random walk.

In the case where $\pi$ is restricted over a convex body $\mathcal{K}$ there are two versions of HMC that can sample from the target distribution. The first is the Reflective HMC that employs boundary reflections in each leapfrog step so that the random walk stays inside the body. The second is the Riemannian HMC that uses the log-barrier of the convex body to satisfy the constraints in each step.

\subsection{Reflective NUTS HMC (ReNHMC)}

The Reflective HMC using a leapfrog integrator was introduced for log-concave sampling from linear and non-linear convex bodies~\cite{afshar2015, Chalkis23b}. The mixing time is unknown and an open problem. ReNHMC is a Reflective HMC sampler that employs the NUTS 
criterion~\cite{Gelman14}. NUTS exploits the fact that a Hamiltonian trajectory tends to fill the space by forming a spiral. Thus, the optimal number of leapfrog steps is that one when the trajectory starts making a U-turn and approaches the initial point. NUTS is a heuristic that checks the inner products of the position and momenta between the current and the initial position to declare stopping.

\paragraph{Step overview.}
When the position is updated, i.e., $p_{i + 1} = p_i + \eta \hat v_{i + 1}$, it might lie outside of $\mathcal{K}$. In this case, ReHMC reflects the ray $\{ p_i + t \hat v_{i + 1}\ |\ t \geq 0 \}$ so that the new proposal is inside $\mathcal{K}$. Notice, that the ray may need to be reflected more than once in the case where the step length $\eta$ again leads outside of $\mathcal{K}$ after a certain reflection. The proposal is again accepted after applying a Metropolis filter. In each leapfrog step we check the NUTS criterion given in~\cite{Gelman14}.

\paragraph{Practical performance.}
In~\cite{Chalkis23b} they show that ReHMC outperforms both HaR and CDHR for log-concave sampling from convex polytopes and spectrahedra in thousands of dimensions. ReNHMC has never been implemented for truncated sampling before this paper.

\paragraph{Software.}
ReNHMC is implemented in the C++ library \texttt{volesti}.

\subsection{Constrained Riemannian HMC (CoRHMC)}

CoRHMC introduced in~\cite{kook22} can be used for log-concave sampling over a convex polytope $\mathcal{P}$.
In general, Riemannian HMC uses a Riemannian manifold over the convex body $\mathcal{K}$. For this, it defines a local distance and integrate it in the Hamiltonian function, i.e.,
\begin{equation}
    H(p,v) = -\log\pi(p) + \frac{1}{2} v^T M(p)^{-1} v + \frac{1}{2}\log \text{det}M(p),
\end{equation}
where $M(p) \in \R^{n\times n}$ is a position-dependent positive definite matrix. The, the ODE solved in each step becomes,
\begin{equation}
\label{eq:hmc_ode_rihmc}
    \begin{split}
	& \frac{dp}{dt} = \frac{\partial H(p,v)}{\partial v}\ = M(p)^{-1}v \\
	& \frac{dv}{dt} = -\frac{\partial H(p,v)}{\partial p} = -\nabla\log\pi(p) - \frac{1}{2}\text{Tr}\bigg[ M(p)^{-1}\frac{\partial M(p)}{\partial p}\bigg]  + \frac{1}{2} p^TM(p)^{-1} \frac{\partial M(p)}{\partial p} M(p)^{-1} p ,
	\end{split}
\end{equation}
where $\text{Tr}[\cdot]$ is the regular matrix trace. Unfortunately, the discrete leapfrog integrator does not guarantee convergence to the target distribution. 
To solve this ODE one should use a symplectic numerical integrator~\cite{Girolami11}. However, this task seems quite challenging in practice. The main difficulty is to achieve the numerical accuracy required when the Markov point is close enough to the boundary so that the computations would not lead to numerical overflow. 

CoRHMC is a specialized sampler for convex polytopes. In~\cite{kook23} they provide efficient methods to solve the ODE in Eq.~(\ref{eq:hmc_ode_rihmc}) by achieving a cost per step of $\OO(n^{3/2})$. In~\cite{kook23} they show that CoRHMC mixes after $\sO(mn^3)$ steps which is worse than the mixing time of both HaR and BaW for rounded distributions. However, this sampler performs significantly well in practice by outperforming both HaR and CDHR as shown in~\cite{kook22}. Its main advantage is that its mixing rate does not depend on the roundness of the distribution; thus, CoRHMC does not require any rounding preprocess for skinny distributions (or skinny polytopes in the case of uniform distribution). Thus, it can sample efficiently from hard instances in thousands of dimensions within minutes~\cite{kook22}. 
Notably, it has never been performed a comparison between CoRHMC and the MMCS with BiW given in~\cite{ChalkisMetabolic21} for uniform sampling over a convex polytope.

CoRHMC is implemented in the MATLAB package \texttt{PolytopeSamplerMatlab}~\cite{RiHMCsampler}. 
\texttt{volesti} also contains an implementation of CoRHMC in C++.

\section{Empirical Study}
\label{sec:experiments}

We employ geometric random walks to explore factor anomalies in constrained long-only portfolios. To do so, we follow the investment guidelines of a prominent market index which we believe accurately represents the typical framework followed by many mutual funds and institutional investors. The index that we look at is the MSCI Diversified Multiple-Factor index\footnote{See \url{https://www.msci.com/diversified-multi-factor-index}.} (hereinafter referred to as the DMF index) for the U.S.\ market. 
The DMF index aims to create systematic exposure to the four factors: value, momentum, quality, and size, while maintaining a risk profile similar to that of the underlying capitalization-weighted parent index, which is the MSCI USA index. The allocation of the DMF index is determined through an optimization which maximizes the exposure to a combination of the four factors, subject to a risk tolerance set equal to the ex-ante variance of the parent index at the time of calculation.

MSCI's prospectus explains that the DMF index is designed to systematically capture the additional sources of systematic return associated with the four factor investing styles. However, the MSCI approach significantly differs from the typical academic literature on factor premia (as the performance difference between a long position in a portfolio of companies with the most favorable factor characteristics and a short position in a basket of assets with the least favorable factor characteristic) in that their model operates in a long-only setup with stringent constraints on asset weights, sector allocations, risk tolerance, and bandwidths on additional factor exposures, all relative to the parent index. In particular, asset weights are bound to the allocation of the parent index by a bandwidth of $\pm 2$\%-points per stock and of $\pm 5$\%-points on sector level. Within those limits, MSCI tries to harvest the premia by tilting the capitalization-weighted allocation towards stocks with elevated factor exposure. 

A pertinent question arises concerning the possibility of capturing academically investigated factor premia within such a strongly constrained setup. To find out we employ a geometric random walk to sample portfolios from within the set of constraints imposed by the DMF index and analyze whether the sampled portfolio's out-of-sample\footnote{Out-of-sample means that, at every point of the back-testing procedure, we only use information that was effectively available at that point in time.} performances varies in a systematic way with their factor scores. Do portfolios with higher value-, momentum-, quality-, size- or volatility-scores  deliver higher, lower, or equal return and risk than their low-scoring counterparts?



The traditional method of factor analysis involves forming percentile portfolios based on a sorting of companies idiosyncratic factor scores. Backtests are then conducted where each percentile portfolio is weighting it's constituents equally or according to capitalization. Factor anomalies are identified through the outperformance of the top percentile portfolio relative to the bottom percentile. Interaction with other factors is examined through regression analysis. In contrast, our RP-based approach offers anomaly detection within investor-defined constraints, and factor interactions can be directly controlled by incorporating factor constraints, eliminating the need for indirect regression methods.

\paragraph{Backtesting framework.}

In Section~\ref{sec:random_portfolio_generation}, we outlined the process of constructing a RP by initially defining investment constraints geometrically and then deriving the distribution of performance statistics either analytically or through sampling and simulation. However, this construction assumes the availability of clean time series data for all stocks, of equal length, to estimate market parameters. Consequently, the described RPs represent snapshots. For an extended analysis of a stock universe over time, adjustments must consider changes in index composition and corporate actions affecting individual stocks (such as splits, mergers, and dividends). These factors complicate matters as both the number of investable assets and the geometric constraints evolve over time. Therefore, akin to quintile portfolios, RPs must undergo regular rebalancings, precluding analytical solutions even for the naive case.

The method which we pursue is the following: Every month we construct a new RP by casting the constraints employed by the DMF index to a convex body. The linear constraints, which include the lower and upper bounds on asset, sector, and factor exposures, establish a polytope, and the constraint on variance shapes an ellipsoid. Notice that these bounds shift from one month to the next, adapting to changes in the allocation and risk level of the parent index. Then, we sample from the convex body using a geometric random walk. Because of the quadratic constraint, the choice of sampler is limited to either HaR, CDHR, BaW, BiW or ReNHMC. Among those, only HaR, CDHR, BaW and ReNHMC allow for a distribution other than the uniform. Since ReNHMC is considerably faster than HaR, CDHR and BaW 
(see \cite{Chalkis23b,bib:CousinsVempala2016}), ReNHMC is our method of choice (without the variance constraint we would use CoRHMC). As probability model we use the Dirichlet distribution with a parameter vector equal to the weights of the  capitalization-weighted parent index, i.e., $\alpha = \omega_{bm}$. 
This generates samples which have $\mathds{E}(\omega) = \omega_{bm}$, yet are close to the boundary of the constraints set\footnote{Recall that sampling with a parameter vector $\alpha$ with sum less than $n$ results in weights towards the boundaries. We want points to be close to the boundary since the solution of the optimization problem that the DMF index is solving is always a boundary point.}. 

For each sampled portfolio, we then simulate the cumulative returns path for the subsequent month using the returns of the portfolio's constituents. At the next rebalancing, the process is repeated and the simulated price paths are concatenated to consecutive time series spanning the entire period of analysis. 

Concatenation from one period to the next follows a similar logic to the formation of quintile portfolios. Portfolios are matched based on factor scores, where the portfolio with the highest exposure to a certain factor in one period is combined with the portfolio scoring highest in the following period, and so forth, down to the lowest scoring pair. Sorting the time-series by factor exposure allows for an analysis of the correlation between factor rank and performance over a long-term period. For example, if there is a monotonic relation between momentum score and return, this should be evident in the backtests through a decreasing out-of-sample return with the simulation index. In other words, the first simulation, forming a concatenation of best-momentum portfolios, should yield the largest cumulative return over the entire backtesting period, with performance gradually decreasing for subsequent simulations.

To evaluate the influence of the MSCI constraints, we iterate the backtesting exercise twice, initially applying only the simplex condition before introducing the MSCI constraints in a second round.

\subsection{Data}\label{sec:data}


The implementation of the MSCI Multifactor framework requires a comprehensive dataset of stock price series from companies covered by the MSCI USA index, which includes large and mid-cap equities traded in the U.S\footnote{See \url{https://www.msci.com/our-solutions/indexes/developed-markets}.}. As the composition of the MSCI USA index is proprietary, we suggest utilizing data from the Center for Research in Securities Prices (CRSP) as an alternative data source. We select a subset of stocks listed on the NYSE, AMEX, and NASDAQ indexes, focusing solely on ordinary common shares of companies incorporated in the U.S. (CRSP shares code 10 and 11). Additionally, we exclude foreign shares, certificates, American depository receipts, shares of beneficial interest, depository units, American trust components, closed-end funds, and real estate investment trusts from our dataset. We found that a good approximation to mimick the MSCI universe is to filter for the largest stocks listed on the three exchanges such that in aggregate they cover 85\% of total market capitalization. On average over the backtesting period, this results in about $600$ securities per rebalancing.
Though this approach does not precisely emulate the MSCI methodology, a comparative analysis using both datasets indicates that the results are qualitatively identical. The ultimate aim is not an exact replication but to provide a representative framework. Through the utilization of CRSP data, all our findings can be reproduced and validated.

Furthermore, to be included in our study, stocks need a consistent price history of five years without any gaps larger than two weeks. This is needed to compute the covariance matrix in the quadratic constraint bounding portfolio variance to that of the parent index. Moreover, illiquid stocks are removed from the investable universe. As a threshold, we require a median trading volume over the previous 252 trading days to be above USD 1.5 million. We do this because, on the one hand, such illiquid stocks are not easily tradable and therefore would lead to a large implementation shortfall (i.e., the difference between a simulated performance and one obtained from real investments) and on the other hand, such companies display artificially low volatility due to a lack of trading and not because they are not risky. 
The cleaning process is necessary to ensure that at every point in time, the investable universe only contains information that was effectively available at that point in time and to avoid any positive survivorship bias. 

All estimations are based on discrete daily total\footnote{Returns, i.e., the percentage changes in prices from time $t-1$ to $t$, are termed \emph{total} when adjusted for dividends (i.e., dividends are re-invested).} returns using closing prices denoted in USD, covering the period 01.01.1995 to 31.12.2022. Backtests begin on 03.01.2000. All descriptive statistics are calculated on annualized discrete monthly returns, as is customary in the financial industry.

\paragraph{Factors.}

The DMF index actively manages its exposure to specific target and non-target style factors relative to the parent index, not only via a weighting in the objective function, but also by setting benchmark-relative constraints. Book-to-Price, Earnings Yield, Earnings Quality, Investment Quality, Profitability, and Momentum exposures are required to fall within the range of $[0.1, 0.6]$, Earnings Variability, Leverage, and Size within $[-0.6, -0.1]$. Moreover, the index limits its exposure to non-target style factors, namely Beta, Residual Volatility, Growth, and Liquidity to remain within a range of $\pm 0.1$ standard deviations relative to the parent index. 

For value and quality, MSCI computes scores by a weighting of sub-scores. Value, for instance, consists of one-third Price-to-Book and two-thirds Earnings Yield. The scores are further sector standardized using the GICS\footnote{See \url{https://www.msci.com/our-solutions/indexes/gics}} classification system. Details can be taken from the MSCI methodology paper. 

Table ~\ref{tab:mapping_msci_to_jkp} in Appendix ~\ref{sec:appendix_factor_mapping} shows how we have mapped the MSCI factors to JKP factors. All scores (i.e., exposures) are ultimately standardized to z-scores and winsorized at $\pm3$. Note that the standardization is a monotonous transformation which does not affect the ordering of the scores. Winsorization, on the other hand, does have an influence. 

On a portfolio level, the factor score i.e., the exposure of a portfolio $\omega$ to some factor $f_s$ is simply taken to be the weighted sum of the portfolio weights times the company factor scores, i.e., $\sum_i \omega_i \beta_{i,s}$. We notice that, because of the winsorization step., the weighted sum method is not exactly the same as when we would regress the portfolio returns on the factors.

\subsection{Implementation}
\label{sec:implementation}

We use the C++ library \volesti~\cite{Chalkis21,volesti24}, an open-source library for high dimensional MCMC sampling and volume approximation. \texttt{volesti} provides the implementations of all the geometric random walks in Section~\ref{sec:geometric_random_walks}. 
The core of its implementation is in C++ to optimize performance while the user interface is implemented in R. The library employs
\eigen~\cite{eigenweb} for linear algebra and  \boost~\cite{boostrandom} for random number generation. It has been successfully used for analyzing and sampling steady states of metabolic networks~\cite{ChalkisMetabolic21}, for volume estimation of convex bodies~\cite{Chalkis23} and economical crises detection~\cite{bib:CalesChalkisEmirisFisikopoulos2018} as well as volatility anomaly detection in stock markets~\cite{bib:BachelardEtAl2023}.

\subsection{Results}
\label{sec:results}

\begin{figure}[h!] 
    \centering
    \includegraphics[width=1\linewidth]{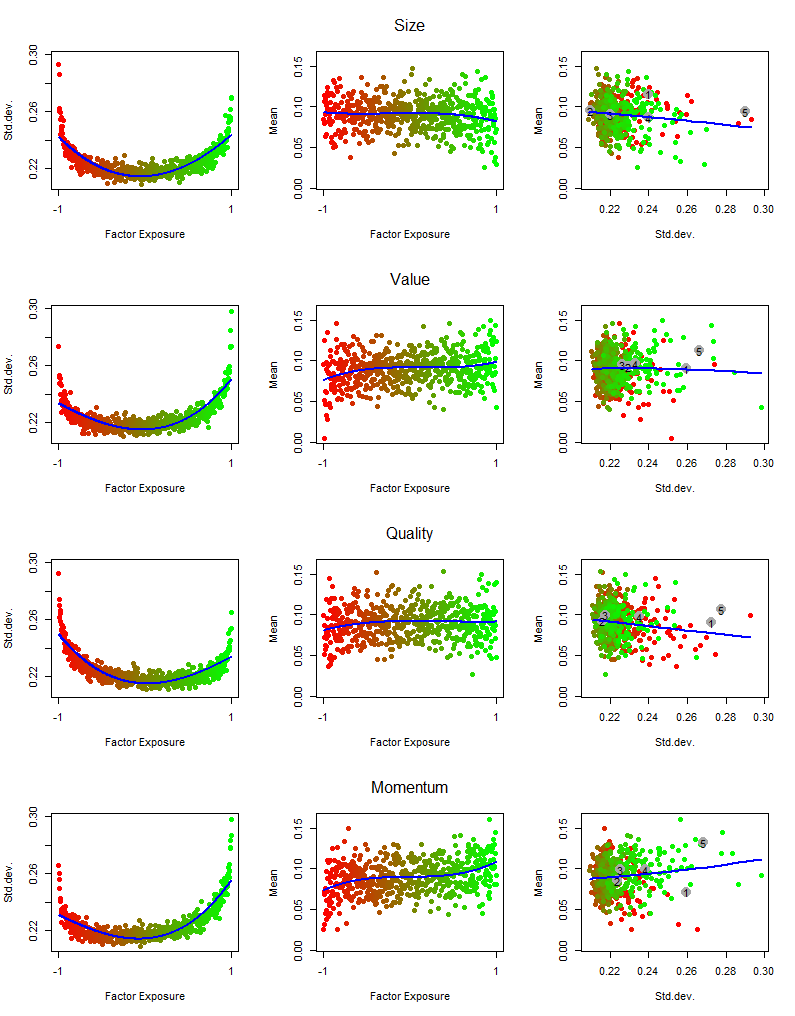}
    \caption{{\small Performance vs. factor exposure - Long Only. 
    \label{fig:factor_exposures_performance_unconstrained}}}
\end{figure}

\begin{figure}[h!] 
    \centering
    \includegraphics[width=1\linewidth]{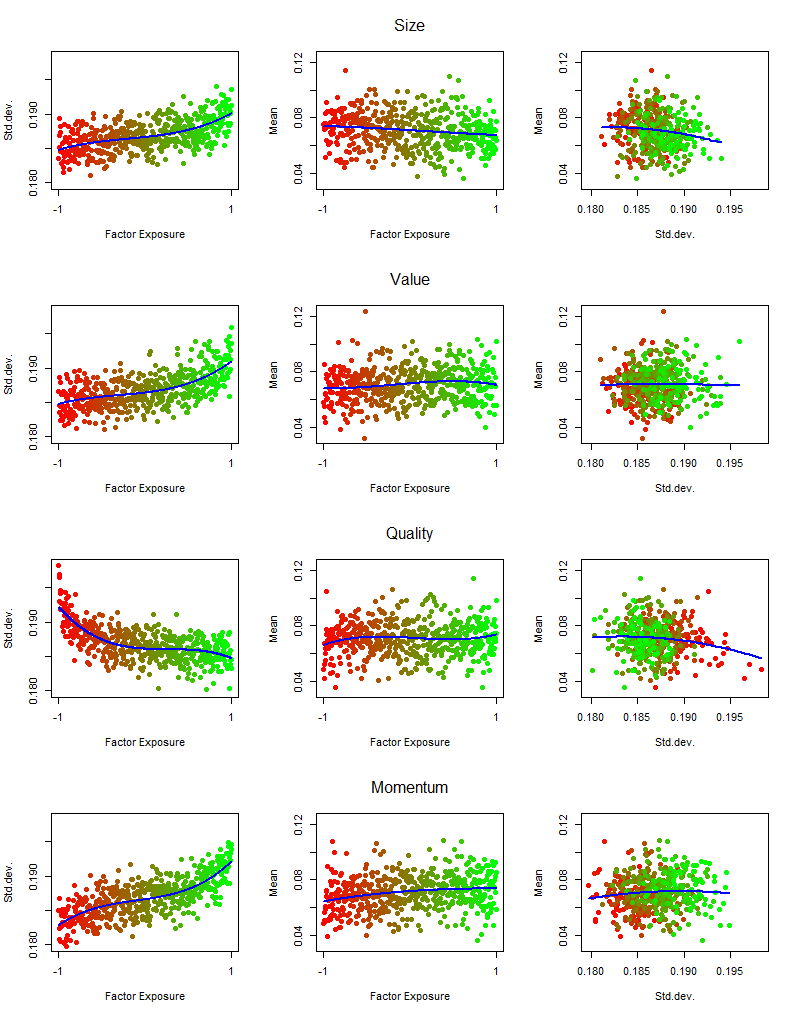}
    \caption{{\small Performance vs. factor exposure - Within the constraints of the MSCI Diversified Multifactor Index. 
    \label{fig:factor_exposures_performance_constrained}}}
\end{figure}

Figures ~\ref{fig:factor_exposures_performance_unconstrained} and ~\ref{fig:factor_exposures_performance_constrained} contain scatter-plots depicting the relation between portfolio factor exposure and portfolio risk (first column), portfolio factor exposure and portfolio return (second column) and portfolio risk and portfolio return (third column). Factor exposure levels are scaled for all factors to live on the interval [-1, 1]. Figure ~\ref{fig:factor_exposures_performance_unconstrained} reflects the case where no constraints are applied other than the simplex condition. The plots are overlayed by grey points, labeled one to five, indicating the performance statistics of five sorting-based quintile portfolios. The blue line indicates the fit of a polynomial regression (of degree two).
Figure ~\ref{fig:factor_exposures_performance_constrained} shows the results when the constraints imposed by the DMF index are met.

We note the following observations. In the first backtests, where only the simplex constraint is applied, the results confirm the presence of a positive relation between portfolio factor scores and returns in all factors except size. For size, the relation is slightly negative. This may be due to the fact the asset universe considered by MSCI contains only large and medium-sized companies. Value and momentum display a non-linearity with slopes becoming steeper at both ends.

The relation between portfolio factor score and risk is U-shaped meaning that both, portfolios with particularly high and particularly low factor exposure exhibit higher variance than their unexposed counterparts. This pattern is seen with all four factors. 

The results are mostly in line with classical sorting-based simulations which are overlayed in Figure ~\ref{fig:factor_exposures_performance_unconstrained} as grey dots labeled $1$ to $5$ indicating the corresponding (equally-weighted) quintile portfolios. The primary benefit of the sampling-based approach over the sorting-based method lies in its independence from the weighting within percentile portfolios, eliminating potential strong influences on the results. Additionally, the sampling method provides greater granularity by illustrating the uncertainty in the outcomes of percentile portfolios. Also, it can be seen that some of the quantile portfolio are rather outlying relative to the conditional performance distributions.

However, we view this initial application more as a proof-of-concept. The true advantage emerges when additional constraints are introduced, rendering sorting-based methods impractical.

The addition of the benchmark-relative bounds used by the DMF index has different effects. For momentum, the relation between factor exposure and return remains clearly positive with a correlation coefficient of 0.229, compared to 0.251 for the long-only case. For value, the correlation remains positive but weakens from initially 0.266 to 0.128. For quality, the positive correlation of 0.189 vanishes to 0.038. For size, where the relation was negative, the effect is further strengthened, showing in correlations dropping from initially -0.106 to -0.168. 

The non-linear relationships between factor scores and risk are significantly influenced. The graphs indicate that imposing constraints limits allocations 
in such a way that one ends up in one of the two legs of the previously identified U-shaped patterns. For size, value, and momentum, this corresponds to the right leg, exhibiting positive (near-linear) relations with correlations of 0.553, 0.693, and 0.813, respectively. Conversely, for quality, it corresponds to the left leg, displaying a negative correlation of -0.580.

Given the observable patterns in the constrained setup we draw the following conclusion: Tilts towards value and momentum are rewarded with higher returns but come at the price of increased volatility. The tilt towards quality has a risk reducing effect without impacting returns, thus increasing overall efficiency. The size tilt should not be actively sought as the relation between size exposure and both, risk and return, is unfavorable. However, we have to keep in mind that, due to the concentration of the capitalization-weighted allocation, any deviation from it will likely increase the size exposure. Value, momentum and quality tilts each imply a size tilt, though without the performance drag of an isolated size tilting. Overall, factor patterns appear to be present even within the very strong bounds implied by the DMF index and an active investment approach trying to harvest them seems justified.

Our method for sampling portfolios within given constraints would allow us to analyse the impact of specific constraints in isolation. For instance, it would be interesting to repeat the analysis after removing one of the constraints, like the bound on variance, or a group of constraints, like the control on non-target factor exposures. We leave this idea for future applied research.

To finalize, let us challenge once more the option to use the RP for statistical testing; here, to test the outperformance of the DMF index relative to its parent. With an Information ratio (IR) of 0.42, the DMF index resides at the 0.025 quantile of the distribution of IR's measured from the RP backtests. We could consider this as indication of skill for MSCI's factor team. However, the quantile value is in direct dependence of the parameter of the Dirichlet distribution which we chose to sample weights from. Although our choice of parametrization is economically justified, the sensitivity on inferential results is hardly acceptable. Conversely, the relation between factor exposure and performance is robust to a wide range of parameter values (we have tested $\alpha = \omega_{bm} \times \{0.01, 0.1, 1, 5, 10\}$, all giving qualitatively similar results. Only with a scaling factor $>10$ do the resulting samples become too homogenous).


\subsection{Conclusion}
\label{sec:conclusion}

We discussed the use of randomized control techniques in the context of empirical asset pricing and performance evaluation. We examined the scenario where the performance distribution of a random portfolio , serving as the random control group, forms the basis for assessing the performance of an elaborated investment strategy relative to the distribution of performances under chance. A random portfolio was defined as a portfolio whose weights constitute a random vector with a specific probability distribution supported on a bounded space, reflecting the investor's investment constraints, such as bounds on asset weights or risk limits. 

We explained that commonly employed random portfolios do not offer a statistically acceptable experimental design for performance comparisons and cautioned the reader against such a use case with the aid of an illustrative example. However, under consideration of the random portfolios construction methodology, we found random portfolios to be a helpful tool for visual and descriptive perform analysis. Furthermore, we suggested a novel use case, involving the investigation of the relationship between portfolio factor scores and performance in a constrained setup, which allows to empirically test for the presence of factor anomalies under stringent investment guidelines.

We proposed a categorization of random portfolios into three groups — naive, basic, and regularized — based on weight distribution and geometric structure of the distribution's support. Additionally, we introduced geometric random walks, a class of Markov chain Monte Carlo  methods, as a novel and powerful paradigm to sample random portfolio weights adhering to linear and quadratic constraints under various models. We provided a survey of all existing geometric random walks, explaining their strengths, limitations, and computational complexities.

Ultimately, we performed experiments on real-world datasets using the open-source software project GeomScale which implements various geometric random walk samplers. We replicated the investment guidelines of the MSCI Diversified Multi-Factor index, a well-known index in the industry. This index aims to harvest sources of active returns by tilting the allocation of the capitalization-weighted parent index towards the four factors: value, quality, size, and momentum, while respecting bounds on risk, weights, and factor exposures relative to the parent index.

Our study revealed diverse impacts of DMF index constraints on factor-exposure-performance relations, compared to an otherwise unconstrained long-only framework. Momentum maintained a positive relation to returns, nearly as strong as in the long-only case. Value showed a diminishing positive relation, while quality's positive relation vanished under constraints. Size exhibited a negative relation, which intensified with constraints. Regarding risk, there was a truncation of the initial U-shaped relation, with momentum, value, and size showing increased volatility with higher factor exposures, while quality exhibited a negative correlation with volatility.

In summary, tilts towards value and momentum offered higher returns with increased volatility, quality tilts reduced risk without affecting returns, and size tilts led to lower returns and higher risk. Our findings suggest that factor patterns persist even within stringent constraints, justifying an active investment approach to capitalize on the positive ones.

\bibliography{references.bib}

\begin{thebibliography}{100}

\bibitem{bib:BAI1968}
Measuring the investment performance of pension funds for the purpose of
  inter-fund comparison.
\newblock {\em Bank Administration Institute}, 1968.

\bibitem{afshar2015}
H.M. Afshar and J.~Domke.
\newblock {Reflection, refraction, and Hamiltonian Monte Carlo}.
\newblock In {\em Advances in neural information processing systems}, pages
  3007--3015, 2015.

\bibitem{bib:Ali1973}
M.~M. Ali.
\newblock Content of the frustum of a simplex.
\newblock {\em Pacific Journal of Mathematics}, 49:313--322, 1973.

\bibitem{Anstreicher02}
K.~M. Anstreicher.
\newblock Improved complexity for maximum volume inscribed ellipsoids.
\newblock {\em SIAM Journal on Optimization}, 13(2):309--320, 2002.

\bibitem{bib:ArnottHsuKalesnikTindall2013}
R.~Arnott, J.~Hsu, V.~Kalesnik, and P.~Tindall.
\newblock The surprising alpha from malkiel's monkey and upside-down
  strategies.
\newblock {\em The Journal of Portfolio Management}, 4:91--105, 2013.

\bibitem{bib:AsnessFrazziniPedersen2019}
C.~S. Asness, A.~Frazzini, and L.~H. Pedersen.
\newblock Quality minus junk.
\newblock {\em Review of Accounting Studies}, 24:34--112, 2019.

\bibitem{bib:Bachelard2024}
C.~Bachelard.
\newblock On the geometry of the bootstrap.
\newblock {\em Unpublished Manuscript}, 2024.

\bibitem{bib:BachelardEtAl2023}
C.~Bachelard, A.~Chalkis, and V.~Fisikopoulos~E. Tsigaridas.
\newblock Randomized geometric tools for anomaly detection in stock markets.
\newblock In Francisco Ruiz, Jennifer Dy, and Jan-Willem van~de Meent, editors,
  {\em Proceedings of The 26th International Conference on Artificial
  Intelligence and Statistics}, volume 206 of {\em Proceedings of Machine
  Learning Research}, pages 9400--9416. PMLR, 25--27 Apr 2023.

\bibitem{bib:Banz1981}
R.~W. Banz.
\newblock The relationship between return and market value of common stocks.
\newblock {\em Journal of Financial Economics}, 9:3--18, 1981.

\bibitem{Berbee1987}
H.~C.~P. Berbee, C.~G.~E. Boender, A.~H.~G. Rinnooy~Ran, C.~L. Scheffer, R.~L.
  Smith, and J.~Telgen.
\newblock Hit-and-run algorithms for the identification of nonredundant linear
  inequalities.
\newblock {\em Mathematical Programming}, 37(2):184--207, Jun 1987.

\bibitem{Bertsimas04}
D.~Bertsimas and S.~Vempala.
\newblock Solving convex programs by random walks.
\newblock {\em J. ACM}, 51(4):540–556, July 2004.

\bibitem{bib:BickelGotzeVanZwet1997}
P.~J. Bickel, F.~Götze, and W.~R. van Zwet.
\newblock Resampling fewer than n observations.
\newblock {\em Statistica Sinica}, 7:1--31, 1997.

\bibitem{bib:BillioEtAl2011}
M.~Billio, L.~Calès, and D.~Guegan.
\newblock A cross-sectional score for the relative performance of an
  allocation.
\newblock {\em International Review of Applied Financial Issues and Economics},
  3, 2011.

\bibitem{bib:BrinsonFachler1985}
G.~P. Brinsan and N.~Fachler.
\newblock Measuring non-us. equity portfolio performance.
\newblock 11:73--76, 1985.

\bibitem{bib:BrinsonHoodBeebower1986}
G.~P. Brinsan, L.~R.~Hood Jr., and G.~L. Beebower.
\newblock Determinants of portfolio performance.
\newblock {\em Financial Analysts Journal}, 41:39--44, 1986.

\bibitem{bib:Burns2007}
P.~Burns.
\newblock Random portfolios for performance measurement.
\newblock In E.~Kontoghiorghes and C.~Gatu, editors, {\em Optimisation,
  Econometric, and Financial Analysis}, pages 227--249, New York, 2007.
  Springer.

\bibitem{Smith93}
C.~J.~P. Bélisle, H.~E. Romeijn, and R.~L. Smith.
\newblock Hit-and-run algorithms for generating multivariate distributions.
\newblock {\em Mathematics of Operations Research}, 18(2):255--266, 1993.

\bibitem{bib:CalesChalkisEmiris2021}
L.~Cales, A.~Chalkis, and I.~Emiris.
\newblock On the cross-sectional distribution of portfolios' returns.
\newblock {\em arXiv:2105.06573}, 2021.

\bibitem{bib:CalesChalkisEmirisFisikopoulos2018}
L.~Cales, A.~Chalkis, I.~Emiris, and V.~Fisikopoulos.
\newblock Practical volume computation of structured convex bodies, and an
  application to modeling portfolio dependencies and financial crises.
\newblock 2018.

\bibitem{bib:Carhart1997}
M.~M. Carhart.
\newblock On persistence in mutual fund performance.
\newblock {\em The Journal of Finance}, 52:57--82, 1997.

\bibitem{bib:ChalkisEtAl2021}
A.~Chalkis, E.~Christoforou, I.~Z. Emiris, and T.~Dalamagas.
\newblock Modeling asset allocation strategies and a new portfolio performance
  score.
\newblock {\em Digital Finance}, 3:333--371, 2021.

\bibitem{Chalkis23}
A.~Chalkis, I.~Z. Emiris, and V.~Fisikopoulos.
\newblock A practical algorithm for volume estimation based on billiard
  trajectories and simulated annealing.
\newblock {\em ACM J. Exp. Algorithmics}, 28, may 2023.

\bibitem{Chalkis22}
A.~Chalkis, I.~Z. Emiris, V.~Fisikopoulos, P.~Repouskos, and E.~Tsigaridas.
\newblock Efficient sampling in spectrahedra and volume approximation.
\newblock {\em Linear Algebra and its Applications}, 648:205--232, 2022.

\bibitem{Chalkis21}
A.~Chalkis and V.~Fisikopoulos.
\newblock {volesti: Volume Approximation and Sampling for Convex Polytopes in
  R}.
\newblock {\em {The R Journal}}, 13(2):642--660, 2021.

\bibitem{Chalkis23b}
A.~Chalkis, V.~Fisikopoulos, M.~Papachristou, and E.~Tsigaridas.
\newblock Truncated log-concave sampling for convex bodies with reflective
  hamiltonian monte carlo.
\newblock {\em ACM Trans. Math. Softw.}, 49(2), jun 2023.

\bibitem{volesti24}
A.~Chalkis, V.~Fisikopoulos, and E.~Tsigaridas.
\newblock volesti.
\newblock \url{https://github.com/GeomScale/volesti}, 2024.

\bibitem{ChalkisMetabolic21}
A.~Chalkis, V.~Fisikopoulos, E.~Tsigaridas, and H.~Zafeiropoulos.
\newblock {Geometric Algorithms for Sampling the Flux Space of Metabolic
  Networks}.
\newblock In Kevin Buchin and \'{E}ric Colin~de Verdi\`{e}re, editors, {\em
  37th International Symposium on Computational Geometry (SoCG 2021)}, volume
  189 of {\em Leibniz International Proceedings in Informatics (LIPIcs)}, pages
  21:1--21:16, Dagstuhl, Germany, 2021. Schloss Dagstuhl -- Leibniz-Zentrum
  f{\"u}r Informatik.

\bibitem{ChePioCaz18}
A.~Chevallier, S.~Pion, and F.~Cazals.
\newblock {Hamiltonian Monte Carlo with boundary reflections, and application
  to polytope volume calculations}.
\newblock Research Report RR-9222, {INRIA Sophia Antipolis, France}, 2018.

\bibitem{bib:Chiarawongse2012}
A.~Chiarawongse, S.~Kiatsupaibul, S.~Tirapat, and B.~Van Roy.
\newblock Portfolio selection with qualitative input.
\newblock {\em Journal of Banking \& Finance}, 36:489--496, 2012.

\bibitem{bib:ClareMotsonThomas2013}
A.~Clare, N.~Motson, and S.~Thomas.
\newblock An evaluation of alternative equity indices - part 2: Fundamental
  weighting schemes.
\newblock {\em Cass Consulting, March 2013}, 2013.

\bibitem{bib:CohenPogue1967}
K.~J. Cohen and J.~A. Pogue.
\newblock An empirical evaluation of alternative portfolio-selection models.
\newblock {\em The Journal of Business}, 40:166--193, 1967.

\bibitem{Cohen19}
M.~B. Cohen, B.~Cousins, Y.~T. Lee, and X.~Yang.
\newblock A near-optimal algorithm for approximating the john ellipsoid.
\newblock In Alina Beygelzimer and Daniel Hsu, editors, {\em Proceedings of the
  Thirty-Second Conference on Learning Theory}, volume~99 of {\em Proceedings
  of Machine Learning Research}, pages 849--873. PMLR, 25--28 Jun 2019.

\bibitem{CousinsThesis17}
B.~Cousins.
\newblock {\em Efficient high-dimensional sampling and integration}.
\newblock PhD thesis, Georgia Institute of Technology, Georgia, U.S.A., 2017.

\bibitem{CousinsV14}
B.~Cousins and S.~Vempala.
\newblock Bypassing {KLS: Gaussian} cooling and an ${O}^*(n^3)$ volume
  algorithm.
\newblock In {\em Proc. ACM STOC}, pages 539--548, 2015.

\bibitem{bib:CousinsVempala2016}
B.~Cousins and S.~Vempala.
\newblock A practical volume algorithm.
\newblock {\em Mathematical Programming Computation}, 8:133--160, 2016.

\bibitem{bib:DawsonYoung2003}
R.~Dawson and R.~Young.
\newblock Near-uniformly distributed, stochastically generated portfolios.
\newblock 2003.

\bibitem{bib:LopezDePrado2022}
M.~Lopez de~Prado.
\newblock Causal factor investing: Can factor investing become scientific?
\newblock {\em ADIA Research Lab}, 2022.

\bibitem{bib:Dietz1966}
P.~O. Dietz.
\newblock Pension funds: Measuring investment performance.
\newblock 1966.

\bibitem{DyerFrKa91}
M.~Dyer, A.~Frieze, and R.~Kannan.
\newblock A random polynomial-time algorithm for approximating the volume of
  convex bodies.
\newblock {\em J. ACM}, 38(1):1--17, 1991.

\bibitem{bib:Efron1979}
B.~Efron.
\newblock The bootstrap: Another look at the jacknife.
\newblock {\em The Annals of Statistics}, 7:1--26, 1979.

\bibitem{Emiris14}
{I.Z.} Emiris and V.~Fisikopoulos.
\newblock Practical polytope volume approximation.
\newblock {\em ACM Transactions of Mathematical Software, 2018},
  44(4):38:1--38:21, 2014.
\newblock Conf. version: Proc. Symp. Comp. Geometry, 2014.

\bibitem{bib:FamaFrench1993}
E.~F. Fama and K.~R. French.
\newblock Common risk factors in the returns on stocks and bonds.
\newblock {\em Journal of Financial Economics}, 33:3--56, 1993.

\bibitem{bib:FamaFrench2015}
E.~F. Fama and K.~R. French.
\newblock A five-factor asset pricing model.
\newblock {\em Journal of Financial Economics}, 116:1--22, 2015.

\bibitem{bib:FamaMacBeth1973}
E.~F. Fama and J.~D. MacBeth.
\newblock Risk, return, and equilibrium: Empirical tests.
\newblock {\em Journal of Political Economy}, 81:607--636, 1973.

\bibitem{bib:FrigyikGuptaChen2010}
B.~Frigyik, M.~Gupta, and Y.~Chen.
\newblock Shadow dirichlet for restricted probability modeling.
\newblock {\em Advances in Neural Information Processing Systems 23}, pages
  613--621, 2010.

\bibitem{eigenweb}
Ga\"{e}l G., Beno\^{i}t J., et~al.
\newblock {\em {Eigen} v3}, 2010.

\bibitem{HitAndRunCranPkg}
{G. van Valkenhoef, T. Tervonen}.
\newblock {\em hitandrun: "Hit and Run" and "Shake and Bake" for Sampling
  Uniformly from Convex Shapes}.
\newblock R Foundation for Statistical Computing, 2022.

\bibitem{Gelman92}
A.~Gelman and D.~B. Rubin.
\newblock Inference from {Iterative} {Simulation} {Using} {Multiple}
  {Sequences}.
\newblock {\em Statistical Science}, 7(4):457--472, 1992.
\newblock Publisher: Institute of Mathematical Statistics.

\bibitem{Girolami11}
M.~Girolami and B.~Calderhead.
\newblock Riemann manifold langevin and hamiltonian monte carlo methods.
\newblock {\em Journal of the Royal Statistical Society: Series B (Statistical
  Methodology)}, 73(2):123--214, 2011.

\bibitem{Gretton12}
A.~Gretton, K.~M. Borgwardt, M.~J. Rasch, B.~Sch{{\"o}}lkopf, and A.~Smola.
\newblock A kernel two-sample test.
\newblock {\em Journal of Machine Learning Research}, 13(25):723--773, 2012.

\bibitem{bib:GryazinaPolyak2014}
E.~Gryazina and B.~Polyak.
\newblock Random sampling: Billiard walk algorithm.
\newblock {\em European Journal of Operational Research}, 238:497--504, 2014.

\bibitem{bib:GustafsonNarayanan2018}
A.~Gustafson and H.~Narayanan.
\newblock John's walk.
\newblock {\em arXiv preprint arXiv:1803.02032}, page~NA, 2018.

\bibitem{haraldsdottir2017chrr}
H.~S. Haraldsd{\'o}ttir, B.~Cousins, I.~Thiele, R.~M.~T. Fleming, and
  S.~Vempala.
\newblock {CHRR}: coordinate hit-and-run with rounding for uniform sampling of
  constraint-based models.
\newblock {\em Bioinformatics}, 33(11):1741--1743, 2017.

\bibitem{Hastings70}
W.~K. Hastings.
\newblock Monte carlo sampling methods using markov chains and their
  applications.
\newblock {\em Biometrika}, 57(1):97--109, 1970.

\bibitem{cobraToolbox}
L.~Heirendt, S.~Arreckx, T.~Pfau, S.~N. Mendoza, A.~Richelle, A.~Heinken, H.~S.
  Haraldsdottir, J.~Wachowiak, S.~M. Keating, V.~Vlasov, S.~Magnusdottir, C.~Y.
  Ng, G.~Preciat, A.~Zagare, S.~H.J. Chan, M.~K. Aurich, C.~M. Clancy,
  J.~Modamio, J.~T. Sauls, A.~Noronha, A.~Bordbar, B.~Cousins, D.~C.~El Assal,
  L.~V. Valcarcel, I.~Apaolaza, S.~Ghaderi, M.~Ahookhosh, M.~B. Guebila,
  A.~Kostromins, N.~Sompairac, H.~M. Le, D.~Ma, Y.~Sun, L.~Wang, J.~T.
  Yurkovich, M.~A.P. Oliveira, P.~T. Vuong, L.~P.~El Assal, I.~Kuperstein,
  A.~Zinovyev, H.~S. Hinton, W.~A. Bryant, F.~J.~A. Artacho, F.~J. Planes,
  E.~Stalidzans, A.~Maass, S.~Vempala, M.~Hucka, M.~A. Saunders, C.~D. Maranas,
  N.~E. Lewis, T.~Sauter, B.~Ø. Palsson, I.~Thiele, and R.~M.T. Fleming.
\newblock Creation and analysis of biochemical constraint-based models: The
  cobra toolbox v3.0.
\newblock {\em Nature Protocols}, 14:639--702, 2019.

\bibitem{Gelman14}
M.~D. Hoffman and A.~Gelman.
\newblock The no-u-turn sampler: Adaptively setting path lengths in hamiltonian
  monte carlo.
\newblock {\em J. Mach. Learn. Res.}, 15(1):1593–1623, January 2014.

\bibitem{bib:HuangMehrotra2013}
K-L. Huang and S.~Mehrotra.
\newblock An empirical evaluation of walk-and-round heuristics for mixed
  integer linear programs.
\newblock {\em Computational Optimization and Applications}, 55:545--570, 2013.

\bibitem{john48}
Fritz J.
\newblock Extremum {Problems} with {Inequalities} as {Subsidiary} {Conditions}.
\newblock In Giorgio Giorgi and Tinne~Hoff Kjeldsen, editors, {\em Traces and
  {Emergence} of {Nonlinear} {Programming}}, pages 197--215. Springer, Basel,
  2014.

\bibitem{HOPS20}
J.~F Jadebeck, A.~Theorell, S.~Leweke, and K.~Nöh.
\newblock {HOPS: high-performance library for (non-)uniform sampling of
  convex-constrained models}.
\newblock {\em Bioinformatics}, 37(12):1776--1777, 10 2020.

\bibitem{bib:JegadeeshTitman1993}
N.~Jegadeesh and S.~Titman.
\newblock Returns to buying winners and selling losers: Implications for stock
  market efficiency.
\newblock {\em The Journal of Finance}, 48:65--91, 1993.

\bibitem{bib:JensenKellyPedersen2023}
T.~I. Jensen, B.~Kelly, and L.~H. Pedersen.
\newblock Is there a replication crisisi in finance?
\newblock {\em The Journal of Finance}, 78:2465--2518, 2023.

\bibitem{Jia21}
H.~Jia, A.~Laddha, Y.~T. Lee, and S.~Vempala.
\newblock Reducing isotropy and volume to {KLS}: An ${O}^*(n^3\psi^2)$ volume
  algorithm.
\newblock In {\em Proceedings of the 53rd Annual ACM SIGACT Symposium on Theory
  of Computing}, STOC 2021, page 961–974, New York, NY, USA, 2021.
  Association for Computing Machinery.

\bibitem{bib:JobsonKorkie1981}
J.~D. Jobson and B.~M. Korkie.
\newblock Performance hypothesis testing with the sharpe and treynor measures.
\newblock {\em The Journal of Finance}, 36:889--908, 1981.

\bibitem{Kalai06}
A.~T. Kalai and S.~Vempala.
\newblock Simulated annealing for convex optimization.
\newblock {\em Mathematics of Operations Research}, 31(2):253--266, 2006.

\bibitem{bib:KannanLovaszSimonovits1997}
R.~Kannan, L.~Lov\'{a}sz, and M.~Simonovits.
\newblock Random walks and an {O}($n^5$) volume algorithm for convex bodies.
\newblock {\em Random Structures and Algorithms}, 11:1 -- 50, 1997.

\bibitem{bib:KannanNarayanan2012}
R.~Kannan and H.~Narayanan.
\newblock Random walks on polytopes and an affine interior point method for
  linear programming.
\newblock {\em Mathematics of Operations Research}, 37:1 -- 20, 2012.

\bibitem{bib:KasakPohlmeier2019}
E.~Kasak and W.~Pohlmeier.
\newblock Testing out-of-sample portfolio performance.
\newblock {\em International Journal of Forecasting}, 35:540--554, 2019.

\bibitem{kaufman98}
D.~E. Kaufman and R.~L. Smith.
\newblock Direction choice for accelerated convergence in hit-and-run sampling.
\newblock {\em Operations Research}, 46(1):84--95, 1998.

\bibitem{Khachiyan93}
L.~G. Khachiyan and M.~J. Todd.
\newblock On the complexity of approximating the maximal inscribed ellipsoid
  for a polytope.
\newblock {\em Mathematical Programming}, 61(1):137--159, August 1993.

\bibitem{bib:KimLee2016}
W.~C. Kim and Y.~Lee.
\newblock A uniformly distributed random portfolio.
\newblock {\em Quantitative Finance}, 16:297--307, 2016.

\bibitem{RiHMCsampler}
Y.~Kook, Y.~T. Lee, R.~Shen, and S.~Vempala.
\newblock Polytopesamplermatlab.
\newblock \url{https://github.com/ConstrainedSampler/PolytopeSamplerMatlab},
  2021.

\bibitem{kook22}
Y.~Kook, Y.~T. Lee, R.~Shen, and S.~Vempala.
\newblock Sampling with riemannian hamiltonian monte carlo in a constrained
  space.
\newblock In S.~Koyejo, S.~Mohamed, A.~Agarwal, D.~Belgrave, K.~Cho, and A.~Oh,
  editors, {\em Advances in Neural Information Processing Systems}, volume~35,
  pages 31684--31696. Curran Associates, Inc., 2022.

\bibitem{kook23}
Y.~Kook, Y.~T. Lee, R.~Shen, and S.~Vempala.
\newblock Condition-number-independent convergence rate of riemannian
  hamiltonian monte carlo with numerical integrators.
\newblock In Gergely Neu and Lorenzo Rosasco, editors, {\em Proceedings of
  Thirty Sixth Conference on Learning Theory}, volume 195 of {\em Proceedings
  of Machine Learning Research}, pages 4504--4569. PMLR, 12--15 Jul 2023.

\bibitem{Kumar05}
P.~Kumar and E.~A. Yildirim.
\newblock Minimum-volume enclosing ellipsoids and core sets.
\newblock {\em Journal of Optimization Theory and Applications}, 126(1):1--21,
  Jul 2005.

\bibitem{Laddha20b}
A.~Laddha, Y.~T. Lee, and S.~Vempala.
\newblock Strong self-concordance and sampling.
\newblock In {\em Proceedings of the 52nd Annual ACM SIGACT Symposium on Theory
  of Computing}, STOC 2020, page 1212–1222, New York, NY, USA, 2020.
  Association for Computing Machinery.

\bibitem{Laddha23}
A.~Laddha and S.~Vempala.
\newblock Convergence of gibbs sampling: Coordinate hit-and-run mixes fast.
\newblock {\em Discrete {\&} Computational Geometry}, Apr 2023.

\bibitem{bib:Lasserre2015}
J.~B. Lasserre.
\newblock Volume of slices and sections of the simplex in closed form.
\newblock {\em Optimization Letters}, 9:1263--1269, 2015.

\bibitem{bib:LedoitWolf2008}
O.~Ledoit and M.~Wolf.
\newblock Robust performance hypothesis testing with the sharpe ratio.
\newblock {\em Journal of Empirical Finance}, 15:850--859, 2008.

\bibitem{bib:LeeEtAl2018}
Y.~Lee, D-G Kwon, W.~C. Kim, and F.~J. Fabozzi.
\newblock An alternative approach for portfolio performance evaluation:
  enabling fund evaluation relative to peer group via malkiel's monkey.
\newblock {\em Applied Economics}, 50:4318--4327, 2018.

\bibitem{lee14}
Y.~T. Lee and A.~Sidford.
\newblock Path {Finding} {Methods} for {Linear} {Programming}: {Solving}
  {Linear} {Programs} in Õ(vrank) {Iterations} and {Faster} {Algorithms} for
  {Maximum} {Flow}.
\newblock In {\em 2014 {IEEE} 55th {Annual} {Symposium} on {Foundations} of
  {Computer} {Science}}, pages 424--433, October 2014.
\newblock ISSN: 0272-5428.

\bibitem{Lee17}
Y.~T. {Lee} and S.~{Vempala}.
\newblock Eldan's stochastic localization and the {KLS} hyperplane conjecture:
  An improved lower bound for expansion.
\newblock In {\em 2017 IEEE 58th Annual Symposium on Foundations of Computer
  Science (FOCS)}, pages 998--1007, Oct 2017.

\bibitem{LeeVemGeodesic}
Y.~T. Lee and S.~Vempala.
\newblock Geodesic walks in polytopes.
\newblock In {\em Proceedings of the 49th Annual ACM SIGACT Symposium on Theory
  of Computing}, STOC 2017, pages 927--940, New York, NY, USA, 2017. ACM.

\bibitem{Lee18c}
Y.~T. Lee and S.~Vempala.
\newblock Convergence rate of riemannian hamiltonian monte carlo and faster
  polytope volume computation.
\newblock In {\em Proceedings of the 50th Annual ACM SIGACT Symposium on Theory
  of Computing}, STOC 2018, page 1115–1121, New York, NY, USA, 2018.
  Association for Computing Machinery.

\bibitem{bib:LeeVempala2018}
Y.~T. Lee and S.~Vempala.
\newblock The {K}annan-{L}ovasz-{S}imonovits conjecture.
\newblock {\em arXiv preprint arXiv:1807.03465}, page~NA, 2018.

\bibitem{bib:Liang1999}
B.~Liang.
\newblock Price pressure: Evidence from the dartboard column.
\newblock {\em The Journal of Business}, 72:119--134, 1999.

\bibitem{bib:Lintner1965}
J.~Lintner.
\newblock Security prices, risk and maximal gains from diversification.
\newblock {\em The Journal of Finance}, 20:587--615, 1965.

\bibitem{bib:Lisi2009}
F.~Lisi.
\newblock Dicing with the market: Randomized procedures for evaluation of
  mutual funds.
\newblock {\em Quantitative Finance}, 11:163--172, 2009.

\bibitem{bib:LovaszSimonovits1990}
L.~Lov\'{a}sz and M.~Simonovits.
\newblock Mixing rate of markov chains, an isoperimetric inequality, and
  computing volume.
\newblock {\em Proceedings 31st Annual Symposium on Foundations of Computer
  Science, 22.-24. Oct. 1990}, pages 346--355, 1990.

\bibitem{bib:LovaszSimonovits1993}
L.~Lov\'{a}sz and M.~Simonovits.
\newblock Random walks in a convex body and an improved volume algorithm.
\newblock {\em Random Structures and Algorithms}, 4:359--412, 1993.

\bibitem{bib:LovaszVempala2003}
L.~Lov\'{a}sz and S.~Vempala.
\newblock Hit and run is fast and fun.
\newblock {\em Technical Report MSR-TR-2003-05, Microsoft Research, Redmond.},
  page~NA, 2003.

\bibitem{LovaszVempala03}
L.~Lov\'{a}sz and S.~Vempala.
\newblock Logconcave functions: geometry and efficient sampling algorithms.
\newblock In {\em 44th Annual IEEE Symposium on Foundations of Computer
  Science, 2003. Proceedings.}, pages 640--649, 2003.

\bibitem{bib:LovaszVempala2006}
L.~Lov\'{a}sz and S.~Vempala.
\newblock Fast algorithms for logconcave functions: Sampling, rounding,
  integration and optimization.
\newblock {\em Proceedings 47th Annual IEEE Symposium on Foundations of
  Computer Science, 21- 24. Oct 2006}, 2006.

\bibitem{Lovasz06}
L.~Lov\'{a}sz and S.~Vempala.
\newblock Hit-and-run from a corner.
\newblock {\em SIAM J. Comp.}, 35(4):985--1005, 2006.

\bibitem{LovVem}
L.~Lovász and S.~Vempala.
\newblock Simulated annealing in convex bodies and an {O}$^*(n^4)$ volume
  algorithms.
\newblock {\em J. Computer and System Sciences}, 72:392--417, 2006.

\bibitem{bib:Malkiel1973}
B.~G. Malkiel.
\newblock A random walk down wall street.
\newblock 1973.

\bibitem{Mangoubi17}
O.~Mangoubi and A.~Smith.
\newblock Rapid mixing of hamiltonian monte carlo on strongly log-concave
  distributions, 2017.

\bibitem{Mangoubi19}
O.~Mangoubi and N.K. Vishnoi.
\newblock Faster algorithms for polytope rounding, sampling, and volume
  computation via a sublinear "ball walk".
\newblock In {\em FOCS}, 2019.

\bibitem{boostrandom}
J.~Maurer and S.~Watanabe.
\newblock Boost random number library.
\newblock Software, 2017.

\bibitem{Megchelenbrink14}
W.~Megchelenbrink, M.~Huynen, and E.~Marchiori.
\newblock optgpsampler: An improved tool for uniformly sampling the
  solution-space of genome-scale metabolic networks.
\newblock {\em PLOS ONE}, 9(2):1--8, 02 2014.

\bibitem{bib:Memmel2003}
C.~Memmel.
\newblock Performance hypothesis testing with the sharpe ratio.
\newblock {\em Finance Letters}, 1:21--23, 2003.

\bibitem{bib:MetcalfMalkiel1994}
G.~E. Metcalf and B.~G. Malkiel.
\newblock The wall street journal contests: the experts, the darts, and the
  efficient market hypothesis.
\newblock {\em Applied Financial Economics}, 4:371--374, 1994.

\bibitem{bib:Mossin1966}
J.~Mossin.
\newblock Equilibrium in a capital asset market.
\newblock {\em Econometrica}, 34:768--783, 1966.

\bibitem{narayanan16}
H.~Narayanan.
\newblock Randomized interior point methods for sampling and optimization.
\newblock {\em Annals of Applied Probability}, 26(1):597--641, February 2016.
\newblock Publisher: Institute of Mathematical Statistics.

\bibitem{Narayanan20}
H.~Narayanan and P.~Srivastava.
\newblock On the mixing time of coordinate hit-and-run, 2020.

\bibitem{Neal11}
R.~M. Neal.
\newblock {\em MCMC Using Hamiltonian Dynamics}, chapter~5.
\newblock CRC Press, 2011.

\bibitem{Nemirovski99}
A.~Nemirovski.
\newblock On self-concordant convex-concave functions.
\newblock {\em Optimization Methods \& Software - OPTIM METHOD SOFTW}, 11, 05
  1999.

\bibitem{bib:Nielsen2016}
S.~P. Nielsen.
\newblock {ll-ACHRB}: a {S}calable {A}lgorithm for {S}ampling the {F}easible
  {S}olution {S}pace of {M}etabolic {N}etworks.
\newblock {\em Bioinformatics}, 32:2330--2337, 2016.

\bibitem{bib:PettengillClark2001}
G.~N. Pettengill and J.~M. Clark.
\newblock Estimating expected returns in an event study framework: Evidence
  from the dartboard column.
\newblock {\em Quarterly Journal of Business and Economics}, 40:3--21, 2001.

\bibitem{bib:RosenbergReidLanstein1985}
B.~Rosenberg, K.~Reid, and R.~Lanstein.
\newblock ersuasive evidence of market inefficiency.
\newblock {\em The Journal of Portfolio Management}, 11:9--16, 1985.

\bibitem{Roy20}
V.~Roy.
\newblock {Convergence Diagnostics for Markov Chain Monte Carlo}.
\newblock {\em Annual Review of Statistics and Its Application}, 7(1):387--412,
  2020.

\bibitem{bib:Rubin1981}
R.~Rubin.
\newblock The bayesian bootstrap.
\newblock {\em The Annals of Statistics}, 9:130--134, 1981.

\bibitem{Rudelson99}
M.~Rudelson.
\newblock Random vectors in the isotropic position.
\newblock {\em J. Funct. Anal}, pages 60--72, 1999.

\bibitem{Saa16}
P.~A. Saa and L.~K. Nielsen.
\newblock {ll-ACHRB: A Scalable Algorithm for Sampling the Feasible Solution
  Space of Metabolic Networks}.
\newblock {\em Bioinformatics}, 32(15):2330--2337, 03 2016.

\bibitem{Sachdeva2016}
S.~Sachdeva and N.~K. Vishnoi.
\newblock The mixing time of the dikin walk in a polytope—a simple proof.
\newblock {\em Operations Research Letters}, 44(5):630--634, 2016.

\bibitem{bib:Sharpe1964}
W.~F. Sharpe.
\newblock Capital asset prices: A theory of market equilibrium under conditions
  of risk.
\newblock {\em The Journal of Finance}, 19:425--444, 1964.

\bibitem{bib:Sharpe1991}
W.~F. Sharpe.
\newblock The arithmetic of active management.
\newblock {\em The Financial Analysts' Journal}, 47:7 -- 9, 1991.

\bibitem{smith84}
R.~L. Smith.
\newblock Efficient monte carlo procedures for generating points uniformly
  distributed over bounded regions.
\newblock {\em Operations Research}, 32(6):1296--1308, 1984.

\bibitem{Song2022}
Z.~Song, X.~Yang, Y.~Yang, and T.~Zhou.
\newblock Faster algorithm for structured john ellipsoid computation, 2022.

\bibitem{bib:Stein2014}
R.~Stein.
\newblock Not fooled by randomness: Using random portfolios to analyse
  investment funds.
\newblock {\em Investment Analysts Journal}, 43:1--15, 2014.

\bibitem{bib:Surz1994}
R.~J. Surz.
\newblock Portfolio opportunity distributions.
\newblock {\em The Journal of Investing}, 3:36--41, 1994.

\bibitem{bib:Surz2006}
R.~J. Surz.
\newblock A fresh look at investment performance evaluation: Unifying best
  practices to improve timeliness and reliability.
\newblock {\em The Journal of Portfolio Management}, 32:54--65, 2006.

\bibitem{bib:TianEtAl2009}
G.-L. Tian, H.-B. Fang, M.~Tan, H.~Qin, and M.-L. Tang.
\newblock Uniform distributions in a class of convex polyhedrons with
  applications to drug combination studies.
\newblock {\em Journal of Multivariate Analysis}, 100:1854--1865, 2009.

\bibitem{Todd07}
M.~J. Todd and E.~A. Yıldırım.
\newblock On khachiyan's algorithm for the computation of minimum-volume
  enclosing ellipsoids.
\newblock {\em Discrete Applied Mathematics}, 155(13):1731--1744, 2007.

\bibitem{vaidya96}
P.~M. Vaidya.
\newblock A new algorithm for minimizing convex functions over convex sets
  {\textbar} {SpringerLink}, 1996.

\bibitem{MeerscheXsample}
K.~Van~den Meersche, K.~Soetaert, and D.~Van~Oevelen.
\newblock xsample(): An r function for sampling linear inverse problems.
\newblock {\em Journal of Statistical Software, Code Snippets}, 30(1):1–15,
  2009.

\bibitem{bib:Varsi1973}
G.~Varsi.
\newblock The multidimensional content of the frustum of the simplex.
\newblock {\em Pacific Journal of Mathematics}, 46:303--314, 1973.

\bibitem{bib:Vempala2005}
S.~Vempala.
\newblock Geometric random walks: A survey.
\newblock {\em Combinatorial and Computational Geometry}, 52:577--616, 2005.

\bibitem{walkr20}
A.~Yao.
\newblock walkr.
\newblock \url{https://github.com/andyyao95/walkr}, 2020.

\bibitem{Chen21}
C.~Yuansi.
\newblock An almost constant lower bound of the isoperimetric coefficient in
  the {KLS} conjecture.
\newblock {\em Geometric and Functional Analysis}, 31(1):34--61, Feb 2021.

\bibitem{chen_vaidya_17}
C.~Yuansi, R.~Dwivedi, M.~J. Wainwright, and Y.~Bin.
\newblock Vaidya walk: {A} sampling algorithm based on the volumetric barrier.
\newblock In {\em 2017 55th {Annual} {Allerton} {Conference} on
  {Communication}, {Control}, and {Computing} ({Allerton})}, pages 1220--1227,
  October 2017.

\bibitem{chen18}
C.~Yuansi, R.~Dwivedi, M.~J. Wainwright, and Y.~Bin.
\newblock Fast {MCMC} {Sampling} {Algorithms} on {Polytopes}.
\newblock {\em Journal of Machine Learning Research}, 19(55):1--86, 2018.

\bibitem{PolytopeWalk21}
C.~Yuansi, R.~Dwivedi, M.~J. Wainwright, and Y.~Bin.
\newblock Polytopewalk.
\newblock \url{https://github.com/yuachen/polytopewalk}, 2020.

\bibitem{Chen2022}
C.~Yuansi and R.~Eldan.
\newblock Hit-and-run mixing via localization schemes, 2022.

\bibitem{Zhang03}
Y.~Zhang and L.~Gao.
\newblock On numerical solution of the maximum volume ellipsoid problem.
\newblock {\em SIAM Journal on Optimization}, 14(1):53--76, 2003.

\end{thebibliography}

\newpage
\appendix
\section{Appendix}
\label{sec:appendix}
\hfill

\subsection{Appendix A}
\label{sec:appendix_varsi}
\hfill \break

\begin{algorithm}[!h]
	\caption{Varsi's algorithm}
	\label{algo:varsi}
	Compute $u_i = z_i - \gamma$, $i=1, ...,n$.\\
	Label non-negative $u_i$ as $u_1^+, ..., u_F^+$ and negatives as $u_1^-, ..., u_H^-$.\\
	Initialize $A_0=1, A_1 = A_2 = ... = A_F = 0$.\\
	For $h=1,...,H$ repeat $A_f = \frac{u_f^+ A_f - u_h^- A_{f-1}}{u_f^+ - u_h^-}$, $f = 1,...,F$.\\
	Then, for $h=H$, $A_f = \frac{V(\mathcal{S}^{n-1} \cap H(z, \gamma))}{V(\mathcal{S}^{n-1})}$.	
\end{algorithm}

\subsection{Appendix B}
\label{sec:appendix_factor_mapping}
\hfill \break

\begin{table}[ht]
\centering
\begin{tabular}{|c|c|}
\hline
\textbf{MSCI factor} & \textbf{JKP factor}\\
\hline
Book-to-Price & Book-to-market equity\\
\hline
Earnings Yield & Earnings-to-price\\
\hline
Earnings Quality & Total accruals\\
\hline
Investment Quality & Return on equity\\
\hline
Profitability & Gross profits-to-assets\\
\hline
Momentum & Price momentum t-12 to t-1\\
\hline
Earnings Variability & Earnings variability\\
\hline
Leverage & Book leverage\\
\hline
Size & Market Equity\\
\hline
Beta & Market Beta\\
\hline
Residual Volatility & Idiosyncratic volatility from the CAPM (252 days)\\
\hline
Growth & Sales Growth (1 year)\\
\hline
Liquidity & Liquidity of market assets\\
\hline
\end{tabular}
\caption{Mapping of MSCI factors to JKP factors.}
\label{tab:mapping_msci_to_jkp}
\end{table}

\end{document}